# Construction and properties of a topological index for periodically driven time-reversal invariant 2D crystals

D. Carpentier [1], P. Delplace, M. Fruchart *, K. Gawędzki [2], C. Tauber

*Laboratoire de Physique, École Normale Supérieure de Lyon, Université de Lyon, 47 allée d'Italie, 69007 Lyon, France*

**Abstract**

We present mathematical details of the construction of a topological invariant for periodically driven two-dimensional lattice systems with time-reversal symmetry and quasienergy gaps, which was proposed recently by some of us. The invariant is represented by a gap-dependent $\mathbb{Z}_2$-valued index that is simply related to the Kane–Mele invariants of quasienergy bands but contains an extra information. As a byproduct, we prove new expressions for the two-dimensional Kane–Mele invariant relating the latter to Wess–Zumino amplitudes and the boundary gauge anomaly.

## 1. Introduction

A characteristic feature of gapped systems is that they have no low energy excitations. Yet, the interface between two gapped systems may host some kind of protected low-energy excitations, which are impossible to get rid of. This phenomenon, first encountered in the quantum Hall effect [29], happens to be ubiquitous and appears in domains such as mechanical systems [31], optics [22,18,20], microwave networks [2,21], cold atoms [25] or electrical networks [26,1]. It arises from the non-trivial topology of the ground state of the system, usually characterized by a topological invariant. In condensed matter, where such topological phases were first discovered, electronic systems can often be described by tight-binding models on an infinite lattice. The seminal paper by Kane and Mele [32] made apparent the importance of symmetries such as time reversal in the appearance of the so-called symmetry protected topological phases. This led to a classification of noninteracting fermionic systems according to those symmetries [45,47,30], the applicability of which exceeds the sole field of condensed matter. Topological insulators have also sparked interest from mathematicians: the $\mathbb{Z}_2$-valued Kane–Mele invariant was from the start identified in [32] as taking values in specific $K$-theory group, see [30,11], and understood as a geometric obstruction [12] as well as in terms of an equivariant cohomology [7].

* Corresponding author.
*E-mail address:* michel.fruchart@ens-lyon.fr (M. Fruchart).

[1] Member of CNRS.

[2] Researcher emeritus.

With the goal of achieving a better control of topological phase transitions in electronic systems, it was proposed to produce out-of-equilibrium (but stationary) topological phases through a periodic driving, such as the application of light, in semiconductor quantum wells [35] and graphene [33], following pioneering works on photoinduced properties of 2d electron gases [24,40]. With the idea in mind that a periodic driving will allow to simulate equilibrium phases, the same method was used to realize archetypal models of topological phases in artificial systems such as (shaken) cold atom lattices [25] and arrangements of helical waveguides [46] (in this system, the periodic evolution in time is replaced by a continuous periodic modulation in the third dimension of real space along which propagation occurs). Yet, out-of-equilibrium phases display richer topological features than equilibrium phases, as was first pointed out by Kitagawa et al. [28] and soon after observed in optical experiments [27]. A new framework to fully describe the topological properties of the unitary evolution of a two-dimensional periodically driven system without specific symmetry was developed by Rudner, Lindner, Berg and Levin [44]. In particular, this new characterization correctly accounted for the existence of chiral edge states at the boundary of a finite sample.

The aim of the present work is to characterize the topological properties of periodically forced systems that are invariant under time-reversal symmetry, namely of driven analogues of the equilibrium Kane–Mele two-dimensional insulators. A brief account of our results which focused on the physical aspects, backed by numerical simulations of a toy model, was recently published [5]. The present paper aims at providing mathematical proofs of our claims.

We study fermions on a two-dimensional lattice which are periodically driven in time. We assume that unspecified relaxation processes eventually lead to a steady state described by a unitary evolution. As we shall see, the family of evolution operators $(t, k) \mapsto U(t, k)$ of such a system over one period $T$ is more convenient to work with than the Hamiltonian $H(t, k)$. Here, $k$ is the quasimomentum, taking values in the Brillouin zone BZ assimilated with a two-dimensional torus and $t$ is the time. Topological insulators are remarkable examples of band insulators, and as such have an *energy gap* separating bands which hold specific topological properties. However, energy is not conserved in driven systems. In the replacement for the energy spectrum, a periodically driven system has a quasienergy spectrum, defined as the spectrum of the Floquet operator $U(T)$ describing the unitary evolution over one period. Pictorially, the system is coupled to a bath of energy quanta multiple of $\hbar\omega$, where $\omega = 2\pi/T$ is the driving angular frequency, so although the energy $E$ of the system is ill-determined, its value modulo $\hbar\omega$ remains well-defined. The quasienergy spectrum of a driven system, lying on the circle, is periodic: this is a fundamental difference with the energy spectrum of static systems. However, the quasienergy spectrum can still display bands separated by gaps and we are interested in such instances.

As shown by Rudner and collaborators [44], the periodicity of the quasienergy spectrum can lead to situations where the usual invariants, such as the first Chern numbers of the bands, fail to completely characterize the topological features of the system. To overcome this difficulty, they constructed an integer-valued invariant $W_\epsilon[U]$, where $\epsilon$ designates a quasienergy gap, capturing the additional topological information contained in the unitary evolution of a periodically driven system without specific symmetry. The construction uses a gap-dependent periodization of the family $U(t,k)$ of evolution operators. However, when the evolution is time-reversal invariant then $W_\epsilon[U]$ always vanishes. This is in analogy to the static situation, where the first Chern numbers of bands vanish in a time-reversal invariant system and have to be replaced by the finer Kane–Mele invariant [32]. It is therefore natural to expect that in a periodically forced time-reversal invariant systems the $W_\epsilon[U]$ index would also be replaced by a finer invariant.

As will be shown in the present paper, such an invariant, that we denote $K_\epsilon[U]$, indeed exists [5]. It is a $\mathbb{Z}_2$-valued quantity, as the Kane–Mele index. Besides, $K_\epsilon[U]$ is linked to the Kane–Mele invariant of the quasienergy band $\mathcal{E}_{\epsilon,\epsilon'}$ delimited by the gaps around $\epsilon$ and $\epsilon'$ by the relation

$$K_{\epsilon'}[U] - K_\epsilon[U] = \mathrm{KM}(\mathcal{E}_{\epsilon,\epsilon'}) \tag{1.1}$$

analogous to the link shown in [44] between the $W$ index and the first Chern number of the quasienergy band in-between the two gaps for the case without time-reversal symmetry. Finally, the proof of Eq. (1.1) sheds a new light on the Kane–Mele index of the vector bundle $\mathcal{E}$ of states spanned by a family of projectors $P(k)$, relating it to a carefully defined square root of the Wess–Zumino amplitude [49] of the associated family of unitary operator $U_P(k) = I - 2P(k)$. The relation takes the form of the identity

$$(-1)^{\mathrm{KM}[\mathcal{E}]} = \Big(\exp\left[\mathrm{i}S_{\mathrm{WZ}}(U_P)\right]\Big)^{1/2}. \tag{1.2}$$

Those three claims: the existence of index $K_\epsilon[U]$ as a $\mathbb{Z}_2$-valued quantity, its relation to the Kane–Mele invariant, and the links between the Kane–Mele invariant and Wess–Zumino amplitudes, are the subject of this paper.

In Section 2 that follows Introduction, we review the useful bits of the Floquet theory and of the homotopy theory which are needed to construct the invariants $W_\epsilon[U]$ of [44] (without particular symmetry) and $K_\epsilon[U]$ of [5] (with time-reversal invariance). The implications of time-reversal symmetry on the Floquet theory needed in the second case are briefly summarized. In Section 3, we actually define $K_\epsilon[U]$, show that it is well-determined as a $\mathbb{Z}_2$-valued quantity, and discuss its properties. A key point in the construction is that the periodized evolution operator at the half-period can be contracted to the identity while keeping its time-reversal symmetry, and that $K_\epsilon[U]$ does not depend on the choice of the contraction. We show the existence of time-reversal invariant contractions in Section 3.2. The arbitrariness in the choice of the contraction is the reason why $K_\epsilon[U]$ is a $\mathbb{Z}_2$-valued quantity rather than a $\mathbb{Z}$-valued one, as shown in Section 3.3. The rest of the paper is devoted to a proof of the link (1.1) between index $K_\epsilon[U]$ and the Kane–Mele invariant. Our proof, whose main part is exposed in Section 4, is based on a new representation for the Kane–Mele invariant (or, more precisely, for its form given by Fu and Kane in [9]), summarized by identity (1.2). It consists of showing that the right-hand side of that identity localizes on contributions from time-reversal invariant quasimomenta in the Brillouin torus. For illustration, we explain in Section 4.3 how this works in the case of a particular tight-binding model with both time-reversal and inversion symmetries. In the key step in the proof of representation (1.2), spread over Sections 4.4 to 4.7, we relate the right-hand side of Eq. (1.2), written as a specific square root of the Wess–Zumino amplitude of field $U_P$, to a ratio of two genuine Wess–Zumino amplitudes of auxiliary fields taking values in unitary matrices of rank reduced to that of the rank of projectors $P(k)$. The remaining argument of the proof of representation (1.2) computes explicitly such auxiliary Wess–Zumino amplitudes. For pedagogical reasons, this is first done in Section 4.8 for the case when projectors $P(k)$ have rank 2, as this instance may be treated with fewer technicalities. The computation for general rank, presented in Section 5, requires a more developed arsenal of techniques. It is based on a boundary gauge anomaly formula for Wess–Zumino amplitudes, of possible independent interest, whose proof is given in Section 6. Appendices A–D include some accessory material related to the main text.

Our proof of the representation (1.2) for the Kane–Mele invariant contains some steps that bear resemblance to the arguments scattered in the literature, in particular, in [9,36,34,43,45], but our point of view and the detailed mathematical argument seem new.

## 2. General background

### 2.1. Floquet theorem

We consider smooth periodic families of self-adjoint Hamiltonians $H : \mathbb{R} \times BZ \to M_N(\mathbb{C})$ acting on $\mathbb{C}^N$, with

$$H(t,k) = H(t+T,k) \tag{2.1}$$

for $k$ on the Brillouin torus $\mathrm{BZ} = \mathbb{R}^2/(2\pi\mathbb{Z})$ and $T$ the driving period. Such families are obtained from the Fourier transform of time-periodic lattice Hamiltonians, describing e.g. a periodically driven crystal in tight-binding approximation. The identification of the Brillouin torus with $\mathbb{R}^2/(2\pi\mathbb{Z})$ is obtained by an appropriate choice of a basis for the reciprocal lattice. The internal degrees of freedom in this description, in finite number $N$, include unit cell positions, orbitals, and spin. The corresponding evolution operators $U(t,k)$ belonging to the unitary group $U(N)$ solve the equation

$$\mathrm{i}\,\dot{U}(t,k) = H(t,k)U(t,k) \tag{2.2}$$

with initial condition $U(0,k) = I$, where $I$ is the identity operator. The time-periodicity of $H$ gives rise to the property

$$U(t+T,k) = U(t,k)U(T,k) \tag{2.3}$$

so that the whole information about the evolution is contained in the first period of time. In particular, $U(-T,k) = U(T,k)^{-1}$. Operators $U(t,k)$ define a smooth mapping from $[0,T] \times \mathrm{BZ}$ to $U(N)$. A natural invariant characterizing the topological properties of smooth maps between two manifolds is their homotopy class [6,4], which is, however, trivial for $U$.[3] A periodic map from $S^1 \times \mathrm{BZ}$ to $U(N)$, however, could have a nontrivial homotopy class. One may periodize $U$ in a natural way using Floquet theory if unitary operators $U(T,k)$ have a common spectral gap [44].

To do so, consider the spectral decomposition of the unitary operators $U(T,k)$

$$U(T,k) = \sum_n \lambda_n(k)|\psi_n(k)\rangle\langle\psi_n(k)|, \tag{2.4}$$

[3] Only the windings of the determinant of $U(t,k)$ around nontrivial 1-cycles of the Brillouin zone could provide nontriviality. However, they are trivial for all $t$ by continuity of the map, as $U(0,k) = I$.

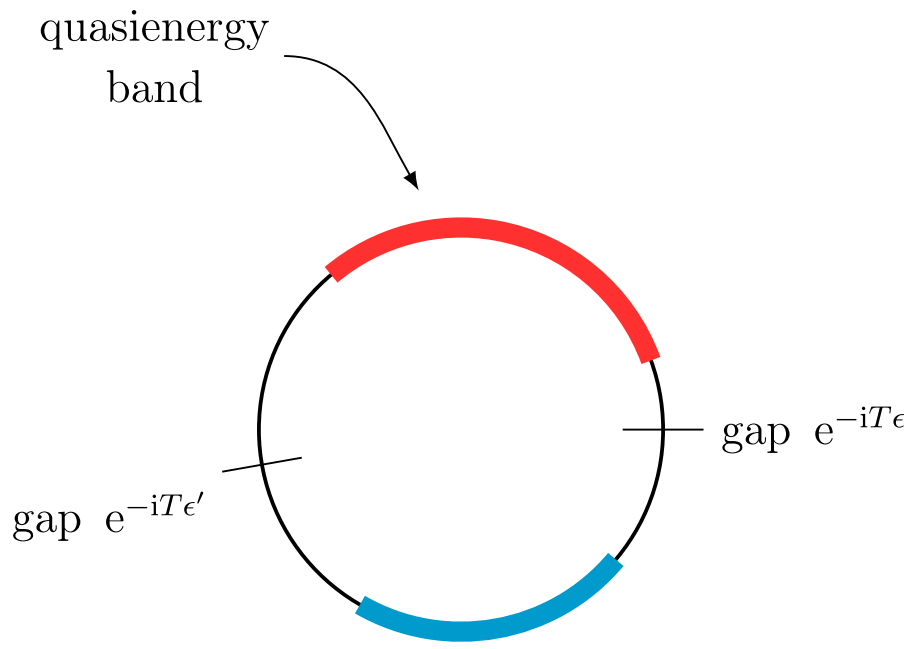

Fig. 1. The spectrum of the unitary evolution operator $U(T)$, or, equivalently, *quasienergy spectrum*, lives on the circle $U(1)$. As in the case of energy spectra of tight-binding Hamiltonians, the reunion of the (discrete) spectra of the operators $U(T,k)$ for $k$ in the Brillouin torus forms continuous quasienergy bands separated by spectral gaps (around $e^{-iT\epsilon}$ and $e^{-iT\epsilon'}$ in the sketch). Corresponding eigenstates form vector bundles over the Brillouin torus.

where $|\lambda_n(k)| = 1$. It is usual to parameterize the eigenvalues of $U(T,k)$ by real *quasienergies* writing $\lambda_n(k) = e^{-iT\epsilon_n(k)}$, where $\epsilon_n(k)$ are defined modulo $2\pi/T$ so that the quasienergy spectrum repeats itself with that period. If real number $\epsilon$ is in the gap of the quasienergy spectrum (i.e. $e^{-iT\epsilon} \neq \lambda_n(k)$ for all $n$ and $k$), see Fig. 1, then we may define the operator

$$H_\epsilon^{\text{eff}}(k) = \frac{i}{T}\sum_n \ln_{-T\epsilon} \lambda_n(k)|\psi_n(k)\rangle\langle\psi_n(k)|, \tag{2.5}$$

where

$$\ln_\phi(re^{i\varphi}) = \ln r + i\varphi \quad \text{for} \quad r > 0 \quad \text{and} \quad \phi - 2\pi < \varphi < \phi \tag{2.6}$$

is a branch of the logarithm. $H_\epsilon^{\text{eff}}(k)$ depends smoothly on $k$ and does not change when $\epsilon$ changes within the same quasienergy gap. Besides

$$H_{\epsilon+2\pi/T}^{\text{eff}}(k) = H_\epsilon^{\text{eff}}(k) + \tfrac{2\pi}{T} I. \tag{2.7}$$

Because of the relation

$$U(T,k) = e^{-iTH_\epsilon^{\text{eff}}(k)}, \tag{2.8}$$

$H_\epsilon^{\text{eff}}(k)$ are called *effective Hamiltonian*. They allow to define periodized versions of $U(t,k)$

$$V_\epsilon(t,k) = U(t,k)\, e^{itH_\epsilon^{\text{eff}}(k)} \tag{2.9}$$

that satisfy the relations

$$V_\epsilon(t+T,k) = V_\epsilon(t,k), \tag{2.10}$$

$$V_\epsilon(0,k) = I = V_\epsilon(T,k), \tag{2.11}$$

$$V_{\epsilon+2\pi/T}(t,k) = e^{\frac{2\pi i t}{T}} V_\epsilon(t,k). \tag{2.12}$$

Maps $V_\epsilon$, explicitly dependent on the quasienergy gap, may therefore be considered as defined on the 3-torus $S^1 \times \text{BZ}$. They take values in the unitary group $U(N)$. For $\epsilon' > \epsilon$,

$$H_{\epsilon'}^{\text{eff}}(k) - H_\epsilon^{\text{eff}}(k) = \tfrac{2\pi}{T} \tilde{P}_{\epsilon,\epsilon'}(k) \tag{2.13}$$

where

$$\tilde{P}_{\epsilon,\epsilon'}(k) = \sum_{\epsilon<\varepsilon<\epsilon'} \sum_{\substack{n \\ \lambda_n(k)=\mathrm{e}^{-\mathrm{i}\varepsilon T}}} |\psi_n(k)\rangle\langle\psi_n(k)|. \tag{2.14}$$

Note that only a finite number of quasienergies $\varepsilon$ contributes to the sum above. As already noticed, formula (2.13) holds true if there is no spectral value of quasienergy between $\epsilon$ and $\epsilon'$ so that $\tilde{P}_{\epsilon,\epsilon'}(k) = 0$. Suppose now that there is exactly one spectral value $\varepsilon$ of quasienergies between $\epsilon$ and $\epsilon'$. It has then to satisfy the inequalities $\epsilon < \varepsilon < \epsilon + \frac{2\pi}{T}$ (if the right inequality failed then $\varepsilon - \frac{2\pi}{T}$ would be another spectral value of quasienergy in the interval $(\epsilon, \epsilon')$). It follows that the contribution from $\varepsilon$ to $H^{\mathrm{eff}}_{\epsilon}(k)$ is

$$\varepsilon \sum_{\substack{n \\ \lambda_n(k)=\mathrm{e}^{-\mathrm{i}\varepsilon T}}} |\psi_n(k)\rangle\langle\psi_n(k)|. \tag{2.15}$$

On the other hand, one must have $\epsilon' < \varepsilon + \frac{2\pi}{T} < \epsilon' + \frac{2\pi}{T}$ (again, if the left inequality failed then $\varepsilon + \frac{2\pi}{T}$ would be another quasienergy spectral value in the interval $(\epsilon, \epsilon')$). Thus the contribution from $\varepsilon$ to $H^{\mathrm{eff}}_{\epsilon'}(k)$ is

$$(\varepsilon + \tfrac{2\pi}{T}) \sum_{\substack{n \\ \lambda_n(k)=\mathrm{e}^{-\mathrm{i}\varepsilon T}}} |\psi_n(k)\rangle\langle\psi_n(k)| \tag{2.16}$$

so that relation (2.13) follows in the case under consideration. The general case is proven gradually increasing $\epsilon'$.

It will be convenient to extend the definition of $\tilde{P}_{\epsilon,\epsilon'}(k)$ to gap quasienergies in general position by setting

$$\begin{aligned} \tilde{P}_{\epsilon,\epsilon'}(k) &= 0 \qquad \text{if} \qquad \epsilon = \epsilon', \\ \tilde{P}_{\epsilon,\epsilon'}(k) &= -\tilde{P}_{\epsilon',\epsilon}(k) \qquad \text{if} \qquad \epsilon > \epsilon'. \end{aligned} \tag{2.17}$$

With such an extension we have the relation

$$V_{\epsilon'}(t,k) = V_{\epsilon}(t,k)\,\mathrm{e}^{\frac{2\pi\mathrm{i}t}{T}\tilde{P}_{\epsilon,\epsilon'}(k)} \tag{2.18}$$

for arbitrary gap quasienergies $\epsilon, \epsilon'$. Note that

$$\tilde{P}_{\epsilon,\epsilon'+\frac{2\pi}{T}}(k) = \tilde{P}_{\epsilon-\frac{2\pi}{T},\epsilon'}(k) = \tilde{P}_{\epsilon,\epsilon'}(k) + I\,. \tag{2.19}$$

If $\epsilon < \epsilon'$ ($\epsilon > \epsilon'$) and if $\mathrm{e}^{-\mathrm{i}\epsilon T}$ and $\mathrm{e}^{-\mathrm{i}\epsilon' T}$ lie in two different spectral gaps of $U(T,k)$ then, up to an integer multiple of $I$, $\tilde{P}_{\epsilon,\epsilon'}(k)$ is equal to the orthogonal projector $P_{\epsilon,\epsilon'}(k)$ on the subspace spanned by the eigenvectors $|\psi_n(k)\rangle$ of $U(T,k)$ with the eigenvalues $\lambda_n(k)$ in the subinterval of the circle joining $\mathrm{e}^{-\mathrm{i}\epsilon T}$ to $\mathrm{e}^{-\mathrm{i}\epsilon' T}$ clock-wise (counter-clockwise). If $\mathrm{e}^{-\mathrm{i}\epsilon T}$ and $\mathrm{e}^{-\mathrm{i}\epsilon' T}$ lie in same spectral gap of $U(T,k)$ then $\tilde{P}_{\epsilon,\epsilon'}(k)$ is an integer multiple of $I$ and we shall set $P_{\epsilon,\epsilon'}(k) = 0$. The ranges of projectors $P_{\epsilon,\epsilon'}(k)$ form a vector bundle over the Brillouin torus BZ that we shall denote $\mathcal{E}_{\epsilon,\epsilon'}$.

### 2.2. Time reversal

In the following, we will consider time-reversal invariant systems of free fermions with half-integer spin. The number of internal degrees of freedom is therefore necessarily even, with $N = 2M$. The anti-unitary time-reversal operator

$$\Theta = \mathrm{e}^{\mathrm{i}\pi S_y/\hbar} C, \tag{2.20}$$

where $S_y$ is the $y$-component of the spin operator and $C$ represents complex conjugation, can be written in an adequate basis[4] as

$$\Theta = \Omega C \qquad \text{with} \qquad \Omega = \begin{pmatrix} 0 & I_M \\ -I_M & 0 \end{pmatrix}. \tag{2.21}$$

The time reversal acts on the group $U(2M)$ by the involution

$$U \mapsto \Theta U \Theta^{-1}. \tag{2.22}$$

The fixed-point set of this involution forms the symplectic subgroup $Sp(2M)$:

$$\{U \in U(2M) \,|\, \Theta U \Theta^{-1} = U\} = Sp(2M) \equiv \{U \in U(2M) \,|\, U^t \Omega U = \Omega\} \tag{2.23}$$

which is connected and simply connected [23]. For $M = 1$, $Sp(2) = SU(2)$. We will use later the fact that $\det(U) = 1$ for $U \in Sp(2M)$.

In a time dependent system, the time-reversal invariance of the Hamiltonian $H$ means that

$$\Theta\, H(t,k)\, \Theta^{-1} = H(-t,-k) \tag{2.24}$$

The above symmetry implies for the evolution operators that

$$\Theta\, U(t,k)\, \Theta^{-1} = U(-t,-k) \tag{2.25}$$

as both sides satisfy the same first-order equation with the same initial condition. In particular,

$$\Theta\, U(T,k)\, \Theta^{-1} = U^{-1}(T,-k) \tag{2.26}$$

which implies, by spectral decomposition,

$$\Theta\, H_\epsilon^{\mathrm{eff}}(k)\, \Theta^{-1} = H_\epsilon^{\mathrm{eff}}(-k) \tag{2.27}$$

for the effective Hamiltonian (2.5). For the periodized evolution operators $V_\epsilon(t,k)$ the time-reversal invariance entails the relation

$$\Theta\, V_\epsilon(t,k)\, \Theta^{-1} = V_\epsilon(-t,-k) = V_\epsilon(T-t,-k)\,. \tag{2.28}$$

In particular,

$$\Theta\, V_\epsilon(T/2,k)\, \Theta^{-1} = V_\epsilon(T/2,-k)\,. \tag{2.29}$$

Eq. (2.13) implies that in the time-reversal symmetric case

$$\Theta\, \tilde{P}_{\epsilon,\epsilon'}(k)\, \Theta^{-1} = \tilde{P}_{\epsilon,\epsilon'}(-k)\,, \qquad \Theta\, P_{\epsilon,\epsilon'}(k)\, \Theta^{-1} = P_{\epsilon,\epsilon'}(-k)\,, \tag{2.30}$$

[4] For a spin-$j$ representation, the matrix $\Omega$ is expressed in the basis with magnetic numbers $j$, $-(j-1)$, $j-2$, ..., $-j$, $j-1$, $-(j-2)$, ... (e.g. 5/2, −3/2, 1/2, −5/2, 3/2, −1/2 for $j = 5/2$).

where the second equality follows from the first one. Taking $k = 0$ we see that the range of $P_{\epsilon,\epsilon'}(0)$ is preserved by $\Theta$ so that its dimension has to be even implying that vector bundles $\mathcal{E}_{\epsilon,\epsilon}$ have even rank in the time-reversal invariant case. For general $k$, operator $\Theta$ maps the range of $P_{\epsilon,\epsilon'}(k)$ to that $P_{\epsilon,\epsilon'}(-k)$ defining on the bundle $\mathcal{E}_{\epsilon,\epsilon}$ an antilinear involution $\theta$ that projects to the map $\vartheta_2 : k \mapsto -k$ on the base-space torus BZ.

### *2.3. Homotopy classes of maps from the 3-torus to the unitary group*

Continuous maps $V$ from the 3-torus $T^3$ to the unitary group $U(N)$ with $N \geq 2$, like $V_\epsilon$, may be classified up to homotopy [13] by considering 3 elements of $\pi_1(U(N)) \simeq \mathbb{Z}$ as well as one element of $\pi_3(U(N)) \simeq \mathbb{Z}$ (the second homotopy group being trivial).

We may always assume that $V$ is smooth since the homotopy class of continuous $V$ necessarily contains smooth maps. The first 3 homotopy invariants are obtained by restricting $V$ to three independent non-trivial 1-cycles $\mathcal{C}_i$ around the torus and by considering the winding numbers of the determinant of $V$ around zero in the complex plane,

$$w_{\mathcal{C}_i}[V] = \frac{1}{2\pi \mathrm{i}} \int\limits_{\mathcal{C}_i} \mathrm{tr}(V^{-1} \mathrm{d}V) = \frac{1}{2\pi \mathrm{i}} \int\limits_{\mathcal{C}_i} \mathrm{d} \log \det V \,. \tag{2.31}$$

To obtain the element of $\pi_3(U(N))$, consider the loop $L = \mathcal{C}_1\mathcal{C}_2\mathcal{C}_3\mathcal{C}_1^{-1}\mathcal{C}_2^{-1}\mathcal{C}_3^{-1}$ on the 3-torus that is contractible. One quotients the torus into a sphere $S^3$ by collapsing $L$ to a point: the map $V$ can always be deformed to a map constant on $L$ which descends to the quotient and whose homotopy class belongs to $\pi_3(U(N))$.

The value in $\mathbb{Z} \simeq \pi_3(U(N))$ corresponding to $V$ may be given by the integral [3]

$$\frac{1}{24\pi^2} \int\limits_{T^3} V^*\chi \equiv \deg(V) \,, \tag{2.32}$$

where[5]

$$\chi = \mathrm{tr}(u^{-1}\mathrm{d}u)^3 \tag{2.33}$$

is a real closed 3-form on the group $U(N)$ and $V^*\chi$ denotes the pullback of $\chi$ by $V$. Somewhat abusively, we shall call $\deg(V)$ the degree of $V$. The normalization of the integral in (2.32) is chosen so that the degree is equal to 1 for the embedding of $SU(2) \cong S^3$ by block matrices into $U(N)$ (see e.g. [39]).

If $V_1$ and $V_2$ are two maps from $T^3$ to $U(N)$ and $V_1V_2$ is their point-wise multiplication, then

$$\deg(V_1V_2) = \deg(V_1) + \deg(V_2). \tag{2.34}$$

In particular, $\deg(V^{-1}) = -\deg(V)$ and $\deg(Vu) = \deg(uV) = \deg(V)$ if $u$ is a fixed element of $U(N)$. This follows from the formula (A.1) for 3-forms on $U(N) \times U(N)$ from Appendix A which implies that

$$(V_1V_2)^*\chi = V_1^*\chi + V_2^*\chi \, + \text{ an exact 3-form.} \tag{2.35}$$

---

[5] We omit the exterior product signs $\wedge$.

In the particular case when $V_2 = V_1^{-1}$, the exact form contribution vanishes implying that

$$(V^{-1})^\star \chi = -V^\star \chi \,. \tag{2.36}$$

If $D$ is a diffeomorphism of $T^3$ then

$$\deg(V \circ D) = \pm \deg(V)\,, \tag{2.37}$$

where the positive (negative) sign applies to orientation-preserving (orientation-reversing) diffeomorphisms.

When $N = 2M$ so that the time-reversal operator $\Theta = U_\Theta C$ can be defined, where $U_\Theta \in U(2M)$, then

$$\deg(\Theta V \Theta^{-1}) = \deg(V) \tag{2.38}$$

Indeed, $\Theta V \Theta^{-1} = U_\Theta \overline{V} U_\Theta^{-1}$, where the overline denotes the complex conjugation. Hence $\deg(\Theta V \Theta^{-1}) = \deg(\overline{V}) = \deg(V)$, where the last equality follows from the reality of 3-form $\chi$.

For $V = V_\epsilon$, where $V_\epsilon : T^3 \to U(N)$ was constructed in the previous section, the winding numbers $w_\mathcal{C}(V_\epsilon)$ of the determinant of $V_\epsilon$ vanish for 1-cycles in the Brillouin torus, i.e. for $\mathcal{C} \subset \{t\} \times \mathrm{BZ}$. Indeed, such winding numbers depend continuously on $t$ and $V_\epsilon(0, k) = I$. On the other hand, for 1-cycle $[0, T], \times\{0\}$, the winding numbers satisfy the relation

$$w_\mathcal{C}(V_{\epsilon+2\pi/T}) = w_\mathcal{C}(V_\epsilon) + N \tag{2.39}$$

due to (2.12). Taken modulo $N$, such winding numbers define an invariant $w_\epsilon[U] \in \mathbb{Z}_N$ of periodically driven crystals that depends on the spectral gap of $U[T]$ satisfies the relation

$$w_{\epsilon'}[U] - w_\epsilon[U] = \mathrm{rk}(\mathcal{E}_{\epsilon,\epsilon'}) \bmod N\,, \tag{2.40}$$

where $\mathrm{rk}(\mathcal{E}_{\epsilon,\epsilon'})$ is the rank (i.e. the dimension of the fibers) of the vector bundle $\mathcal{E}_{\epsilon,\epsilon'}$ of states between two spectral gaps of $U(T)$ introduced at the end of Section 2.1. Identity (2.40) follows from Eq. (2.18). In the time-reversal invariant case, relation (2.28) implies that $V_\epsilon(T/2, 0) \in Sp(2M)$, so that $\det V_\epsilon(T/2, 0) = 1$ and that $\overline{\det V_\epsilon(t, 0)} = \det V_\epsilon(T - t, 0)$ implying that the winding numbers of $\det V_\epsilon(t, 0)$ when $t$ runs from 0 to $T/2$ and from $T/2$ to $T$ are equal. It follows that $w_\epsilon[U]$ takes even values in $\mathbb{Z}_{2M}$ in that case and one may consider $\frac{1}{2} w_\epsilon[U] \in \mathbb{Z}_M$ as the winding index. Recall from the end of Section 2.2 that $\mathrm{rk}(\mathcal{E}_{\epsilon,\epsilon'})$ is also even in the time-reversal symmetric case, in agreement with (2.40).

In [44] it was proposed to consider the degree $\deg(V_\epsilon) = \deg(V_{\epsilon+2\pi/T})$ (see Eq. (2.12)) as another gap-dependent topological invariant, denoted $W_\epsilon[U]$, of periodically driven crystals.[6] It was shown there that

$$W_{\epsilon'}[U] - W_\epsilon[U] = c_1(\mathcal{E}_{\epsilon,\epsilon'}) \tag{2.41}$$

where $c_1(\mathcal{E}_{\epsilon,\epsilon'})$ is the first Chern number of vector bundle $\mathcal{E}_{\epsilon,\epsilon'}$. However, in the time-reversal invariant case when the relation (2.28) holds,

$$\deg(V_\epsilon) = \deg(\Theta V_\epsilon \Theta^{-1}) = \deg(V_\epsilon \circ \vartheta_3) = -\deg(V_\epsilon) \tag{2.42}$$

where $\vartheta_3$ is an orientation-reversing diffeomorphism of $T^3$ induced by the map $(t, k) \mapsto (T - t, -k)$. Hence $W_\epsilon[U] = \deg(V_\epsilon) = 0$ in that case.

[6] In fact, Ref. [44] used a slightly different but equivalent periodization of the evolution operator, see Appendix C.

## 3. *K* index for periodically driven crystals

### 3.1. Definition and properties

Consider a periodically driven time-reversal invariant tight-binding system described by a family of evolution operators $U(t,k)$, as described in part 2. Assume that $e^{-iT\epsilon}$ lies in a spectral gap of $U(T,k)$ and consider the periodized evolution operators $V_\epsilon(t,k)$.

Time-reversal invariance leads to the vanishing of $\deg(V_\epsilon)$, essentially because the contributions of Kramers partners cancel [5]. To circumvent such cancellations, one may keep only the first half of the periodized time evolution $V_\epsilon(t,k)$, which contains all the information without redundancy. One has then to find a way to extract the topological information from the half-period evolution. This will be done by extending it in a specific way to a periodic map whose degree is defined and may be computed. Since $V_\epsilon(0,k)=I$, such an extension will be provided by a contraction of $V_\epsilon(T/2,\cdot)$ to the constant map equal to $I$. The contraction will preserve the topological information if it keeps the time-reversal invariance constraint (2.29) all along.

Assuming the existence of such a contraction, that may be always chosen smooth, and taking its parameter with values in the interval $[T/2,T]$, we obtain a continuous map $\widehat{V}_\epsilon$ from $[0,T]\times$ BZ to $U(N)$ such that

$$\widehat{V}_\epsilon(t,k)=V_\epsilon(t,k) \qquad \text{for } 0\le t\le T/2, \tag{3.1}$$

$$\Theta\,\widehat{V}_\epsilon(t,k)\,\Theta^{-1}=\widehat{V}_\epsilon(t,-k) \qquad \text{for } T/2\le t\le T\,, \tag{3.2}$$

$$\widehat{V}_\epsilon(T,k)=I=\widehat{V}_\epsilon(0,k)\,. \tag{3.3}$$

Besides, $\widehat{V}_\epsilon$ is smooth on $[0,T/2]\times$ BZ and $[T/2,T]\times$ BZ. The $\mathbb{Z}_2$-valued index $K$ is then defined by setting

$$K_\epsilon[U]=\deg(\widehat{V}_\epsilon)\ \bmod 2 \tag{3.4}$$

(recall that $V_\epsilon$ was obtained from $U$).

We shall prove the existence of contractions of $V_\epsilon(T/2,\cdot)$ with the desired symmetry property in Section 3.2. For now, let us address the question whether the $\mathbb{Z}_2$-valued quantity $K_\epsilon[U]$ in independent of the choice of the contraction, i.e., whether for two different contractions, the corresponding maps $\widehat{V}_{\epsilon,1}$ and $\widehat{V}_{\epsilon,2}$ from $[0,T]\times$ BZ satisfy

$$\deg(\widehat{V}_{\epsilon,1})-\deg(\widehat{V}_{\epsilon,2})\in 2\mathbb{Z}\,, \tag{3.5}$$

implying that the index defined in (3.4) is the same when computed from $\widehat{V}_{\epsilon,1}$ or $\widehat{V}_{\epsilon,2}$. According to (3.1), the two maps coincide for $t\in[0,T/2]$ so that

$$\begin{aligned}\deg(\widehat{V}_{\epsilon,1})-\deg(\widehat{V}_{\epsilon,2}) &= \frac{1}{24\pi^2}\bigg(\int\limits_{[T/2,T]\times\mathrm{BZ}}\widehat{V}_{\epsilon,1}^{*}\chi\ -\int\limits_{[T/2,T]\times\mathrm{BZ}}\widehat{V}_{\epsilon,2}^{*}\chi\bigg)\\ &= \frac{1}{24\pi^2}\int\limits_{[0,T]\times\mathrm{BZ}}\widehat{V}_{\epsilon,12}^{*}\chi\,,\end{aligned} \tag{3.6}$$

where the map $V_{\epsilon,12}:[0,T]\times$ BZ is given by

$$\widehat{V}_{\epsilon,12}(t,k)=\begin{cases}\widehat{V}_{\epsilon,2}(T-t,k) & \text{if} \quad 0\le t\le T/2\\ \widehat{V}_{\epsilon,1}(t,k) & \text{if} \quad T/2\le t\le T\end{cases} \tag{3.7}$$

which is well defined since

$$\widehat{V}_{\epsilon,2}(T/2,k) = V_\epsilon(T/2,k) = \widehat{V}_{\epsilon,1}(T/2,k)\,. \tag{3.8}$$

Besides

$$\widehat{V}_{\epsilon,12}(0,k) = I = \widehat{V}_{\epsilon,12}(T,k)\,. \tag{3.9}$$

We infer that

$$\deg(\widehat{V}_{\epsilon,1}) - \deg(\widehat{V}_{\epsilon,2}) = \deg(\widehat{V}_{\epsilon,12}) \tag{3.10}$$

Moreover, $\widehat{V}_{\epsilon,12}$ has the symmetry

$$\Theta\,\widehat{V}_{\epsilon,12}(t,k)\,\Theta^{-1} = \widehat{V}_{\epsilon,12}(t,-k) \tag{3.11}$$

for every $t \in [0,T]$. We shall prove in Section 3.3 that the degree of a such map is always even, which will show that (3.5) holds.

Index $K_\epsilon[U]$ coincides for quasienergies corresponding to the same spectral gap of $U(T)$. Indeed, it does not change if $\epsilon$ does not cross any quasienergy $\varepsilon_n$ because $V_\epsilon$ does not change then, see (2.18). Besides

$$K_{\epsilon+2\pi/T}[U] = K_\epsilon[U]\,. \tag{3.12}$$

The last equality requires an argument. We may take for $\widehat{V}_{\epsilon+2\pi/T}$ the map $\widehat{V}_\epsilon \widehat{U}_0$, where

$$\widehat{U}_0(t) = \begin{cases} \mathrm{e}^{\frac{2\pi i t}{T}} I & \text{for } 0 \le t \le T/2, \\ \begin{pmatrix} \mathrm{e}^{\frac{2\pi i t}{T}} I_M & 0 \\ 0 & \mathrm{e}^{-\frac{2\pi i t}{T}} I_M \end{pmatrix} & \text{for } T/2 \le t \le T \end{cases} \tag{3.13}$$

which for $\Theta$ given by (2.21) satisfies $\Theta\widehat{U}_0(t)\Theta^{-1} = \widehat{U}_0(t)$ when $T/2 \le t \le T$. For dimensional reasons ($\widehat{U}_0$ depends on only one variable), $\deg(\widehat{U}_0) = 0$ implying that (3.12) holds.

In the time-reversal invariant case, relation (2.41), whose both sides become trivial, will be replaced by the identity

$$K_{\epsilon'}[U] - K_\epsilon[U] = \mathrm{KM}(\mathcal{E}_{\epsilon,\epsilon'})\,, \tag{3.14}$$

where $\mathrm{KM}(\mathcal{E}_{\epsilon,\epsilon'})$ is the $\mathbb{Z}_2$-valued Kane–Mele index [32,9] of the vector bundle $\mathcal{E}_{\epsilon,\epsilon'}$ equipped with an antilinear involution $\theta$ defined by the restriction of $\Theta$ to its fibers. We shall prove relation (3.14) in Section 4, establishing on the way a new representation for the Kane–Mele invariant of the unforced case.

Note that the winding number $w_{\mathcal{C}}(\widehat{V}_\epsilon)$ along 1-cycle $[0,T]\times 0$ is equal to the winding number of $\det V(t,0)$ on the interval $[0,T/2]$ since $\widehat{V}_\epsilon(t,0) \in Sp(2M)$ for $t \in [T/2,T]$ and thus has determinant 1. Besides, it follows from (2.12) that

$$w_{\mathcal{C}}(\widehat{V}_{\epsilon+2\pi/T}) = w_{\mathcal{C}}(\widehat{V}_\epsilon) + M\,. \tag{3.15}$$

Hence for the 1-cycle $\mathcal{C}$ winding around the time period,

$$w_{\mathcal{C}}(\widehat{V}_\epsilon) \ \mathrm{mod}\ M \ = \ \tfrac{1}{2} w_\epsilon[U] \in \mathbb{Z}_M\,, \tag{3.16}$$

see the discussion in Section 2.3.

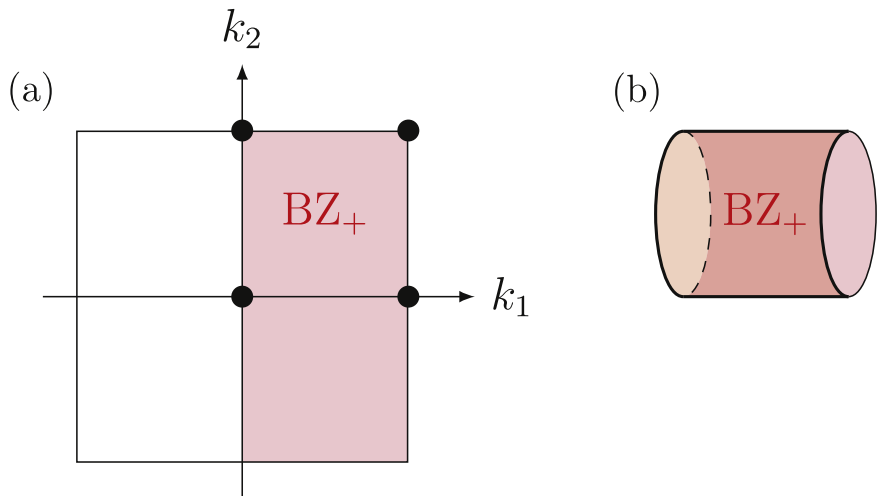

Fig. 2. (a) We consider one half $BZ_+$ of the Brillouin zone (in red). The time-reversal invariant momenta (TRIM) where $k = -k + G$, $G$ being a reciprocal lattice vector, are marked as black dots. (b) $BZ_+$ may be viewed as a cylinder because of the periodicity in direction $k_2$. (For interpretation of the colors in this figure, the reader is referred to the web version of this article.)

### 3.2. Existence of the contraction

We construct here an explicit contraction of $V_\epsilon(T/2, \cdot)$ to the constant map equal to $I$, conserving the symmetry property (3.2) for all values of the contraction parameter.

More formally, consider any continuous map $V$ from the Brillouin torus BZ to the unitary group $U(2M)$ such that

$$\Theta V(k)\Theta^{-1} = V(-k) \tag{3.17}$$

$$w_{\mathcal{C}}[V] = 0 \qquad \text{for all 1-cycles of BZ}. \tag{3.18}$$

In general, the obstructions to the existence of a contraction of a continuous map from BZ to $U(2M)$ to a constant map are precisely the non-trivial winding numbers of $\det V$. If such windings are trivial then one may deform the map to an $SU(N)$-valued one by multiplying it by $\exp[-\frac{r}{N} \ln\det V(k)]$ for $r \in [0, 1]$ and the $SU(N)$-valued maps may be already contracted to $I$.

Such a contraction will generally not have symmetry (3.17) for the intermediate values of the contraction parameter, however, and cannot be used here. We want to establish then the existence of a continuous map

$$[0, 1] \times \text{BZ} \ni (r, k) \longmapsto \widetilde{V}(r, k) \in U(2M) \tag{3.19}$$

with the properties

$$\widetilde{V}(0, k) = V(k), \qquad \widetilde{V}(1, k) = I, \tag{3.20}$$

$$\Theta\, \widetilde{V}(r, k)\, \Theta^{-1} = \widetilde{V}(r, -k). \tag{3.21}$$

This will be done in several steps, but the general idea is as follows. Property (3.17) allows to consider only $V(k)$ restricted to the half $BZ_+$ of the Brillouin torus[7] that corresponds to $0 \le k_1 \le \pi$, to build a contraction of this restriction, and finally to extend it to BZ using (3.21). However, $BZ_+$ is not a closed surface anymore, so its boundary should be treated carefully. Note that

$$V(k_1, -\pi) = V(k_1, \pi), \tag{3.22}$$

which means that the restriction of $V$ to $BZ_+$ is still $k_2$-periodic (Fig. 2), and that

$$\Theta\, V(0, k_2)\, \Theta^{-1} = V(0, -k_2), \qquad \Theta\, V(\pi, k_2)\, \Theta^{-1} = V(\pi, -k_2)\,. \tag{3.23}$$

[7] The half Brillouin zone $BZ_+$ is an *effective Brillouin zone* as defined by Moore and Balents [36].

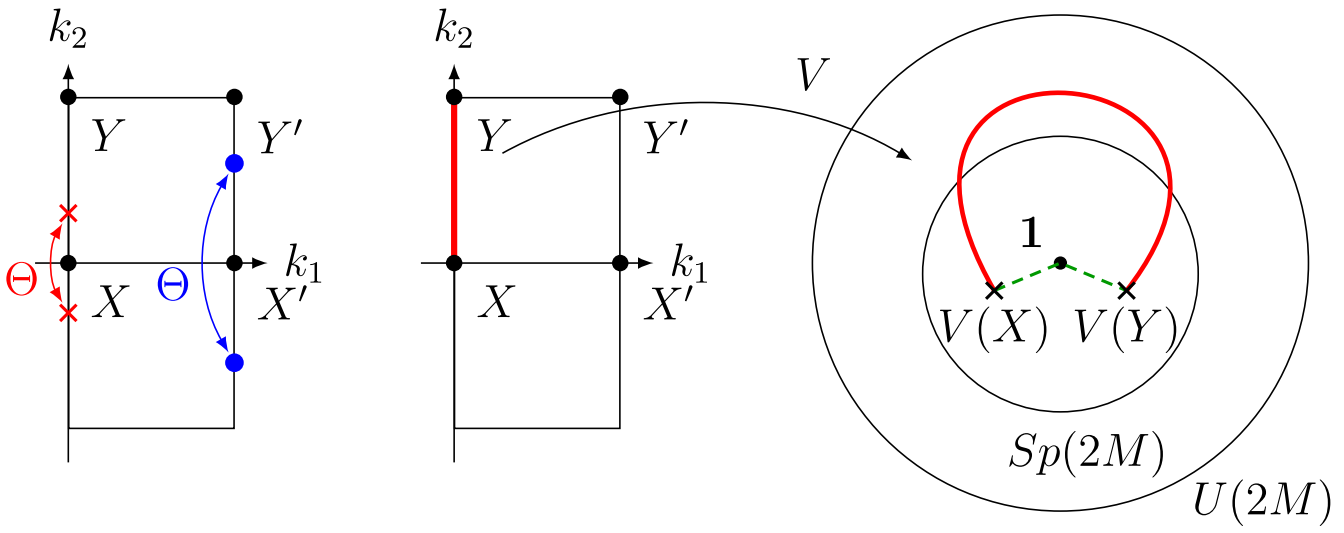

Fig. 3. Time reversal can be restricted to one component of the boundary of the effective Brillouin zone $BZ_+$, e.g. at $k_1 = 0$, where it relates the point $(0, k_2)$ to $(0, -k_2)$. Consider only one half on that boundary component (connecting points $X$ and $Y$). The map $V$ takes this set to its image in the unitary group $U(2M)$ (drawn in red), with the special properties that end-points $V(X)$ and $V(Y)$ are symplectic elements. Hence, there are paths (in dashed green) inside the symplectic group $Sp(2M)$ connecting $V(X)$ and $V(Y)$ to the unit $I$. (For interpretation of the colors in this figure, the reader is referred to the web version of this article.)

First consider the restriction of $V$ to $k_1 = 0$ and $k_2 \in [0, \pi]$. Due to (2.23),

$$V(0,0),\ V(0,\pi) \in Sp(2M)\,. \tag{3.24}$$

Since $Sp(2M)$ is connected, these two points may be joined to $I$ inside $Sp(2M)$ by continuous curves $[0, \frac{1}{4}] \ni s \mapsto \widetilde{V}_{00}(s)$ and $[0, \frac{1}{4}] \ni s \mapsto \widetilde{V}_{0\pi}(s)$, respectively, such that

$$\widetilde{V}_{0a}(0) = V(0, a), \qquad \widetilde{V}_{0a}(\tfrac{1}{4}) = I \tag{3.25}$$

for $a = 0, \pi$. Then, using $k_2$ as a parameter for $\widetilde{V}_{0a}$, we define a homotopy

$$[0, \tfrac{1}{4}] \times [0, \pi] \ni (r, k_2) \longmapsto \widetilde{V}_0(r, k_2) \in U(2M) \tag{3.26}$$

by the formula

$$\widetilde{V}_0(r, k_2) = \begin{cases} \widetilde{V}_{00}(r - k_2) & \text{for} \quad 0 \le k_2 \le r\,, \\ V(0, \frac{\pi(k_2 - r)}{\pi - 2r}) & \text{for} \quad r \le k_2 \le \pi - r\,, \\ \widetilde{V}_{0\pi}(r - \pi + k_2) & \text{for} \quad \pi - r \le k_2 \le \pi \end{cases} \tag{3.27}$$

assuring that

$$\widetilde{V}_0(0, k_2) = V(0, k_2), \qquad \widetilde{V}_0(\tfrac{1}{4}, 0) = I = \widetilde{V}_0(\tfrac{1}{4}, \pi), \tag{3.28}$$

$$\widetilde{V}_0(r, 0),\ \widetilde{V}_0(r, \pi) \in Sp(2M), \tag{3.29}$$

see Figs. 3 and 4.

At $r = \frac{1}{4}$ we are left with a loop in $U(2M)$. The contraction of this loop to the single point $I$ could be obstructed by a nonzero winding number of $[0, \pi] \ni k_2 \mapsto \det \widetilde{V}_0(\frac{1}{4}, k_2)$. Since inside $Sp(2M)$ the determinant is fixed to 1, this winding number is the same as the one for $[0, \pi] \ni k_2 \mapsto \det V(0, k_2)$. Because of the property (3.17) which implies that $\det V(0, -k_2) = \overline{\det V(0, k_2)}$, the latter winding number is a half of that for $[-\pi, \pi] \ni k_2 \mapsto \det V(0, k_2)$ and it vanishes by assumption (3.18). We infer then that the previously considered loop may be contracted to $I$ so that one can extend the homotopy $\widetilde{V}_0$ from $r \in [0, \frac{1}{4}]$ to $r \in [0, \frac{1}{2}]$ in such a way that

$$\widetilde{V}_0(r, 0) = I = \widetilde{V}_0(r, \pi) \quad \text{for } r \in [\tfrac{1}{4}, \tfrac{1}{2}], \qquad \widetilde{V}_0(\tfrac{1}{2}, k_2) = I\,, \tag{3.30}$$

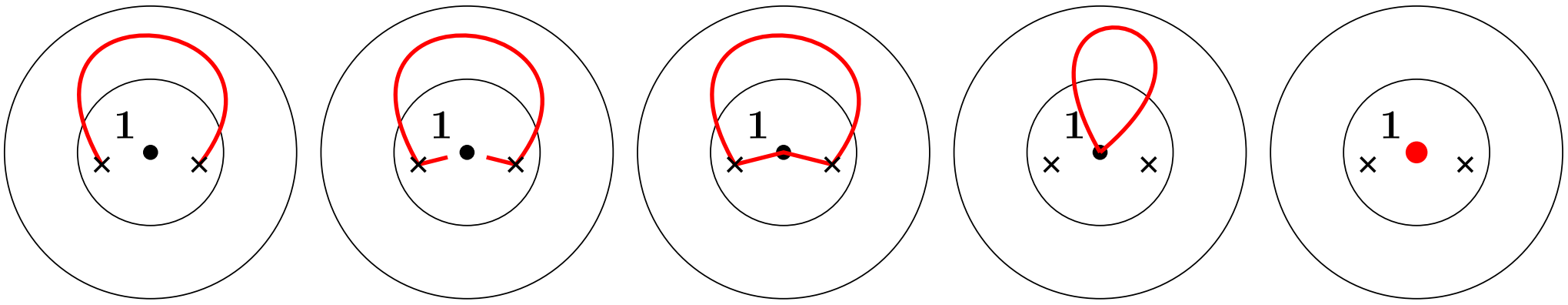

Fig. 4. We contract map $V$ restricted to the half of the boundary component of $BZ_+$ at $k_1 = 0$ to the group unit through the homotopy $\widetilde{V}_0(r, k_2)$ which is visualized in different steps $r$ in this figure (the big circle represents $U(2M)$ and the small one the subgroup $Sp(2M)$, see also Fig. 3). The same construction is done for the other boundary component of $BZ_+$.

see again Fig. 4. Now, setting for $r \in [0, \frac{1}{2}]$ and $k_2 \in [-\pi, 0]$

$$\widetilde{V}_0(r, k_2) = \Theta\, \widetilde{V}_0(r, -k_2)\, \Theta^{-1} \tag{3.31}$$

extends the contraction to the whole line $k_1 = 0,\ k_2 \in [-\pi, \pi]$ resulting in a continuous map

$$[0, \tfrac{1}{2}] \times [-\pi, \pi] \ni (r, k_2) \longmapsto \widetilde{V}_0(r, k_2) \in U(2M) \tag{3.32}$$

such that

$$\widetilde{V}_0(r, -\pi) = \widetilde{V}_0(r, \pi), \qquad \Theta\, \widetilde{V}_0(r, k_2)\, \Theta^{-1} = \widetilde{V}_0(r, -k_2) \tag{3.33}$$

$$\widetilde{V}_0(0, k_2) = V(0, k_2), \qquad \widetilde{V}_0(\tfrac{1}{2}, k_2) = I \tag{3.34}$$

which will provide a contraction of $V$ with the required symmetry restricted to the line $k_1 = 0,\ k_2 \in [-\pi, \pi]$. In the same way, we construct a contraction $\widetilde{V}_\pi$ of $V$ restricted to the line $k_1 = \pi,\ k_2 \in [-\pi, \pi]$ ending up with a map

$$[0, \tfrac{1}{2}] \times [-\pi, \pi] \ni (r, k_2) \longmapsto \widetilde{V}_\pi(r, k_2) \in U(2M) \tag{3.35}$$

such that

$$\widetilde{V}_\pi(r, -\pi) = \widetilde{V}_\pi(r, \pi), \qquad \Theta\, \widetilde{V}_\pi(r, k_2)\, \Theta^{-1} = \widetilde{V}_\pi(r, -k_2) \tag{3.36}$$

$$\widetilde{V}_\pi(0, k_2) = V(\pi, k_2), \qquad \widetilde{V}_\pi(\tfrac{1}{2}, k_2) = I\,. \tag{3.37}$$

This way we have contracted the restriction of $V$ to the boundaries of $BZ_+$. The second step is to extend the contraction to the interior of $BZ_+$. We first define a continuous map

$$[0, \tfrac{1}{2}] \times [0, \pi] \times [-\pi, \pi] \ni (r, k_1, k_2) \longmapsto \widetilde{V}(r, k_1, k_2) \in U(2M) \tag{3.38}$$

by setting

$$\widetilde{V}(r, k_1, k_2) = \begin{cases} \widetilde{V}_0(r - k_1, k_2) & \text{for} \quad 0 \le k_1 \le r\,, \\ V(\frac{\pi(k_1 - r)}{\pi - 2r}, k_2) & \text{for} \quad r \le k_1 \le \pi - r\,, \\ \widetilde{V}_\pi(r - \pi + k_1, k_2) & \text{for} \quad \pi - r \le k_1 \le \pi\,. \end{cases} \tag{3.39}$$

This map is consistently defined and satisfies

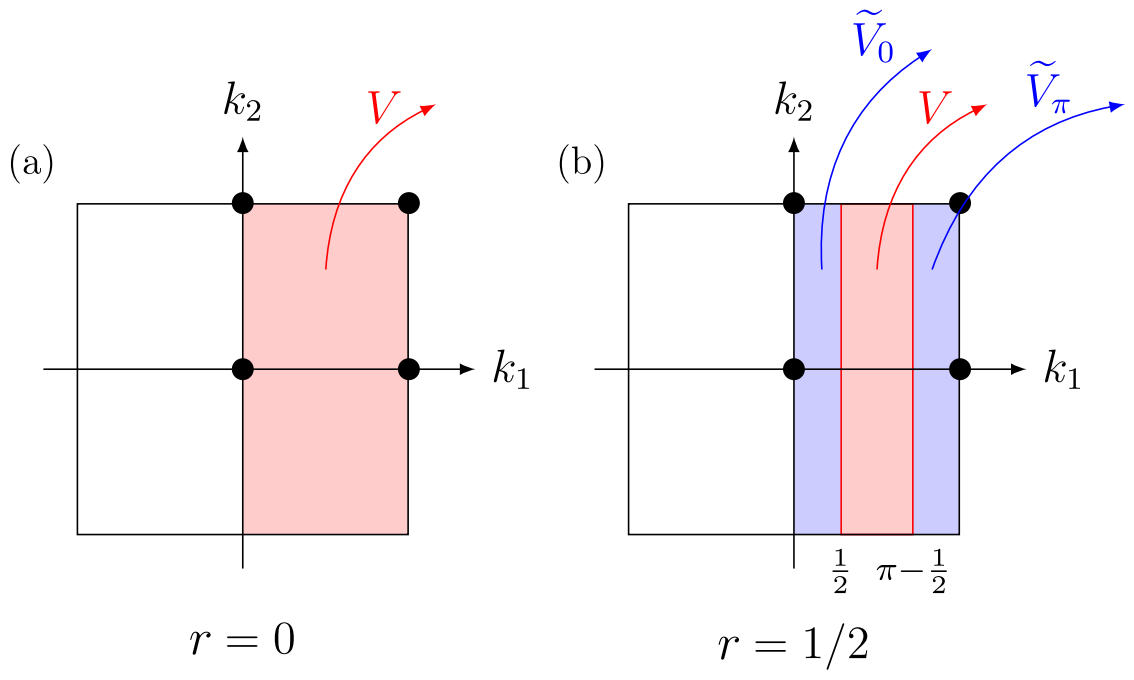

Fig. 5. We have separately constructed maps $\widetilde{V}_0$ and $\widetilde{V}_\pi$, defined only on the boundary components of $BZ_+$, interpolating from the values of $V$ on the boundary to $I$ whilst preserving time-reversal invariance. Those interpolations are put together to construct the map $\widetilde{V}$ which interpolates from the $V = \widetilde{V}(0,\cdot)$ to a map $\widetilde{V}(\frac{1}{2},\cdot)$ which sends the boundary of $BZ_+$ to $I$. As we want the last map to be defined on the half Brillouin zone, we gradually shrink the space allotted to $V$ in order to make room for the interpolations $\widetilde{V}_0$ and $\widetilde{V}_\pi$. (For interpretation of the colors in this figure, the reader is referred to the web version of this article.)

$$\widetilde{V}(0,k_1,k_2) = V(k_1,k_2)\,, \tag{3.40}$$
$$\widetilde{V}(r,k_1,-\pi) = \widetilde{V}(r,k_1,\pi)\,, \tag{3.41}$$
$$\Theta\,\widetilde{V}(r,0,k_2)\,\Theta^{-1} = \widetilde{V}(r,0,-k_2)\,, \tag{3.42}$$
$$\Theta\,\widetilde{V}(r,\pi,k_2)\,\Theta^{-1} = \widetilde{V}(r,\pi,-k_2)\,. \tag{3.43}$$

Note that in the definition (3.39), variable $k_1$ is used as a parameter for $\widetilde{V}_0$ and $\widetilde{V}_\pi$ allowing to “spread” these two contractions on $BZ_+$ in a way that preserves the time-reversal symmetry on the boundary, see Fig. 5. Taking $r = \frac{1}{2}$, we obtain a map

$$[0,\pi]\times[-\pi,\pi] \ni (k_1,k_2) \longmapsto \widetilde{V}(\tfrac{1}{2},k_1,k_2) \in U(2M) \tag{3.44}$$

with the properties

$$\widetilde{V}(\tfrac{1}{2},k_1,-\pi) = \widetilde{V}(\tfrac{1}{2},k_1,\pi)\,, \tag{3.45}$$
$$\widetilde{V}(\tfrac{1}{2},0,k_2) = I = \widetilde{V}(\tfrac{1}{2},\pi,k_2) \tag{3.46}$$

so that we may view $\widetilde{V}(\frac{1}{2},\cdot)$ as defined on a cylinder with the two boundary circles collapsed to the same point which, topologically, may be viewed as a 2-sphere with two opposites points identified. Thus $\widetilde{V}(\frac{1}{2},\cdot)$ is a map of such an “earring” into $U(2M)$, see Fig. 6(c).

The remaining question is whether one may deform the above map to the constant map equal to $I$ keeping the properties (3.45) and (3.46). The only obstruction could come from the winding of the map $[0,\pi] \ni k_1 \mapsto \det\widetilde{V}(\frac{1}{2},k_1,0)$ around zero (note that such windings are the same for any $k_2 \in [-\pi,\pi]$).

Since $\widetilde{V}_0$ and $\widetilde{V}_1$ preserve time-reversal invariance, they belong to $Sp(2M)$ for $k_2 = 0$ with their determinant fixed to 1 there. Thus the winding of $[0,\pi] \ni k_1 \mapsto \det\widetilde{V}(\frac{1}{2},k_1,0)$ is the same as that of the map $[0,\pi] \ni k_1 \mapsto \det V(k_1,0)$ which vanishes by the argument given above for the winding number of $[0,\pi] \ni k_2 \mapsto \det V(0,k_2)$. One may then contract the loop $[0,\pi] \ni k_1 \mapsto \widetilde{V}(\frac{1}{2},k_1,0)$ to the constant loop keeping the values at $k_1 = 0,\pi$ fixed. This is realized by the map

$$[\tfrac{1}{2},\tfrac{3}{4}]\times[0,\pi] \ni r \longmapsto \widetilde{V}(r,k_1,0) \tag{3.47}$$

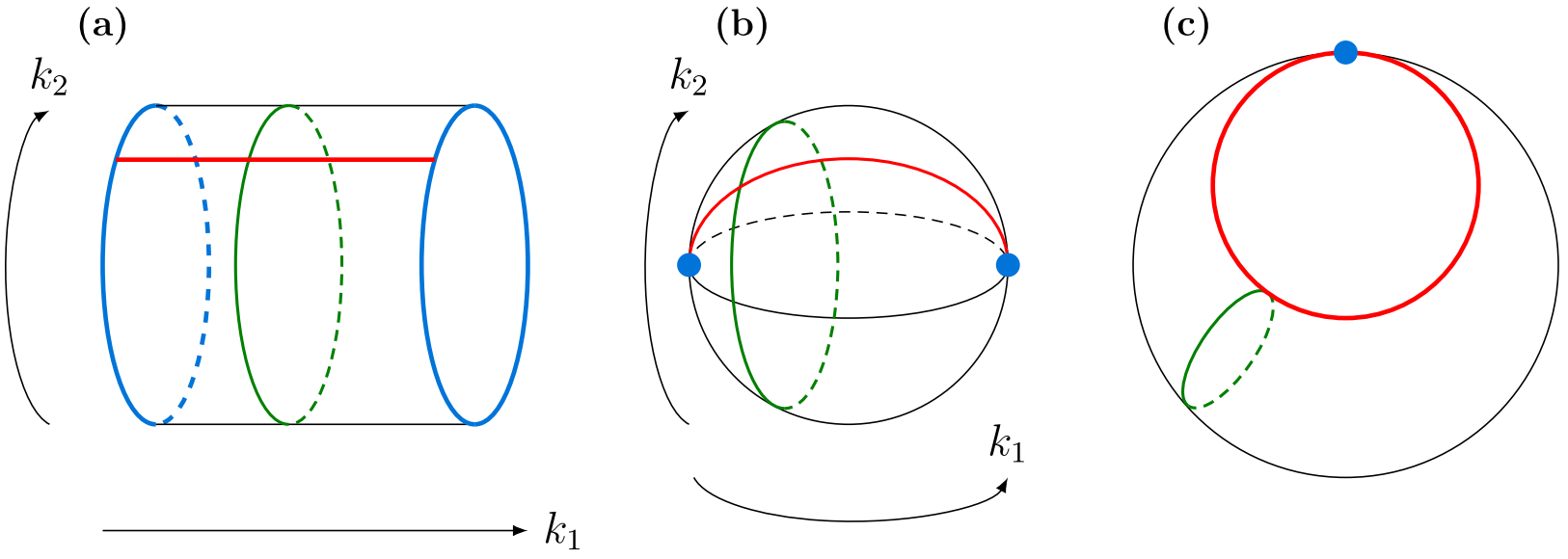

Fig. 6. The half Brillouin zone $BZ_+$ on which $\widetilde{V}(\frac{1}{2},\cdot)$ is defined is topologically a cylinder (a). Map $\widetilde{V}(\frac{1}{2},\cdot)$ sends the boundary $\partial(BZ_+)$ of the cylinder (in blue) to the unit matrix $I$. We want to maintain this property while deforming $\widetilde{V}(\frac{1}{2},\cdot)$ to the constant map equal to $I$. It helps to view $\widetilde{V}(\frac{1}{2},\cdot)$ as a map from the sphere (b) obtained by contracting the boundary circles to points (in blue). Furthermore, those two points corresponding to $k_1=0$ and $k_1=\pi$ should be identified producing an "earring" (a pinched torus) (c). Two kinds of non-trivial paths exist on such a surface. Loops on the original cylinder (in green) remain loops on the earring. Additionally, paths like the red one are also loops on the earring. The desired contraction of $\widetilde{V}(\frac{1}{2},\cdot)$ would also provide a contraction of the restriction of $\widetilde{V}(\frac{1}{2},\cdot)$ to such a red loop whose existence could be obstructed by a nontrivial winding of det $\widetilde{V}(r,\cdot)$ along that loop. Fortunately, the latter winding is trivial. Conversely, we can use a contraction of $\widetilde{V}(\frac{1}{2},\cdot)$ restricted to the red loop to deform $\widetilde{V}(\frac{1}{2},\cdot)$ to a map defined effectively on a 2-sphere, obtained by contracting the red loop on the earing to a marked point that is sent to $I$ at the end of the deformation. Finally, the map on the sphere with one marked point mapped to $I$ may be contracted to the constant map keeping the value at the marked point fixed. (For interpretation of the colors in this figure, the reader is referred to the web version of this article.)

such that

$$\widetilde{V}(r,0,0)=I=\widetilde{V}(r,\pi,0)\,,\qquad \widetilde{V}(\tfrac{3}{4},k_1,0)=I\,. \tag{3.48}$$

We subsequently extend the latter map to $(r,k_1,k_2)\in[\frac{1}{2},\frac{3}{4}]\times[0,\pi]\times[-\pi,\pi]$ by setting

$$\widetilde{V}(r,k_1,k_2)=\begin{cases}\widetilde{V}(r-|k_2|,k_1,0) & \text{for} \quad |k_2|\le r-\frac{1}{2}\,,\\ \widetilde{V}(\frac{1}{2},k_1,\frac{\pi(k_2-r+\frac{1}{2})}{\pi-r+\frac{1}{2}}) & \text{for} \quad r-\frac{1}{2}\le k_2\le\pi\,,\\ \widetilde{V}(\frac{1}{2},k_1,\frac{\pi(k_2+r-\frac{1}{2})}{\pi-r+\frac{1}{2}}) & \text{for} \quad -\pi\le k_2\le\frac{1}{2}-r\,,\end{cases} \tag{3.49}$$

so that for $\frac{1}{2}\le r\le\frac{3}{4}$,

$$\widetilde{V}(r,k_1,\pi)=\widetilde{V}(k_1,-\pi)\,,\qquad \widetilde{V}(r,0,k_2)=I=\widetilde{V}(r,\pi,k_2)\,. \tag{3.50}$$

Now,

$$\widetilde{V}(\tfrac{3}{4},k_1,0)=I \tag{3.51}$$

meaning that $\widetilde{V}(\frac{3}{4},\cdot)$ is effectively defined on a 2-sphere $S^2$ with one marked point obtained from the earring represented in Fig. 6(c) by identifying all the points $(k_1,0)$ on the red loop. Since $\pi_2(U(2M))=0$, $\widetilde{V}(\frac{3}{4},\cdot)$ may be further deformed to $\widetilde{V}(1,\cdot)=I$ in such a way that $\widetilde{V}(r,k_1,0)=I$ for $r\in[\frac{3}{4},1]$.

The above construction provides a continuous extension of $\widetilde{V}$ defined previously on $[0,\frac{1}{2}]\times BZ_+$ to $[\frac{1}{2},1]\times BZ_+$ in such a way that

$$\widetilde{V}(r,k_1,-\pi)=\widetilde{V}(r,k_1,\pi)\,, \tag{3.52}$$

$$\widetilde{V}(1,k_1,k_2) = I\,, \tag{3.53}$$

$$\widetilde{V}(r,0,k_2) = I = \widetilde{V}(r,\pi,k_2) \tag{3.54}$$

for $r \in [\frac{1}{2}, 1]$. Finally, we extend the resulting contraction $\widetilde{V}$ of map $V$ restricted to $\mathrm{BZ}_+$ to the whole BZ by time reversal (note that this is possible only because we have preserved time-reversal invariance on the boundaries of $\mathrm{BZ}_+$ all along the contraction). Setting

$$\widetilde{V}(r,k_1,k_2) = \Theta\,\widetilde{V}(r,-k_1,-k_2)\,\Theta^{-1} \quad \text{for} \quad 0 \le r \le 1 \quad \text{and} \quad -\pi \le k_1 \le 0\,, \tag{3.55}$$

we obtain a map (3.19) with properties (3.20) and (3.21).

To conclude this part, we construct the map $\widehat{V}_\epsilon$ used previously in the definition of index $K$. Taking $V(k) = V_\epsilon(T/2,k)$, we obtain a contraction $(r,k) \mapsto \widetilde{V}(r,k)$. The map

$$\widehat{V}_\epsilon(t,k) = \begin{cases} V_\epsilon(t,k) & \text{for } 0 \le t \le T/2\,, \\ \widetilde{V}\left(\frac{2}{T}\left(t-\frac{T}{2}\right),k\right) & \text{for } T/2 \le t \le T \end{cases} \tag{3.56}$$

has then the required properties.

## 3.3. Independence of the choice of contraction

Consider a map

$$[0,1] \times \mathrm{BZ} \ni (s,k) \longmapsto V(s,k) \in U(2M) \tag{3.57}$$

satisfying

$$V(0,k) = I = V(1,k), \tag{3.58}$$

$$\Theta\,V(s,k)\,\Theta^{-1} = V(s,-k). \tag{3.59}$$

In the following, we prove that the degree of such a map is an even integer.

To do so, we start by cutting BZ in to halves $\mathrm{BZ}_+$ and $\mathrm{BZ}_-$ interchanged by map $\vartheta_2 : k \mapsto -k$ which preserves orientation as BZ is two-dimensional. Hence

$$\begin{aligned}\deg(V) &= \frac{1}{24\pi^2} \int\limits_{[0,1]\times \mathrm{BZ}_+} \left(V^*\chi + (V\circ\vartheta_2)^*\chi\right) \\ &= \frac{1}{24\pi^2} \int\limits_{[0,1]\times \mathrm{BZ}_+} \left(V^*\chi + (\Theta V \Theta^{-1})^*\chi\right).\end{aligned} \tag{3.60}$$

Since $\chi$ is a real form, we end up with

$$\deg(V) = 2 \times \frac{1}{24\pi^2} \int\limits_{[0,1]\times \mathrm{BZ}_+} V^*\chi \tag{3.61}$$

so that $\deg(V)$ is even if and only if

$$\frac{1}{24\pi^2} \int\limits_{[0,1]\times \mathrm{BZ}_+} V^*\chi \;\in \mathbb{Z}\,. \tag{3.62}$$

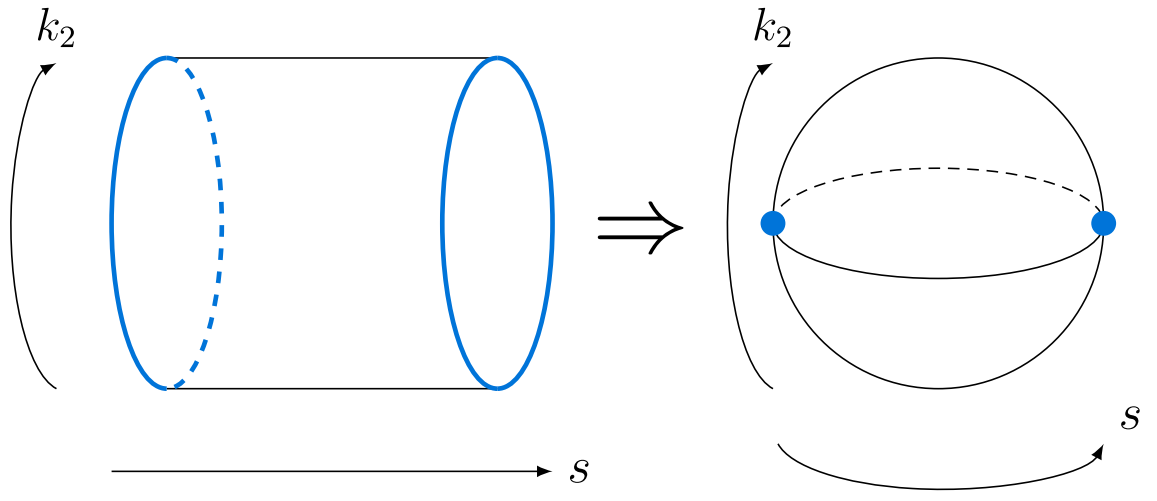

Fig. 7. As the map $V$ restricted to $k_1 = 0$ is periodic in the variable $k_2$, it may be viewed as a map from a cylinder. Moreover, since it is equal to $I$ on the whole boundary of the cylinder (in blue), where $s = 0$ or $s = 1$, it is effectively a map from a 2-sphere obtained by collapsing the boundary to two points (in blue). (For interpretation of the colors in this figure, the reader is referred to the web version of this article.)

To prove the last assertion, the general idea is to construct a map $\widetilde{V}$ extending $V$ to a manifold without boundary with the (integer) degree equal to the previous quantity. First consider the restriction of $V$ to $k_1 = 0$, i.e. the map

$$[0, 1] \times [-\pi, \pi] \ni (s, k_2) \longmapsto V(s, 0, k_2) \in U(2M)\,. \tag{3.63}$$

This is really a map from a 2-sphere $S^2$ to $U(2M)$, see Fig. 7, because of the relations

$$V(s, 0, -\pi) = V(s, 0, \pi)\,, \tag{3.64}$$

$$V(0, 0, k_2) = I = V(1, 0, k_2)\,. \tag{3.65}$$

Since $\pi_2(U(2M)) = 0$, it may be deformed to the constant map equal to $I$. We shall, however, need a very special of such deformations that preserves the time-reversal invariance property (3.59). The problem may look very similar to the previous one about the existence of a contraction discussed in Section 3.2. Here, however, the "horizontal" direction is $s$ instead of $k_1$ (compare Figs. 7 and 6) and we have to preserve the symmetry under $k_2 \mapsto -k_2$. Note that

$$\Theta\, V(s, 0, k_2)\, \Theta^{-1} = V(s, 0, -k_2) \tag{3.66}$$

In particular we have for $k_2 = 0$ and $\pi$

$$V(s, 0, 0),\; V(s, 0, \pi) \in Sp(2M) \subset U(2M) \tag{3.67}$$

so that the map

$$[0, 1] \times [0, \pi] \ni (s, k_2) \longmapsto V(s, 0, k_2) \in U(2M) \tag{3.68}$$

may be viewed as a map of the disc $D$ obtained by contracting the sides $\{0\} \times [0, \pi]$ and $\{1\} \times [0, \pi]$ in the square $[0, 1] \times [0, \pi]$ to points $p_0$ and $p_1$, see Fig. 8. The map sends the boundary $\partial D$ of $D$ to $Sp(2M)$ and $p_0, p_1 \in \partial D$ to the unit $I$ of this group. Since $Sp(2M)$ is simply connected, we may continuously deform the above map of $D$ in such way that the image of $\partial D$ stays in $Sp(2M)$ and those of $p_0$ and $p_1$ stay at $I$ during the deformation, with $\partial D$ mapped to $I$ at the end of the deformation. This may be done by contracting the two loops inside $Sp(2M)$ obtained from the map of $\partial D$ to the constant loop equal to $I$, see Fig. 8. We leave to the reader the task to write a formula for such a deformation of $V|_{k_1=0}$. The map resulting from the deformation may be naturally viewed as defined on a 2-sphere obtained by identifying $\partial D$ with a single marked point sent to $I$. It can be contracted to the constant map equal to $I$ since $\pi_2(U(2M))$ is trivial. To sum up, we have shown the existence of a homotopy

$$[0, 1] \times [0, 1] \times [0, \pi] \ni (r, s, k_2) \longmapsto \widetilde{V}_0(r, s, k_2) \in U(2M) \tag{3.69}$$

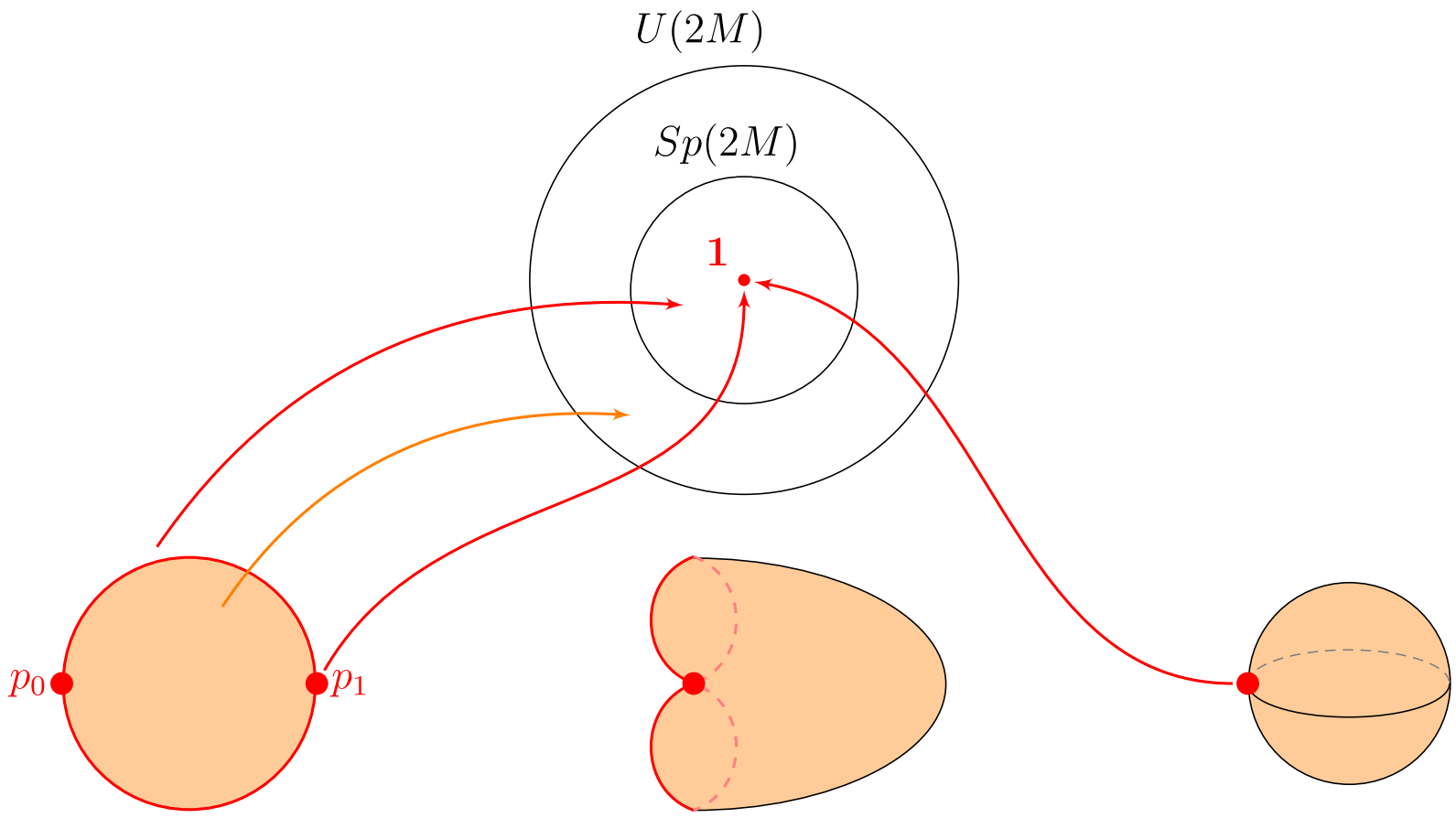

Fig. 8. The image by map (3.68) of the boundary $\partial D$ (at left, in red) of the disk $D$ (at left, in orange) is in $Sp(2M)$. Points $p_0$ and $p_1$ (in red) of the boundary are mapped to the identity of the group. The map is then deformed by contracting the boundary $\partial D$ to a point inside the symplectic group in order to obtain a map defined on a 2-sphere (at the right). (For interpretation of the colors in this figure, the reader is referred to the web version of this article.)

deforming the restriction (3.68) of $V$ to the constant map such that

$$\widetilde{V}_0(r=0,s,k_2)=V(s,0,k_2), \qquad \widetilde{V}_0(1,s;k_2)=I \tag{3.70}$$

and that the boundary conditions are preserved all along the contraction:

$$\widetilde{V}_0(r,0,k_2)=I=\widetilde{V}_0(r,1,k_2)\,, \tag{3.71}$$

$$\widetilde{V}_0(r,s,0),\ \widetilde{V}_0(r,s,\pi)\in Sp(2M)\,. \tag{3.72}$$

We then extend this map to the region $[0,1]\times[0,1]\times[-\pi,0]$ by setting

$$\widetilde{V}_0(r,s,k_2)=\Theta\,\widetilde{V}_0(r,s,-k_2)\,\Theta^{-1} \quad \text{for} \quad -\pi\le k_2\le 0\,, \tag{3.73}$$

which agrees with the previous definition on the square $[0,1]\times[0,1]\times\{0\}$ and satisfies $\widetilde{V}_0(r,s,-\pi)=\widetilde{V}_0(r,s,\pi)$. We end up with a map

$$[0,1]\times[0,1]\times[-\pi,\pi]\ni(r,s,k_2)\longmapsto\widetilde{V}_0(r,s,k_2)\in U(2M) \tag{3.74}$$

which is $2\pi$-periodic in $k_2$ and has properties (3.70), (3.71) and (3.73). This map completes the initial map $V$ restricted to $\mathrm{BZ}_+$ on the side $k_1=0$ if one considers $-r$ as extending the domain of $k_1$, see Fig. 9.

Similarly, we may complete this map on the side $k_1=\pi$ by a map $\widetilde{V}_\pi$ with analogous properties. Gluing $\widetilde{V}_0$, $V$ and $\widetilde{V}_\pi$, we end up with a map $\widetilde{V}$ which completes $V$ restricted to $\mathrm{BZ}_+$ to a continuous map with values in $U(2M)$ that may be viewed as defined on a 3-torus. Moreover,

$$(24\pi^2)\deg(\widetilde{V})=-\int\limits_{\mathcal{I}}\widetilde{V}_0^*\chi+\int\limits_{[0,1]\times\mathrm{BZ}_+}V^*\chi+\int\limits_{\mathcal{I}}\widetilde{V}_\pi^*\chi \tag{3.75}$$

with $\mathcal{I}=[0,1]\times[0,1]\times[-\pi,\pi]$. Now, by time-reversal invariance,

$$\int\limits_{\mathcal{I}}\widetilde{V}_0^*\chi=-\int\limits_{\mathcal{I}}(\widetilde{V}_0\circ\vartheta_1)^*\chi=-\int\limits_{\mathcal{I}}(\Theta\widetilde{V}_0\Theta^{-1})^*\chi=-\int\limits_{\mathcal{I}}\widetilde{V}_0^*\chi=0 \tag{3.76}$$

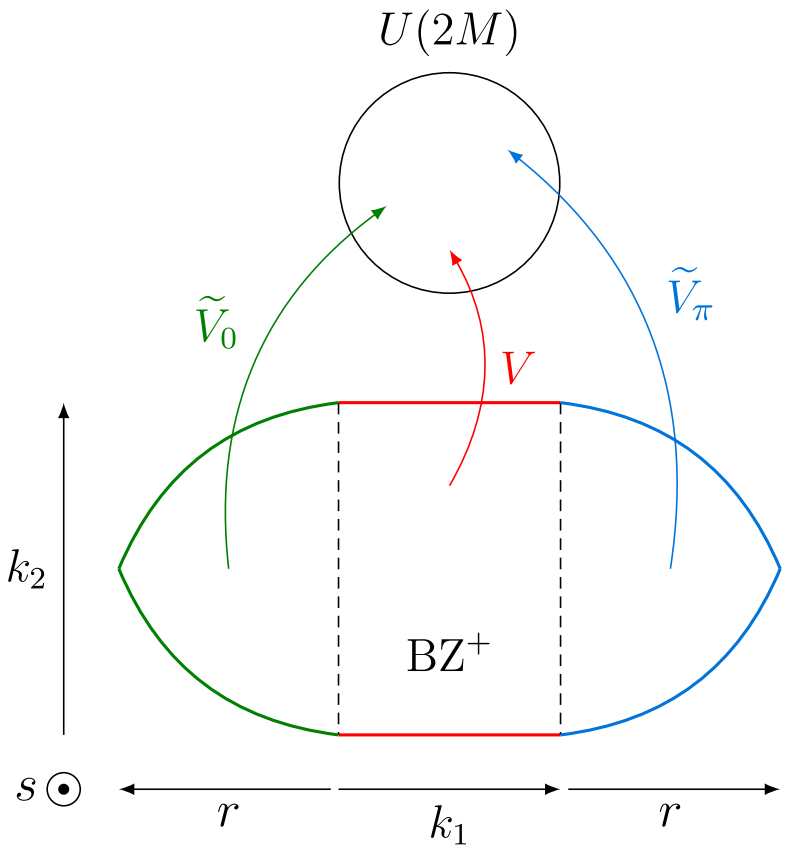

Fig. 9. Sketch of a constant-$s$ cross-section of the space on which $\widetilde{V}$ is defined. It is obtained by gluing the spaces on which $\widetilde{V}_0$, $V$, and $\widetilde{V}_\pi$ live (to be precise, the boundaries facing each other along the broken lines on the sketch are glued together). The top and the bottom boundary components corresponding to $k_2 = \pm\pi$ are also identified. (For interpretation of the colors in this figure, the reader is referred to the web version of this article.)

because $\vartheta_1 : (r, s, k_2) \mapsto (r, s, -k_2)$ changes the orientation. The integral of $\widetilde{V}_\pi$ vanishes for the same reason. We infer that

$$\frac{1}{24\pi^2} \int\limits_{[0,1]\times \mathrm{BZ}_+} V^*\chi = \deg(\widetilde{V}) \tag{3.77}$$

which is an integer by definition of the degree. This completes the argument showing that $\deg(V) \in 2\mathbb{Z}$.

## 4. Relation to the Kane–Mele invariant

### 4.1. Motivations and definition

In the situation where time-reversal invariance is not present, the invariants $W_\epsilon[U]$ are related to the first Chern numbers of quasienergy bands [44], as discussed in Section 2.3, see Eq. (2.41). Both the $W$ invariants and the first Chern numbers vanish when time-reversal invariance is enforced but, instead, one has a similar relation (3.14) between the $K_\epsilon[U]$ indices and the Kane–Mele invariants that we shall establish in this paper. Namely, we shall prove that

$$K_{\epsilon'}[U] - K_\epsilon[U] = \big(\deg(\widehat{V}_{\epsilon'}) - \deg(\widehat{V}_\epsilon)\big) \bmod 2 = \deg(\widehat{V}_\epsilon^{-1}\widehat{V}_{\epsilon'}) \bmod 2 \tag{4.1}$$

is equal to the Kane–Mele invariant $\mathrm{KM}(\mathcal{E}_{\epsilon,\epsilon'})$ [32] of the vector bundle $\mathcal{E}_{\epsilon,\epsilon'}$ of states in the band in-between the two gaps which is equipped with an antilinear involution $\theta$, see the end of Section 2.2. In simple situations, it is readily possible to show that this is indeed the case [5] (see also Appendix D and Section 4.3). A general proof requires some amount of work, but it also provides a relation between the (two-dimensional) Kane–Mele invariant and Wess–Zumino amplitudes.

We shall consider a general situation where a vector bundle $\mathcal{E}$ over the Brillouin torus BZ with fibers $P(k)\mathbb{C}^{2M}$ related to a continuous family of orthogonal projectors $P(k)$ is given and under

the time reversal

$$\Theta\, P(k)\, \Theta^{-1} = P(-k)\,. \tag{4.2}$$

The action of $\Theta$ defines an antilinear involution $\theta$ on $\mathcal{E}$ mapping the fiber over $k$ to that over $-k$. To those data, we associate a periodic family of unitary matrices

$$V(t,k) = \mathrm{e}^{\frac{2\pi \mathrm{i}t}{T} P(k)} = \mathrm{e}^{\frac{2\pi \mathrm{i}t}{T}} P(k) + I - P(k) = V(t+T,k) \tag{4.3}$$

with the time-reversal symmetry

$$\Theta\, V(t,k)\, \Theta^{-1} = V(-t,-k) = V(T-t,-k)\,. \tag{4.4}$$

In particular, at the half-period $V(T/2,k) = I - 2P(k)$ satisfying the relation

$$\Theta\, V(T/2,k)\, \Theta^{-1} = V(T/2,-k)\,. \tag{4.5}$$

We already know from Section 3.2 that there exists a contraction

$$[0,1]\times \mathrm{BZ} \ni (r,k) \mapsto \widetilde{V}(r,k) \in U(2M)\,, \tag{4.6}$$

$$\widetilde{V}(0,k) = I - 2P(k), \qquad \widetilde{V}(1,k) = I, \tag{4.7}$$

that is time-reversal invariant so that for all $r$,

$$\Theta\, \widetilde{V}(r,k)\, \Theta^{-1} = \widetilde{V}(r,-k)\,. \tag{4.8}$$

Proceeding as before when defining the $K_\epsilon[U]$ index, we consider the periodic map

$$\widehat{V}(t,k) = \begin{cases} V(t,k) & \text{for} \quad 0 \le t \le T/2\,, \\ \widetilde{V}((2t-T)/T,k) & \text{for} \quad T/2 \le t \le T\,, \end{cases} \tag{4.9}$$

as in relation (3.56), and we define an invariant associated to vector bundle $\mathcal{E}$ by setting

$$K[\mathcal{E}] = \deg(\widehat{V}) \mod 2\,, \tag{4.10}$$

with the right-hand side independent of the choice of the contraction $\widetilde{V}$ by the same considerations as in Sections 3.1 and 3.3. In the situation at hand, the contribution of times between 0 and $T/2$ for which $\widehat{V} = V$ to the integral for $\deg(\widehat{V})$ vanishes because of the relation $\Theta V(t,k)\Theta^{-1} = V(t,-k)^{-1}$. Indeed,

$$\begin{aligned} \int\limits_{[0,T/2]\times\mathrm{BZ}} V^*\chi &= \int\limits_{[0,T/2]\times\mathrm{BZ}} (\Theta V \Theta^{-1})^*\chi = \int\limits_{[0,T/2]\times\mathrm{BZ}} (V^{-1}\circ(\mathrm{Id}\times\vartheta_2))^*\chi \\ &= \int\limits_{[0,T/2]\times\mathrm{BZ}} (V^{-1})^*\chi = -\int\limits_{[0,T/2]\times\mathrm{BZ}} V^*\chi = 0\,, \end{aligned} \tag{4.11}$$

where the last-but-one equality is due to relation (2.36). It follows that

$$K[\mathcal{E}] = \frac{1}{24\pi^2} \int\limits_{[0,1]\times\mathrm{BZ}} \widetilde{V}^*\chi \mod 2\,. \tag{4.12}$$

We shall show in the remaining part of the paper that $K[\mathcal{E}]$ is equal to the Kane–Mele invariant [32] $\mathrm{KM}[\mathcal{E}]$ as expressed in [9].

Returning to the comparison of indices $K_\epsilon[U]$ and $K_{\epsilon'}[U]$, let us observe that the map $\widehat{V}_{\epsilon'}^{-1}\widehat{V}_\epsilon$ defined on $[0,T]\times\mathrm{BZ}$ satisfies the relations

$$\widehat{V}_\epsilon^{-1}\widehat{V}_{\epsilon'}(t,k) = \begin{cases} V_\epsilon^{-1}V_{\epsilon'}(t,k) = \mathrm{e}^{\frac{2\pi i t}{T}\tilde{P}_{\epsilon,\epsilon'}(k)} & \text{for} \quad 0 \le t \le T/2\,, \\ \Theta\, V_\epsilon^{-1}V_{\epsilon'}(t,-k)\,\Theta^{-1} & \text{for} \quad T/2 \le t \le T \end{cases} \tag{4.13}$$

following from Eqs. (3.1), (2.18) and (3.2). We may modify $\widehat{V}_\epsilon^{-1}\widehat{V}_{\epsilon'}$ by multiplying it by an appropriate power of map $\widehat{U}_0$ of Eq. (3.13) to replace in (4.13) matrices $\tilde{P}_{\epsilon,\epsilon'}(k)$ by orthogonal projectors $P_{\epsilon,\epsilon'}(k)$ differing from $\tilde{P}_{\epsilon,\epsilon'}(k)$ by an integer multiple of the unit matrix. This will not change the degree of $\widehat{V}_\epsilon^{-1}\widehat{V}_{\epsilon'}$. The corrected map $\widehat{V}_\epsilon^{-1}\widehat{V}_{\epsilon'}$ will then have the form (4.9) for the family $P(k) = P_{\epsilon,\epsilon'}(k)$ of projectors and we could rewrite relation (4.1) as the identity

$$K_{\epsilon'}[U] - K_\epsilon[U] = K[\mathcal{E}_{\epsilon,\epsilon'}]\,. \tag{4.14}$$

Hence the identification $K[\mathcal{E}] = \mathrm{KM}[\mathcal{E}]$ will establish the relation (3.14).

On the other hand, we may apply the construction of index $K[\mathcal{E}]$ to crystals with time-independent Hamiltonians $H(k)$ that possess the time-reversal symmetry

$$\Theta\, H(k)\, \Theta^{-1} = H(-k)\,. \tag{4.15}$$

For insulators, where $H(k)$ have a common spectral gap inside which the Fermi energy is located, one may take for $P(k)$ the orthogonal projectors on the valence-bands states with the energy below the gap. Property (4.2) follows then from the symmetry (4.15). The resulting valence-band vector bundle $\mathcal{E}$ has vanishing Chern number and, together with the antilinear involution $\theta$ induced by the action of $\Theta$, is fully characterized by the $\mathbb{Z}_2$-valued Kane–Mele invariant $\mathrm{KM}[\mathcal{E}] \in \mathbb{Z}_2$ [7,12]. Our results identifying the indices $K[\mathcal{E}]$ and $\mathrm{KM}[\mathcal{E}]$ will then provide a new expression for the Kane–Mele invariant in the original context of valence-state bundle for topological insulators with time-reversal invariant time-independent Hamiltonians.

### *4.2. Relation of $K[\mathcal{E}]$ to Wess–Zumino amplitudes*

Formula (4.12) permits to relate index $K[\mathcal{E}]$ to the value of the Wess–Zumino (WZ) action functional of the two-dimensional classical field

$$\mathrm{BZ} \ni k \mapsto I - 2P(k) \equiv U_P(k) \in U(2M)\,. \tag{4.16}$$

The WZ action functional plays an important role in Wess–Zumino–Witten models [49,42] of conformal field theory, see [15] for a more detailed introduction. For a map $G$ from a closed oriented two-dimensional surface $\Sigma$ to the unitary group $U(N)$ for which there exists a smooth homotopy $\widetilde{G} : [0,1] \times \Sigma \to U(N)$ such that

$$\widetilde{G}(1,x) = G(x)\,, \qquad G(0,x) = I \tag{4.17}$$

the WZ-action may be defined by the expression [49]

$$S_{\mathrm{WZ}}(G) = \tfrac{1}{12\pi} \int\limits_{[0,1]\times\Sigma} \widetilde{G}^*\chi\,. \tag{4.18}$$

The dependence on the choice of homotopy makes $S_{\mathrm{WZ}}(G)$ determined only modulo $2\pi$, but the corresponding WZ Feynman amplitude $\exp[\mathrm{i}S_{\mathrm{WZ}}(G)]$ is unambiguous. If the values of $G$ end up in a region of $U(N)$ on which $\chi = dB$ for a 2-form $B$ then, by the Stokes theorem,

$$S_{\mathrm{WZ}}(G) = \tfrac{1}{12\pi} \int\limits_{\Sigma} G^*B\,, \tag{4.19}$$

but in general, the latter expression may be used only locally since, globally, $\chi$ is a 3-form that is closed but not exact. The gluing of such local expressions together consistently involves certain geometric structure on the target group $U(N)$, called a bundle gerbe with connection [16], that permits to define $\exp[\mathrm{i}S_{\mathrm{WZ}}(G)$ even if $\det G$ has nontrivial windings so that contraction $\widetilde{G}$ does not exist.

At the first sight,

$$K[\mathcal{E}] = -\tfrac{1}{2\pi} S_{\mathrm{WZ}}(U_P) \ \bmod 2 \tag{4.20}$$

(the minus sign is due to different orientations, but is immaterial for the modulo 2 value). Nevertheless, since normally $S_{\mathrm{WZ}}(U_P)$ is defined modulo $2\pi$, the right-hand side is not well defined. The problem is solved by imposing restriction (4.8) on the admissible contractions of $U_P$. As follows from our previous analysis, such a restriction makes action $S_{\mathrm{WZ}}(U_P)$ well defined modulo $4\pi$ rather than modulo $2\pi$, which is what is needed to give sense to the right-hand side of (4.20). We may summarize the relation between $K[\mathcal{E}]$ and the WZ action $S_{\mathrm{WZ}}(U_P)$ in the identity

$$(-1)^{K[\mathcal{E}]} = \Big(\exp[\mathrm{i}S_{\mathrm{WZ}}(U_P)]\Big)^{1/2}, \tag{4.21}$$

where the square root of the WZ amplitude appearing on the right-hand side is specified by imposing the time-reversal symmetry (4.8) on possible contractions of $U_P$ (note that the WZ amplitude $\exp[\mathrm{i}S_{\mathrm{WZ}}(U_P)]$ itself is equal to 1). In the language of bundle gerbes, specifying the square root of WZ amplitudes, as in relation (4.21), requires an additional equivariant structure on the gerbe with respect to an involution on the target space given here by $U \mapsto \Theta U \Theta^{-1}$. We shall discuss elsewhere such geometric aspects underlying formula (4.21) and its generalizations, somewhat similar to those governing the construction of orientifold amplitudes [48,17], but with notable distinctions.

### 4.3. *A simple example with $K[\mathcal{E}] = \mathrm{KM}[\mathcal{E}]$*

Let us start by illustrating the equality of indices $K[\mathcal{E}]$ and $\mathrm{KM}[\mathcal{E}]$ on a simple example. Perhaps the most basic class of models exhibiting a nontrivial Kane–Mele phase is obtained by imposing both time-reversal and inversion invariance, as was done by Fu and Kane in [10]. In its simplest version, this class of models is described by the Hamiltonians acting in space $\mathbb{C}^2 \otimes \mathbb{C}^2$

$$H(k) = d_1(k)\Gamma_1 + d_2(k)\Gamma_2 + d_5(k)\Gamma_5\,, \tag{4.22}$$

where $d : \mathrm{BZ} \to \mathbb{R}^3$ depends on the model and

$$\Gamma_1 = I \otimes \sigma_x\,, \qquad \Gamma_2 = I \otimes \sigma_y\,, \qquad \Gamma_5 = s_z \otimes \sigma_z\,, \tag{4.23}$$

with two sets $s_i$ and $\sigma_i$ of Pauli matrices, the first one corresponding to spin degrees of freedom, and the second one to unspecified internal degrees of freedom (e.g. points in the unit cell of a non-Bravais lattice). The time-reversal operator assumes the form

$$\Theta = \mathrm{i}(s_y \otimes I)C \tag{4.24}$$

where $C$ is the complex conjugation and the action of inversion is represented by

$$\Pi = I \otimes \sigma_x\,. \tag{4.25}$$

Hamiltonian $H(k)$ has both time-reversal invariance and inversion symmetry when $k \mapsto d_1(k)$ is an even function and $k \mapsto d_{2,3}(k)$ are odd functions.

It will be convenient to consider the Hamiltonian $H(\boldsymbol{d})$ as a function of the parameter vector $\boldsymbol{d} = (d_1, d_2, d_5)$. The eigenvalues of $H(\boldsymbol{d})$ are $\pm|\boldsymbol{d}|$, each of which two-fold degenerate. The corresponding eigenvectors depend only on the normalized vector

$$\boldsymbol{n} = \frac{\boldsymbol{d}}{|\boldsymbol{d}|} \tag{4.26}$$

which is well-defined as long as $H$ is gapped, i.e. describes a band insulator. This vector can be parameterized as

$$(n_1 + \mathrm{i}n_2, n_5) = \left(\sqrt{1-n_5^2}\,\mathrm{e}^{\mathrm{i}\theta}, n_5\right). \tag{4.27}$$

The Kramers pair of eigenvectors of $H(\boldsymbol{d})$ with eigenvalue $-|\boldsymbol{d}|$ (corresponding to the valence subbundle) which are also eigenvectors of the spin operator $\frac{1}{2}(s_z \otimes I)$ is [8]

$$u_\uparrow(\boldsymbol{n}) = \frac{1}{\sqrt{2}}\begin{pmatrix} -\sqrt{1-n_5} \\ \sqrt{1+n_5}\,\mathrm{e}^{\mathrm{i}\theta} \\ 0 \\ 0 \end{pmatrix} \qquad \text{and} \qquad u_\downarrow(\boldsymbol{n}) = \frac{1}{\sqrt{2}}\begin{pmatrix} 0 \\ 0 \\ -\sqrt{1+n_5} \\ \sqrt{1-n_5}\,\mathrm{e}^{\mathrm{i}\theta} \end{pmatrix} \tag{4.28}$$

(we use the basis of $\mathbb{C}^2 \otimes \mathbb{C}^2 \cong \mathbb{C}^4$ where first two coordinates correspond to spin up and the last two to spin down). The fact that the above vectors form a Kramers pair resides in the relation

$$\Theta u_\uparrow(\boldsymbol{n}) = u_\downarrow(\vartheta\boldsymbol{n}), \tag{4.29}$$

where

$$\vartheta(n_1, n_2, n_5) = (n_1, -n_2, -n_5) \qquad \text{or, equivalently,} \qquad \vartheta(\theta, n_5) = (-\theta, -n_5), \tag{4.30}$$

so that $\boldsymbol{n}(-k) = \vartheta\boldsymbol{n}(k)$. The Kramers pair (4.28) is well-defined and smooth, except at points $(n_1, n_2, n_5) = (0, 0, \pm 1)$ where it is singular:

$$u_\uparrow(\boldsymbol{n}) \underset{n_5 \to 1}{\approx} \begin{pmatrix} 0 \\ \mathrm{e}^{\mathrm{i}\theta} \\ 0 \\ 0 \end{pmatrix}, \qquad u_\downarrow(\boldsymbol{n}) \underset{n_5 \to 1}{\approx} \begin{pmatrix} 0 \\ 0 \\ -1 \\ 0 \end{pmatrix} \tag{4.31}$$

and similarly

$$u_\uparrow(\boldsymbol{n}) \underset{n_5 \to -1}{\approx} \begin{pmatrix} -1 \\ 0 \\ 0 \\ 0 \end{pmatrix}, \qquad u_\downarrow(\boldsymbol{n}) \underset{n_5 \to -1}{\approx} \begin{pmatrix} 0 \\ 0 \\ 0 \\ \mathrm{e}^{\mathrm{i}\theta} \end{pmatrix}. \tag{4.32}$$

The Kramer pair possesses a protected singular behavior only when the system is topologically non-trivial, which is in agreement with the understanding of the Kane–Mele $\mathbb{Z}_2$ invariant as a geometric obstruction [12]. The projectors on these states, however, are always well-defined. We can therefore express the projector on the valence band constituted by eigenvectors of $H(\boldsymbol{d})$ with eigenvalues $-|\boldsymbol{d}|$ as

$$P(\boldsymbol{n}) = |u_\uparrow(\boldsymbol{n})\rangle\langle u_\uparrow(\boldsymbol{n})| + |u_\downarrow(\boldsymbol{n})\rangle\langle u_\downarrow(\boldsymbol{n})|. \tag{4.33}$$

In the matrix form

$$P(\boldsymbol{n})=\frac{1}{2}\begin{pmatrix} 1-n_5 & -\sqrt{1-n_5^2}\,\mathrm{e}^{-\mathrm{i}\theta} & 0 & 0\\ -\sqrt{1-n_5^2}\,\mathrm{e}^{\mathrm{i}\theta} & 1+n_5 & 0 & 0\\ 0 & 0 & 1+n_5 & -\sqrt{1-n_5^2}\,\mathrm{e}^{-\mathrm{i}\theta}\\ 0 & 0 & -\sqrt{1-n_5^2}\,\mathrm{e}^{\mathrm{i}\theta} & 1-n_5 \end{pmatrix} \tag{4.34}$$

which is block-diagonal:

$$P(\boldsymbol{n})=P_\uparrow(\boldsymbol{n})\oplus P_\downarrow(\boldsymbol{n}). \tag{4.35}$$

One can explicitly check time-reversal invariance of this projector

$$\Theta P(\boldsymbol{n})\Theta^{-1}=P(\vartheta\boldsymbol{n}) \tag{4.36}$$

which translates in terms of the blocks to the relation

$$P_\downarrow(\boldsymbol{n})=\overline{P_\uparrow(\vartheta\boldsymbol{n})}. \tag{4.37}$$

Projectors $P(\boldsymbol{n}(k))$ define a vector bundle $\mathcal{E}$ on the 2-torus to which are associated the indices $K[\mathcal{E}]$ and KM$[\mathcal{E}]$.

Following the general scheme of Section 4.1, consider the family of unitary matrices

$$V(\varphi,\boldsymbol{n})=\mathrm{e}^{\mathrm{i}\varphi P(\boldsymbol{n})}=\mathrm{e}^{\mathrm{i}\varphi}P(\boldsymbol{n})+I-P(\boldsymbol{n}) \tag{4.38}$$

which continuously interpolates between the identity matrix and operators

$$U_P(\boldsymbol{n})=I-2P(\boldsymbol{n}). \tag{4.39}$$

as $\varphi$ (representing the parameter $2\pi t/T$ used in Section 4.1) varies from 0 to $\pi$.

$$\Theta\, V(\varphi,\boldsymbol{n})\,\Theta^{-1}=V(-\varphi,\vartheta\boldsymbol{n})=V^{-1}(\varphi,\vartheta\boldsymbol{n})\,. \tag{4.40}$$

We shall explicitly construct the map $\widehat{V}$ of (4.9) and compute its degree modulo 2 (which is by definition the index $K[\mathcal{E}]$) as the Hopf degree modulo 2 [38] of the map $d:\mathrm{BZ}\to S^2$. To achieve this, we set

$$\widehat{V}(\varphi,\boldsymbol{n})=\begin{cases} V(\varphi,\boldsymbol{n}) & \text{for } 0\le\varphi\le\pi\\ (\mathrm{e}^{\mathrm{i}\varphi}P_\uparrow(\boldsymbol{n})+I-P_\uparrow(\boldsymbol{n}))\oplus(\mathrm{e}^{-\mathrm{i}\varphi}P_\downarrow(\boldsymbol{n})+I-P_\downarrow(\boldsymbol{n})) & \text{for } \pi\le\varphi\le 2\pi \end{cases} \tag{4.41}$$

This definition preserves the block structure: $\widehat{V}=\widehat{V}_\uparrow\oplus\widehat{V}_\downarrow$. With this choice, $\widehat{V}$ is well defined at $\varphi=\pi$ where $\widehat{V}(\pi,\boldsymbol{n})=U_P(\boldsymbol{n})$. Besides,

$$\Theta\,\widehat{V}(\varphi,\boldsymbol{n})\,\Theta^{-1}=\widehat{V}(\varphi,\vartheta\boldsymbol{n})\qquad\text{for}\quad \pi\le\varphi\le 2\pi\,. \tag{4.42}$$

One has to compute $\deg(\widehat{V})$ which factorizes over the blocks as

$$\deg(\widehat{V})=\deg(\widehat{V}_\uparrow)+\deg(\widehat{V}_\downarrow). \tag{4.43}$$

The second term $\deg(\hat{V}_\downarrow)$ vanishes, as the contributions on $[0,\pi]$ and on $[\pi,2\pi]$ are opposite and cancel because $\widehat{V}_\downarrow(\varphi,\boldsymbol{n})=\widehat{V}_\downarrow(2\pi-\varphi,\boldsymbol{n})$ so that

$$\deg(\widehat{V})=\deg(\widehat{V}_\uparrow)=\deg(V_\uparrow). \tag{4.44}$$

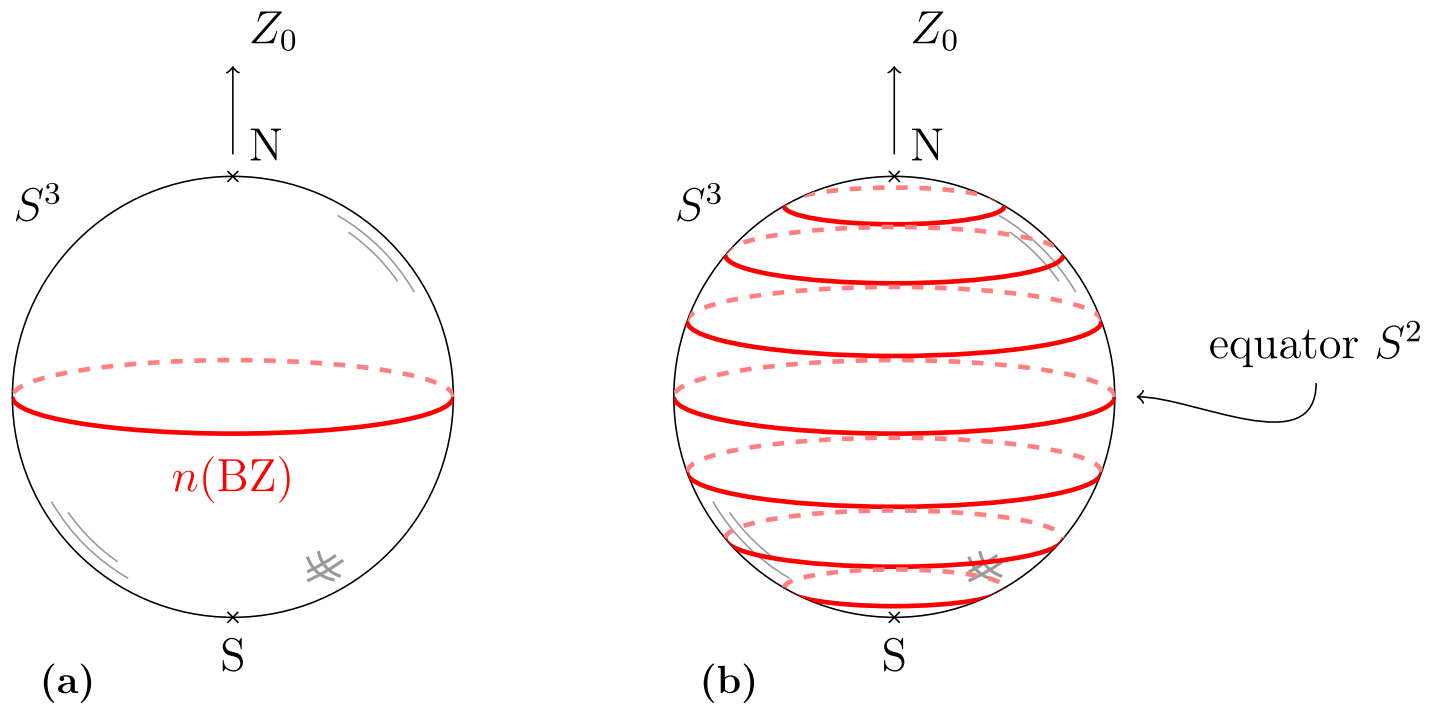

Fig. 10. The image $\boldsymbol{n}$(BZ) in the 2-sphere $S^2$ is spread under the map $Z : [0, 2\pi] \times S^2 \to S^3$ from the equator $Z^0 = 0$ obtained for $\varphi = \pi$ to 2-spheres corresponding other values of $Z^0 = \cos(\varphi/2)$ displayed for several angles $\varphi$. When $\varphi$ runs through $[0, 2\pi]$, the whole 3-sphere is covered as many times as the map $\mathrm{BZ} \ni k \mapsto \boldsymbol{n}(k)$ covers $S^2$ (up to the sign).

To compute this last term, it is useful to decompose the two-by-two unitary matrix $V_\uparrow$ in terms of the Pauli matrices and a phase. Introducing a 4-vector of matrices $(\tau_\mu) = (I, \mathrm{i}\sigma_x, \mathrm{i}\sigma_y, \mathrm{i}\sigma_z)$, we shall write

$$V_\uparrow(\varphi, \boldsymbol{n}) = \mathrm{e}^{\mathrm{i}\varphi/2}\, Z^\mu(\varphi, \boldsymbol{n})\, \tau_\mu\,, \tag{4.45}$$

where

$$Z^0(\varphi, \boldsymbol{n}) = +\cos(\varphi/2)\,, \tag{4.46a}$$

$$Z^1(\varphi, \boldsymbol{n}) = -\sin(\varphi/2)\sqrt{1 - n_5^2}\,\cos\theta\,, \tag{4.46b}$$

$$Z^2(\varphi, \boldsymbol{n}) = -\sin(\varphi/2)\sqrt{1 - n_5^2}\,\sin\theta\,, \tag{4.46c}$$

$$Z^3(\varphi, \boldsymbol{n}) = -\sin(\varphi/2)\, n_5\,. \tag{4.46d}$$

Note that $\Sigma_\mu (Z^\mu)^2 = 1$ so that $Z^\mu \tau_\mu \in SU(2) \cong S^3$. One has

$$V_\uparrow^* \chi = Z^* \chi = 2\,\epsilon_{\mu\nu\rho\sigma} Z^\mu \mathrm{d}Z^\nu \mathrm{d}Z^\rho \mathrm{d}Z^\sigma \tag{4.47}$$

with the form on the right equal to 12 times the standard volume form on $S^3$ (whose total volume is equal to $2\pi^2$). With $Z$ given by Eq. (4.46),

$$2\,\epsilon_{\mu\nu\rho\sigma} Z^\mu \mathrm{d}Z^\nu \mathrm{d}Z^\rho \mathrm{d}Z^\sigma = -6\sin^2(\varphi/2)\,\mathrm{d}\varphi\,\varpi\,, \tag{4.48}$$

where $\varpi = \frac{1}{2}\epsilon_{ijk} n^i \mathrm{d}n^j \mathrm{d}n^k$ is the area form on $S^2$ (with total area $4\pi$). We infer that

$$\deg(V_\uparrow) = \frac{1}{24\pi^2} \int\limits_{[0,2\pi]\times\mathrm{BZ}} V_\uparrow^* \chi = -\frac{1}{4\pi} \int\limits_{\mathrm{BZ}} \boldsymbol{n}^* \varpi\,, \tag{4.49}$$

i.e. that $\deg(V_\uparrow)$ is equal to minus the number of times the map $\mathrm{BZ} \ni k \mapsto \boldsymbol{n}(k) \in S^2$ covers the two-dimensional sphere, see Fig. 10.

We are therefore led to the calculation of this number. As the final result will only need it modulo 2, it is enough to compute the parity of the cardinality $\#\boldsymbol{n}^{-1}(\{x\})$ of the set of preimages of a regular value $x \in S^2$ [38]. To establish a link with the usual expression for the Kane–Mele invariant in this case given in Ref. [10], it is convenient to look at

$$\#\boldsymbol{n}^{-1}(\{(-1,0,0)\}) \quad \text{mod } 2 \tag{4.50}$$

assuming that $(-1,0,0)$ is a regular value (which is the case generically). At first sight, we do not know much about the set $\boldsymbol{n}^{-1}(\{(-1,0,0)\})$. However, we can keep in the calculation only the time-reversal invariant momenta (TRIM) $k^*$, where $k^* = -k^*$ (modulo reciprocal lattice vectors), as other preimages will come in pairs (because of the parity of functions $n_i$ on the Brillouin torus). Besides, at TRIM $k^*$, the components $n_2(k^*)$ and $n_5(k^*)$ vanish so that $n_1(k^*) = \pm 1$. Hence the number of TRIM $k^*$ where $\boldsymbol{n}(k^*) = (-1,0,0)$ is simply

$$\sum_{\substack{\text{TRIM}\\ k^*}} \frac{1-n_1(k^*)}{2}\,. \tag{4.51}$$

We infer then that

$$K[\mathcal{E}] = \deg(V_\uparrow) \text{ mod } 2 = \sum_{\substack{\text{TRIM}\\ k^*}} \frac{1-n_1(k^*)}{2} \quad \text{mod } 2 \tag{4.52}$$

which corresponds to a multiplicative formula

$$(-1)^{K[\mathcal{E}]} = \prod_{\substack{\text{TRIM}\\ k^*}} n_1(k^*). \tag{4.53}$$

On the other hand, this is the expression for KM[$\mathcal{E}$] from Fu and Kane [10] for topological insulators with inversion. Thus we have shown, in this very simple example, that $K[\mathcal{E}] = \text{KM}[\mathcal{E}]$.

### 4.4. $K[\mathcal{E}]$ in a new basis

In order to prove that the two indices are equal in a general situation, we shall rewrite the expression for $K[\mathcal{E}]$ in another basis where the "sewing matrix" used in the formula of Ref. [9] for KM[$\mathcal{E}$] appears explicitly. In the new basis, we shall use a quasi-explicit contraction of $V(T/2)$ which will allow to reduce the expression for $K[\mathcal{E}]$ to a combination of two-dimensional integrals and of Wess–Zumino actions of certain fields. The latter will be computed using local presentations for the corresponding WZ amplitudes first in the case of rank 2 bundle $\mathcal{E}$ and then for a general rank.

Since bundles $\mathcal{E}$ and $\mathcal{E}^\perp$, the latter with fibers $(I - P(k))\mathbb{C}^{2M}$, have trivial Chern numbers, they are trivializable [41]. Thus there exists a continuous $k$-dependent orthonormal base $|e_i(k)\rangle$, $i = 1, \ldots, 2M$, defined globally over BZ such that

$$P(k) = \sum_{i=1}^{2m} |e_i(k)\rangle\langle e_i(k)|\,, \qquad I - P(k) = \sum_{i=2m+1}^{2M} |e_i(k)\rangle\langle e_i(k)|\,, \tag{4.54}$$

where $m$ is the $k$-independent rank of projectors $P(k)$. Besides, $e_i(k)$ may be chosen smooth in their dependence on $k$. Consider the unitary operators

$$R(k) = \sum_{i=1}^{2M} |e_i(k)\rangle\langle f_i| \qquad \text{and} \qquad R(k)^{-1} = \sum_{i=1}^{2M} |f_i\rangle\langle e_i(k)| \tag{4.55}$$

of basis change from the canonical $k$-independent basis $|f_i\rangle,\ i=1,\dots,2M$, of $\mathbb{C}^{2M}$. Clearly,

$$R(k)^{-1}\,P(k)\,R(k)=\sum_{i=1}^{2m}|f_i\rangle\langle f_i|\ \equiv\ P_0, \tag{4.56}$$

$$R(k)^{-1}\,(I-P(k))\,R(k)=\sum_{i=2m+1}^{2M}|f_i\rangle\langle f_i|=I-P_0\,. \tag{4.57}$$

Relation (4.2) implies that

$$W(k)\,P_0\,W(k)^{-1}=P_0 \tag{4.58}$$

for operators

$$W(k)=\sum_{i,j=1}^{2M}W^{ij}(k)|f_i\rangle\langle f_j|, \tag{4.59}$$

with

$$W^{ij}(k)=\langle e_i(-k)|\Theta e_j(k)\rangle=-W^{ji}(-k)\,. \tag{4.60}$$

Note that $W^{ij}(k)=0$ if $i\leq 2m$ and $j\geq 2m+1$ or *vice versa*. Operators $W(k)$ are unitary as

$$\sum_{j=1}^{2M}\overline{W^{ji}(k)}\,W^{jl}(k)=\langle\Theta e_i(k)|e_j(-k)\rangle\langle e_j(-k)|\Theta e_l(k)\rangle=\delta_{il}\,. \tag{4.61}$$

Given a map $\widetilde{V}$ as in (4.6), we transform it to the new basis by defining

$$\widetilde{T}(r,k)=\ R(k)^{-1}\widetilde{V}(r,k)\,R(k)\,. \tag{4.62}$$

The above relation establishes a one to one correspondence between contractions $\widetilde{V}$ with properties (4.7) and (4.8) and maps $\widetilde{T}$ such that

$$\widetilde{T}(1,k)=I-2P_0,\qquad \widetilde{T}(0,k)=I, \tag{4.63}$$

$$W(k)\,\overline{\widetilde{T}(r,k)}\,W(k)^{-1}=\widetilde{T}(r,-k)\,. \tag{4.64}$$

Note that the starting and ending points of contraction $\widetilde{T}$ are $k$-independent but the corresponding time-reversal operator $W$ depends on $k$. Eq. (4.12) takes in the new basis a similar form

$$K[\mathcal{E}]=-\tfrac{1}{24\pi^2}\int\limits_{[0,1]\times\mathrm{BZ}}\widetilde{T}^*\chi\quad\mathrm{mod}\ 2\,, \tag{4.65}$$

but this requires an argument. We use formula (A.4) from Appendix A which implies that

$$\widetilde{V}^*\chi=(R\,\widetilde{T}R^{-1})^*\chi=\widetilde{T}^*\chi\,+\,3\,\mathrm{d}(R,\widetilde{T})^*\beta \tag{4.66}$$

where $\beta$ is a 2-form on $U(N)\times U(N)$ given by Eq. (A.5) in Appendix A. The above identity results in the relation

$$-\tfrac{1}{24\pi^2}\int\limits_{[0,1]\times\mathrm{BZ}}\widetilde{V}^*\chi=-\tfrac{1}{24\pi^2}\int\limits_{[0,1]\times\mathrm{BZ}}\widetilde{T}^*\chi\,+\,\mathcal{B}(\widetilde{T})\,, \tag{4.67}$$

where the boundary term $\mathcal{B}(\widetilde{T})$

$$\mathcal{B}(\widetilde{T}) = \tfrac{1}{8\pi^2}\Big(\int\limits_{\{1\}\times\mathrm{BZ}} - \int\limits_{\{0\}\times\mathrm{BZ}}\Big)\mathrm{tr}\Big(\widetilde{T}R^{-1}(\mathrm{d}R)\widetilde{T}^{-1}R^{-1}(\mathrm{d}R) \\ + R^{-1}(\mathrm{d}R)\big(\widetilde{T}^{-1}(\mathrm{d}\widetilde{T}) + (\mathrm{d}\widetilde{T})\widetilde{T}^{-1}\big)\Big). \tag{4.68}$$

It is easy to see, however, that the boundary integrals vanish because $\widetilde{T}(r,k)$ is $k$-independent for $r = 0, 1$ and satisfies $\widetilde{T}(r,k) = \widetilde{T}(r,k)^{-1}$ for the same values of $r$ implying that

$$\mathrm{tr}\Big(\widetilde{T}R^{-1}(\mathrm{d}R)\widetilde{T}^{-1}R^{-1}(\mathrm{d}R)\Big) = -\,\mathrm{tr}\Big(\widetilde{T}^{-1}R^{-1}(\mathrm{d}R)\widetilde{T}R^{-1}(\mathrm{d}R)\Big) = 0 \tag{4.69}$$

when $r = 0, 1$. This proves formula (4.65).

Due to the block form of $W(k)$ and of $1 - 2P_0$, we may look for a contraction $\widetilde{T}$ such that

$$\widetilde{T}(r,k) = \widetilde{t}(r,k) + (I - P_0)\,, \tag{4.70}$$

where

$$\widetilde{t}(r,k) = \sum_{i,j=1}^{2m} \widetilde{t}^{\,ij}(r,k)|f_i\rangle\langle f_j| \tag{4.71}$$

is a linear map acting in the $2m$-dimensional subspace spanned by $|f_i\rangle$ with $i = 1, \dots, 2m$ and is such that

$$\widetilde{t}(0,k) = -I, \qquad \widetilde{t}(1,k) = I \tag{4.72}$$

$$w(k)\,\overline{\widetilde{t}(r,k)}\,w(k)^{-1} = \widetilde{t}(r,-k)\,, \tag{4.73}$$

for a linear map $w(k)$ that is the $2m$-dimensional block of $W(k)$,

$$w(k) = \sum_{i,j=1}^{2m} W^{ij}(k)|f_i\rangle\langle f_j|\,, \tag{4.74}$$

see relation (4.60). For such a contraction,

$$K[\mathcal{E}] = -\tfrac{1}{24\pi^2}\int\limits_{[0,1]\times\mathrm{BZ}} \widetilde{T}^*\chi \mod 2 = -\tfrac{1}{24\pi^2}\int\limits_{[0,1]\times\mathrm{BZ}} \widetilde{t}^*\chi \mod 2 \tag{4.75}$$

and the calculation is reduced to that for linear maps in the $(2m)$-dimensional subspace.

### *4.5. Sewing matrices*

From now on, we shall make no distinction between linear transformations of $\mathbb{C}^{2m}$ and $(2m) \times (2m)$ matrices. In particular, the unitary operator $w(k)$ will be identified with the "sewing matrix" in the unitary group $U(2m)$ with the entries

$$w^{ij}(k) = \langle e_i(-k)|\Theta e_j(k)\rangle = -w^{ji}(-k)\,, \qquad i, j = 1, \dots 2m\,, \tag{4.76}$$

obtained from the entries of (4.60) by restricting the range of $i, j$. Due to the relation $w(k) = -w(-k)^T$ that implies that $\det w(k) = \det w(-k)$, function $\det w(k)$ has no non-trivial windings on BZ and hence it possesses a continuous $(2m)$-th root $(\det w(k))^{1/(2m)} = (\det w(-k))^{1/(2m)}$ defined up to a global $(2m)$-th root of 1 that will be immaterial in what follows. We shall set

$$\widetilde{w}(k) = \frac{w(k)}{(\det w(k))^{1/(2m)}} . \tag{4.77}$$

Of course, $\det \widetilde{w}(k) = 1$ so that $\widetilde{w}(k) \in SU(2m)$. Note that in relation (4.73), $w(k)$ may be replaced by $\widetilde{w}(k)$. We also still have the relation

$$\widetilde{w}(k) = -\widetilde{w}(-k)^T . \tag{4.78}$$

In the four TRIM in BZ with $k^* = (a, a') = -k^*$ for $a, a' = 0, \pi$, matrix $\widetilde{w}(k^*)$ is antisymmetric. Each antisymmetric matrix $\widetilde{w}_0$ in $SU(2m)$ satisfies

$$\widetilde{w}_0 = u_0 \omega u_0^T \quad \text{with} \quad \omega = \begin{pmatrix} 0 & D_m \\ -D_m & 0 \end{pmatrix} \quad \text{and} \quad D_m = \begin{pmatrix} & & & 1 \\ & & 1 & \\ & \cdot^{\cdot^{\cdot}} & & \\ 1 & & & \end{pmatrix} \tag{4.79}$$

for some $u_0 \in U(2m)$. Necessarily, $\det u_0 = \pm 1$ and is equal to the Pfaffian $\mathrm{pf}(\widetilde{w}_0)$ of the antisymmetric matrix $\widetilde{w}_0$. The fixed-point subgroup of $U(2m)$

$$Sp_{\widetilde{w}_0}(2m) = \big\{ u \in U(2m) = \widetilde{w}_0 \, \bar{u} \, \widetilde{w}_0^{-1} = u \big\} \tag{4.80}$$

is conjugate to the symplectic group $Sp(2m) \subset U(2m)$. For $m = 1$ the situation is particularly simple since, necessarily, $\widetilde{w}_0 = \pm\omega$ with the sign equal to $\mathrm{pf}\, \widetilde{w}_0$ and the subgroups $Sp_{\widetilde{w}_0}(2)$ coincide with $Sp(2) = SU(2)$.

### 4.6. Construction of $\widetilde{t}$

A contraction $\widetilde{t}$ satisfying relations (4.72) and (4.73) may be constructed in essentially the same way as in the dynamical case considered in Section 3.2. Starting from the edges of $\mathrm{BZ}_+$ we first find two maps $\widetilde{t}_0$ and $\widetilde{t}_\pi$ at $k_1 = 0$ and $k_1 = \pi$

$$[0, \tfrac{1}{2}] \times [-\pi, \pi] \ni (r, k_2) \longmapsto \widetilde{t}_a(r, k_2) \in SU(2m) \tag{4.81}$$

for $a = 0, \pi$ such that

$$\widetilde{t}_a(r, \pi) = \widetilde{t}_a(r, -\pi) , \tag{4.82}$$

$$\widetilde{t}_a(0, k_2) = I, \qquad \widetilde{t}_a(\tfrac{1}{2}, k_2) = -I , \tag{4.83}$$

$$\widetilde{w}(a, k_2) \overline{\widetilde{t}_a(r, k_2)} \, \widetilde{w}(a, k_2)^{-1} = \widetilde{t}_a(r, -k_2) . \tag{4.84}$$

For $r \in [\frac{1}{4}, \frac{1}{2}]$, they may be built as in Fig. 4 except that here the final map is constant and equal to $-I$ and that, for later convenience, we replace $r$ by $\frac{1}{2} - r$. We connect $I$ to $-I$ by paths $[0, \frac{1}{4}] \ni s \mapsto \widetilde{t}_{aa'}(s)$ in the corresponding subgroup $Sp_{\widetilde{w}(a,a')} \subset U(2m)$ and set

$$\widetilde{t}_a(r, k_2) = \begin{cases} \widetilde{t}_{a0}(r - \frac{1}{4} + k_2) & \text{for} \quad 0 \le k_2 \le \frac{1}{2} - r , \\ -I & \text{for} \quad \frac{1}{2} - r \le k_2 \le \pi - \frac{1}{2} + r , \\ \widetilde{t}_{a\pi}(r - \frac{1}{4} + \pi - k_2) & \text{for} \quad \pi - \frac{1}{2} + r \le k_2 \le \pi , \end{cases} \tag{4.85}$$

compare to (3.27). The path $[0,\pi]\ni k_2\mapsto \widetilde{t}_a(\frac{1}{4},k_2)$ is a loop in $SU(2m)$ starting and ending in $I$ and it may be contracted to $I$, extending the definition of $\widetilde{t}_a$ to $(r,k_2)\in[0,\frac{1}{2}]\times[0,\pi]$. Now for $(r,k_2)\in[0,\frac{1}{2}]\times[-\pi,0]$, we set

$$\widetilde{t}_a(r,k_2)=\widetilde{w}(a,-k_2)\,\overline{\widetilde{t}_a(r,-k_2)}\,\widetilde{w}(a,-k_2)^{-1}. \tag{4.86}$$

This constructs maps $\widetilde{t}_a$ with the desired properties. For $m=1$, we shall use below an even more specific choice of $\widetilde{t}_a$ which will streamline the calculations.

In the next step, we spread the two edge contractions inside $\mathrm{BZ}_+$ as in Fig. 5, defining for $(r,k)\in[0,\frac{1}{2}]\times\mathrm{BZ}_+$

$$\widetilde{t}(r,k_1,k_2)=\begin{cases}\widetilde{t}_0(\frac{1}{2}-r+k_1,k_2) & \text{for} \quad 0\le k_1\le r\,,\\ -I & \text{for} \quad r\le k_1\le \pi-r\,,\\ \widetilde{t}_\pi(\frac{1}{2}-r+\pi-k_1,k_2) & \text{for} \quad \pi-r\le k_1\le\pi\,,\end{cases} \tag{4.87}$$

compare to (3.39). For $r=\frac{1}{2}$, in particular, we get a map that satisfies

$$\widetilde{t}(\tfrac{1}{2},k_1,\pi)=\widetilde{t}(\tfrac{1}{2},k_1,-\pi)\,,\qquad \widetilde{t}(\tfrac{1}{2},0,k_2)=I=\widetilde{t}(\tfrac{1}{2},\pi,k_2) \tag{4.88}$$

which may be viewed as a map defined on the earring, see Fig. 6. Since $\widetilde{t}(\frac{1}{2},\cdot)$, takes values in $SU(2m)$, there is no obstruction to extend $\widetilde{t}$ to $[0,1]\times\mathrm{BZ}_+$ so that

$$[0,1]\times[0,\pi]\times[-\pi,\pi]\ni(r,k_1,k_2)\longmapsto\widetilde{t}(r,k_1,k_2)\in SU(2m)\,, \tag{4.89}$$

$$\widetilde{t}(r,0,k_2)=I=\widetilde{t}(r,\pi,k_2)\quad\text{for}\quad \tfrac{1}{2}\le r\le 1\,,\quad \widetilde{t}(1,k_1,k_2)=I \tag{4.90}$$

by contracting properly $\widetilde{t}(\frac{1}{2},\cdot)$ to the constant map equal to $I$. Finally, we define

$$\widetilde{t}(r,k_1,k_2)=\widetilde{w}(-k_1,-k_2)\,\overline{\widetilde{t}(r,-k_1,-k_2)}\,\widetilde{w}(-k_1,-k_2)^{-1}\quad\text{for}\quad -\pi\le k_1\le 0 \tag{4.91}$$

obtaining a continuous map well defined on $[0,1]\times\mathrm{BZ}$ with properties (4.72) and (4.73).

### 4.7. Reduction of $K[\mathcal{E}]$ to Wess–Zumino terms

We shall compute $K[\mathcal{E}]$ using (4.75) and the contraction $\widetilde{t}$ constructed above. By splitting the BZ-integral into the one on $\mathrm{BZ}_+$ and $\mathrm{BZ}_-$ and transforming the latter with the use of (4.73) and formulae (A.4) and (A.5) from Appendix A, we obtain

$$\begin{aligned}-\frac{1}{24\pi^2}\int\limits_{[0,1]\times\mathrm{BZ}}\widetilde{t}^*\chi &= -\frac{1}{24\pi^2}\int\limits_{[0,1]\times\mathrm{BZ}_+}\widetilde{t}^*\chi-\frac{1}{24\pi^2}\int\limits_{[0,1]\times\mathrm{BZ}_+}(\widetilde{w}\bar{\widetilde{t}}\widetilde{w}^{-1})^*\chi\\ &= -\frac{1}{12\pi^2}\int\limits_{[0,1]\times\mathrm{BZ}_+}\widetilde{t}^*\chi+\mathcal{B}(\widetilde{t})\,,\end{aligned} \tag{4.92}$$

where the last term comes from the Stokes formula,

$$\mathcal{B}(\widetilde{t})=\frac{1}{8\pi^2}\int\limits_{\partial([0,1]\times\mathrm{BZ}_+)}\Big(\bar{\widetilde{t}}\widetilde{w}^{-1}(\mathrm{d}\widetilde{w})\bar{\widetilde{t}}^{-1}\widetilde{w}^{-1}(\mathrm{d}\widetilde{w})+\widetilde{w}^{-1}(\mathrm{d}\widetilde{w})\big(\bar{\widetilde{t}}^{-1}(\mathrm{d}\bar{\widetilde{t}})+(d\bar{\widetilde{t}})\bar{\widetilde{t}}^{-1}\big)\Big), \tag{4.93}$$

similarly as in (4.67). Note, however, that this time the boundary contains more pieces

$$\partial([0,1]\times \mathrm{BZ}_+) = \{1\}\times \mathrm{BZ}_+ - \{0\}\times \mathrm{BZ}_+ \\ + [0,1]\times\{0\}\times[-\pi,\pi] - [0,1]\times\{\pi\}\times[-\pi,\pi], \tag{4.94}$$

where the signs indicate the orientations. The contributions to the integrals from $r = 0, 1$ vanish for the same reason as in (4.68). The ones from $k_1 = 0, \pi$, however, have to be taken into account. Hence

$$K[\mathcal{E}] = -\frac{1}{12\pi^2}\int_{[0,1]\times \mathrm{BZ}_+} \widetilde{t}^*\chi + \mathcal{B}(\widetilde{t}) \mod 2, \tag{4.95}$$

where

$$\mathcal{B}(\widetilde{t}) = \frac{1}{8\pi^2}\Big(\int_{[0,1]\times\{0\}\times[-\pi,\pi]} - \int_{[0,1]\times\{\pi\}\times[-\pi,\pi]}\Big)\mathrm{tr}\Big(\widetilde{w}^{-1}(\mathrm{d}\widetilde{w})\big(\bar{\widetilde{t}}^{-1}(\mathrm{d}\bar{\widetilde{t}}) + (\mathrm{d}\bar{\widetilde{t}})\bar{\widetilde{t}}^{-1}\big)\Big). \tag{4.96}$$

We dropped the term $\mathrm{tr}\big(\bar{\widetilde{t}}\,\widetilde{w}^{-1}(\mathrm{d}\widetilde{w})\bar{\widetilde{t}}^{-1}\widetilde{w}^{-1}(\mathrm{d}\widetilde{w})\big)$ as it does not contribute for the dimensional reasons.

We now compute (4.95) with the explicit contraction (4.87) defined in the previous section. First note that on the boundaries of $\mathrm{BZ}_+$

$$\widetilde{t}(r,a,k_2) = \begin{cases} \widetilde{t}_a(\frac{1}{2}-r,k_2) & \text{for} \quad 0\le r\le \frac{1}{2}, \\ I & \text{for} \quad \frac{1}{2}\le r\le 1 \end{cases} \tag{4.97}$$

so that the boundary terms appearing in (4.95) will only depend on $\widetilde{t}_a$. Namely,

$$\mathcal{B}(\widetilde{t}) = -\mathcal{B}(\widetilde{t}_0) + \mathcal{B}(\widetilde{t}_\pi) \tag{4.98}$$

with

$$\mathcal{B}(\widetilde{t}_a) = \frac{1}{8\pi^2}\int_{[0,\frac{1}{2}]\times\{a\}\times[-\pi,\pi]} \mathrm{tr}\Big(\widetilde{w}^{-1}(\mathrm{d}\widetilde{w})\overline{\big(\widetilde{t}_a^{\;-1}(\mathrm{d}\widetilde{t}_a) + (\mathrm{d}\widetilde{t}_a)\widetilde{t}_a^{-1}\big)}\Big), \tag{4.99}$$

where we used the fact that $r \in [\frac{1}{2}, 1]$ does not contribute to the latter integrals for dimensional reasons. Using the time-reversal symmetry (4.84) of $\widetilde{t}_a$ and property (4.78), we can split in the integral over $[-\pi,\pi]$ into two halves and express it as twice the same quantity on one half. We end up with

$$\mathcal{B}(\widetilde{t}_a) = \frac{1}{4\pi^2}\int_{[0,\frac{1}{2}]\times\{a\}\times[0,\pi]} \mathrm{tr}\Big((\widetilde{w}^{-1}\mathrm{d}\widetilde{w})\,\overline{\big(\widetilde{t}_a^{-1}(\mathrm{d}\widetilde{t}_a) + (\mathrm{d}\widetilde{t}_a)\widetilde{t}_a^{-1}\big)}\Big). \tag{4.100}$$

These terms depend on the explicit form of $\widetilde{t}_0$ and $\widetilde{t}_\pi$.

For the first term of the right-hand side of (4.95), again by dimensional reasons,

$$-\frac{1}{12\pi^2}\int_{[0,1]\times \mathrm{BZ}_+} \widetilde{t}^*\chi = -\frac{1}{12\pi^2}\int_{[\frac{1}{2},1]\times \mathrm{BZ}_+} \widetilde{t}^*\chi\,. \tag{4.101}$$

Indeed, the contribution of $r \in [0, \frac{1}{2}]$ to the integral vanishes since in this region, $\widetilde{t}$ depends only on two variables, see (4.87). The map $\widetilde{t}$ restricted to $[\frac{1}{2}, 1]\times \mathrm{BZ}_+$, contracting $\widetilde{t}(\frac{1}{2},\cdot)$ to the constant map equal to $I$ may be interpreted as follows. As $\widetilde{t}(\frac{1}{2},\cdot)$ is equal to $I$ on the edges of $\mathrm{BZ}_+$ and periodic in $k_2$, see (4.88), it may be viewed as a map defined on the sphere $S^2$ obtained by identifying the edges at $k_1 = 0$ and $k_1 = \pi$ to two points $p_0$ and $p_1$, see Fig. 11. Since these boundary properties are preserved during the contraction, the map $\widetilde{t}$ restricted to $r \in [\frac{1}{2}, 1]$ may be seen as defined on $[\frac{1}{2}, 1] \times S^2$ satisfying $\widetilde{t}(1,\cdot) = I$. We may then identify, up to a scalar factor, the integral on the right-hand side with the WZ action of the field $U(2m)$-valued field $\widetilde{t}(\frac{1}{2},\cdot)$ viewed as defined on $S^2$:

$$-\frac{1}{12\pi^2} \int\limits_{[\frac{1}{2},1]\times \mathrm{BZ}_+} \widetilde{t}^{\,*}\chi = \frac{1}{\pi} S_{\mathrm{WZ}}(\widetilde{t}(\tfrac{1}{2},\cdot)), \tag{4.102}$$

see (4.18). Recall that $S_{\mathrm{WZ}}$ is defined modulo $2\pi$ making the Feynman amplitudes $\mathrm{e}^{iS_{\mathrm{WZ}}}$ well determined. This agrees with the fact that what enters formula (4.95) is the left-hand side of (4.102) taken modulo 2. No further restrictions on contractions of $\widetilde{t}(\frac{1}{2},\cdot)$ used in (4.18) are needed here.

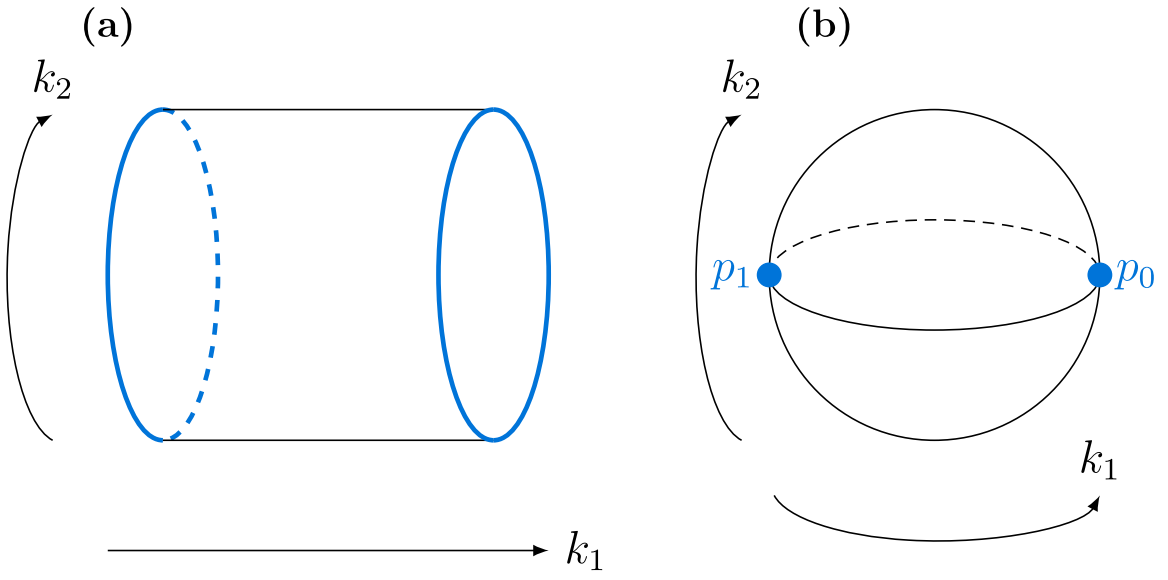

Fig. 11. The map $\widetilde{t}(\frac{1}{2};\cdot)$ sends the edges of $\mathrm{BZ}_+$ to $I$ and is periodic in $k_2$ so it can be seen as a map from a sphere $S^2$, (b) by identifying the edges of the cylinder (a) at $k_1 = 0$ and $k_1 = \pi$ to two points $p_0$ and $p_1$. (For interpretation of the colors in this figure, the reader is referred to the web version of this article.)

Looking more precisely at the map under considered, observe that

$$\widetilde{t}(\tfrac{1}{2}, k_1, k_2) = \begin{cases} \widetilde{t}_0(k_1, k_2) & \text{for} \quad 0 \le k_1 \le \frac{1}{2}, \\ -\mathrm{Id} & \text{for} \quad \frac{1}{2} \le k_1 \le \pi - \frac{1}{2}, \\ \widetilde{t}_\pi(\pi - k_1, k_2) & \text{for} \quad \pi - \frac{1}{2} \le k_1 \le \pi. \end{cases} \tag{4.103}$$

That means that the WZ action on the previous effective sphere $S^2$ can be split into the sum of two contributions coming from $\widetilde{t}_0$ and $\widetilde{t}_\pi$ by identifying the points with intermediate $k_1$ where $\widetilde{t}(\frac{1}{2}, k_1, k_2) = -I$, see Fig. 12, so that

$$S_{\mathrm{WZ}}((\widetilde{t}(\tfrac{1}{2},\cdot)) \bmod 2\pi = S_{\mathrm{WZ}}(\widetilde{t}_0) - S_{\mathrm{WZ}}(\widetilde{t}_\pi) \bmod 2\pi, \tag{4.104}$$

where the minus sign comes from the reversed orientation. At the end of the day we are left with the formula

$$K[\mathcal{E}] = \frac{1}{\pi} S_{\mathrm{WZ}}(\widetilde{t}_0) - \frac{1}{\pi} S_{\mathrm{WZ}}(\widetilde{t}_\pi) - \mathcal{B}(\widetilde{t}_0) + \mathcal{B}(\widetilde{t}_\pi) \quad \bmod 2 \tag{4.105}$$

that may be rewritten in the equivalent multiplicative form as

$$(-1)^{K[\mathcal{E}]} = \mathrm{e}^{\mathrm{i}S_{\mathrm{WZ}}(\widetilde{t}_0)}\, \mathrm{e}^{-\mathrm{i}S_{\mathrm{WZ}}(\widetilde{t}_\pi)}\, \mathrm{e}^{-\pi\mathrm{i}\,\mathcal{B}(\widetilde{t}_0)+\pi\mathrm{i}\,\mathcal{B}(\widetilde{t}_\pi)}. \tag{4.106}$$

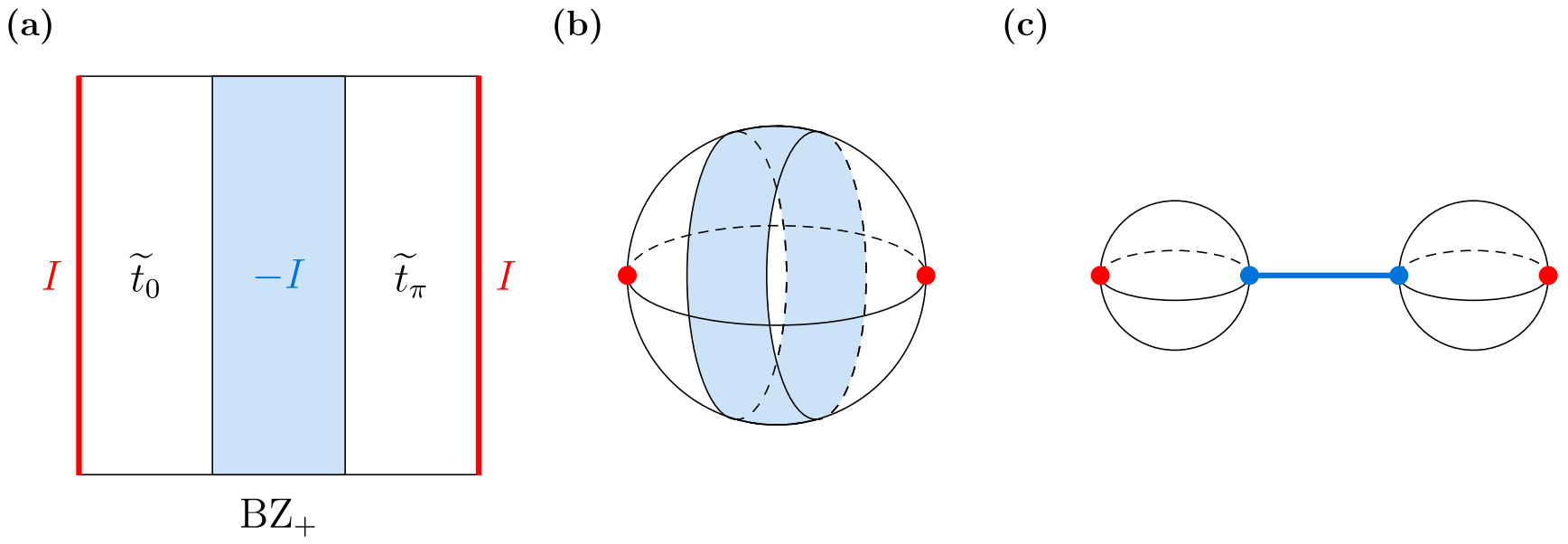

Fig. 12. The map $\widetilde{t}(\frac{1}{2},\cdot)$ is defined on the cylinder $BZ_+$ (a). As it is constant on the boundary $\partial BZ_+$ (in red), it is in fact a map from the sphere (b) obtained by contracting the edges $\partial BZ_+$ to points. Moreover, the map is also constant for $\frac{1}{2} \le k_1 \le \pi - \frac{1}{2}$: all this domain can be contracted to a line (or a point), as shown in (c). This enables to view $\widetilde{t}(\frac{1}{2},\cdot)$ as a map on two spheres. (For interpretation of the colors in this figure, the reader is referred to the web version of this article.)

The index $K[\mathcal{E}]$ may then be fully expressed in terms of two maps $\widetilde{t}_0$ and $\widetilde{t}_\pi$. The last step of our work is then to give explicit expressions for the corresponding WZ amplitudes. We shall treat first the simpler case with $m = 1$ deferring the substantially more difficult general case to separate sections.

### 4.8. Case $m = 1$

When $m = 1$ then

$$\widetilde{w}(a, a') = \pm \begin{pmatrix} 0 & 1 \\ -1 & 0 \end{pmatrix} \quad \text{for} \quad (a, a') = 0, \pi \tag{4.107}$$

and all subgroups $Sp_{\widetilde{w}(a,a')} = SU(2)$. This allows to make a more explicit choice for maps $\widetilde{t}_a$ with properties (4.72) and (4.73) by setting for $(r, k_2) \in [0, \frac{1}{2}] \times [-\pi, \pi]$

$$\widetilde{t}_a(r, k_2) = \begin{cases} e^{-2\pi i\, r\, \sigma_z} & \text{if} \quad 0 \le k_2 \le \pi\,, \\ \widetilde{w}(a, -k_2)\, e^{2\pi i\, r\, \sigma_z}\, \widetilde{w}(a, -k_2)^{-1} & \text{if} \quad -\pi \le k_2 \le 0\,, \end{cases} \tag{4.108}$$

where $\sigma_z$ is the Pauli matrix $\left(\begin{smallmatrix} 1 & 0 \\ 0 & -1 \end{smallmatrix}\right)$. The consistency of the above formula follows from the relation $\widetilde{w}(a, a')\, e^{2\pi i\, r\, \sigma_z}\, \widetilde{w}(a, a') = e^{-2\pi i\, r\, \sigma_z}$. For this choice of $\widetilde{t}_a$, the boundary terms (4.100) become

$$\mathcal{B}(\widetilde{t}_a) = \frac{1}{2\pi i} \int_0^\pi \mathrm{tr}\big(\sigma_z (\widetilde{w}^{-1} d\widetilde{w})(a, k_2)\big)\,. \tag{4.109}$$

The WZ Feynman amplitudes of $SU(2)$-valued fields is particularly easy to compute by gluing local expressions of the type (4.19) rather than by using (4.18) [14,16]. The $SU(2)$ group is a 3-sphere. We shall use a parameterization that decomposes it into the conjugacy classes writing for $g \in SU(2)$,

$$g = \gamma\, e^{2\pi i\, s\, \sigma_z} \gamma^{-1}. \tag{4.110}$$

The parameter $s \in [0, \frac{1}{2}]$ describes the latitude on the 3-sphere, going from $I$ at the north pole $s = 0$ to $-I$ at the south pole $s = \frac{1}{2}$, and the conjugacy classes corresponding to fixed $s$,

$$\mathcal{C}_s = \{\gamma\, e^{2i\pi s\, \sigma_z} \gamma^{-1} \mid \gamma \in SU(2)\}\,, \tag{4.111}$$

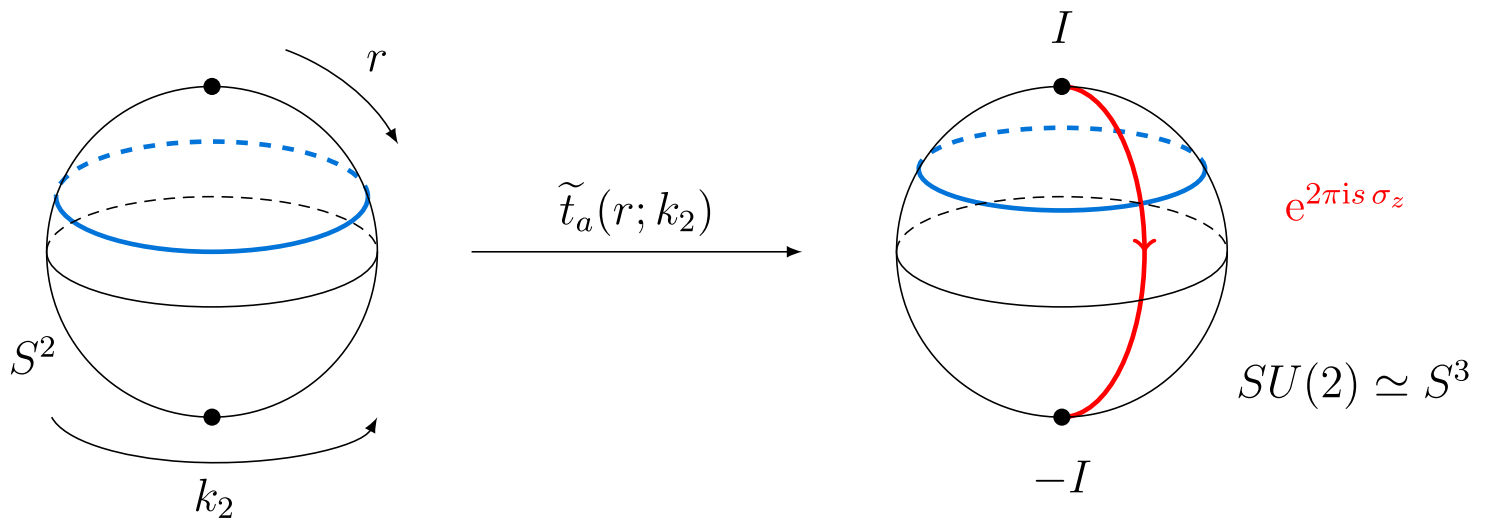

Fig. 13. Maps $\widetilde{t}_a(r;k_2)$ may be viewed as sending a 2-sphere to the special unitary group $SU(2)$, which is a 3-sphere. A circle of fixed latitude on $S^2$ corresponding to constant $r$ (in blue) is mapped into the conjugacy class $\mathcal{C}_s$ in $SU(2)$ for $s=r$ (also in blue). The points $e^{i2\pi s\sigma_z}$ when $s$ varies from 0 to 1/2 are drawn in red. This red curve is deformed along conjugacy classes under the adjoint action of $\gamma(k_2)$. (For interpretation of the colors in this figure, the reader is referred to the web version of this article.)

are 2-spheres for $s \neq 0, 1$, see Fig. 13. The parameterization (4.110) (that is redundant since $\gamma$ and $\gamma\, e^{i\pi x\sigma_z}$ for $x \in \mathbb{R}$ describe the same element in $SU(2)$) will be convenient to compute the WZ action of fields $\widetilde{t}_a$ given by (4.108) from local pieces (4.19). For this end one covers $SU(2)$ by two open subsets

$$O_0 = SU(2) \setminus \{-I\}, \qquad O_1 = SU(2) \setminus \{I\} \tag{4.112}$$

The first one corresponds to $0 \le s < 1/2$ in Eq. (4.110) and the second one to $0 < s \le 1/2$. It can be shown that on these subsets the 3-form $\chi$ is exact with

$$\tfrac{1}{12\pi}\,\chi\big|_{O_i} = \mathrm{d}B_i \tag{4.113}$$

with smooth 2-forms

$$B_0 = \tfrac{1}{4\pi}\mathrm{tr}\Big(\gamma^{-1}(\mathrm{d}\gamma)\, e^{2\pi i\, s\,\sigma_z}\gamma^{-1}(\mathrm{d}\gamma)\, e^{-\pi i\, s\,\sigma_z}\Big) + i\, s\, \mathrm{tr}\Big(\sigma_z\big(\gamma^{-1}(\mathrm{d}\gamma)\big)^2\Big), \tag{4.114}$$

$$\begin{aligned} B_1 = {} & \tfrac{1}{4\pi}\mathrm{tr}\Big(\gamma^{-1}(\mathrm{d}\gamma)\, e^{\pi i\,(s-1)\,\sigma_z}\gamma^{-1}(\mathrm{d}\gamma)\, e^{2\pi i\,(s-1)\,\sigma_z}\Big) \\ & + i(s-\tfrac{1}{2})\,\mathrm{tr}\Big(\sigma_z\big(\gamma^{-1}(\mathrm{d}\gamma)\big)^2\Big) \end{aligned} \tag{4.115}$$

given in terms of parameterization (4.110), see [16]. Moreover, on the intersection $O_0 \cap O_1$, one has

$$B_1 - B_0 = -\tfrac{i}{2}\mathrm{tr}\Big(\sigma_z\big(\gamma^{-1}(\mathrm{d}\gamma)\big)^2\Big) = \tfrac{i}{2}\,\mathrm{d}\,\mathrm{tr}\Big(\sigma_z\gamma^{-1}(\mathrm{d}\gamma)\Big) \equiv \mathrm{d}\alpha \tag{4.116}$$

which is a closed 2-form on $O_0 \cap O_1$ that becomes exact when described in the redundant parameterization (4.110). Formula (4.108) for $\widetilde{t}_a$ is written almost in terms of parameterization (4.110), but not quite. If, however, we define

$$\begin{aligned} & s(r) = r\,, \\ & \gamma(k_2) = \begin{cases} \widetilde{w}(a,0) & \text{for} \quad 0 \le k_2 \le \pi\,, \\ \widetilde{w}(a,-k_2) & \text{for} \quad -\pi \le k_2 \le 0 \quad \text{if} \quad \widetilde{w}(a,\pi) = \widetilde{w}(a,0)\,, \\ \widetilde{w}(a,-k_2)\, e^{-i\,k_2\,\sigma_z} & \text{for} \quad -\pi \le k_2 \le 0 \quad \text{if} \quad \widetilde{w}(a,\pi) = -\widetilde{w}(a,0) \end{cases} \end{aligned} \tag{4.117}$$

then $\gamma(k_2)$ is well defined, $\gamma(-\pi) = \gamma(\pi)$ and

$$\widetilde{t}_a(r,k_2) = \gamma(k_2)\, e^{2\pi i\, s(r)\,\sigma_z}\gamma(k_2)^{-1} \tag{4.118}$$

(recall that matrices $\widetilde{w}(a,a')$ can take only two values differing by sign). Besides, $\widetilde{t}_a(r,k_2)\in O_0$ for $0\le r<\frac{1}{2}$ and $\widetilde{t}_a(r,k_2)\in O_1$ for $0<r\le\frac{1}{2}$. The formula for gluing local expressions of type (4.19) for the WZ amplitude $e^{iS_{WZ}(\widetilde{t}_a)}$ takes then the simple form [16]

$$e^{i\,S_{WZ}(\widetilde{t}_a)}=\exp\Big[\int\limits_{[0,\frac{1}{4}]\times[-\pi,\pi]}\widetilde{t}_a^{\,*}B_0+\int\limits_{[\frac{1}{4},\frac{1}{2}]\times[-\pi,\pi]}\widetilde{t}_a^{\,*}B_1\Big]\mathrm{Hol}_{\mathcal{L}}(\widetilde{t}_a(\tfrac{1}{4},\cdot)),\tag{4.119}$$

where $\mathcal{L}$ is a hermitian line bundle over $O_0\cap O_1\subset SU(2)$ with unitary connection whose curvature is equal to $B_1-B_0$. The holonomy of that connection along loop $[-\pi,\pi]\ni k_2\mapsto\widetilde{t}_a(\frac{1}{4},k_2)$ is given by an integral of the 1-form $\alpha=\frac{i}{2}\,\mathrm{tr}\,\sigma_z\gamma^{-1}d\gamma$, see (4.116),

$$\begin{aligned}\mathrm{Hol}_{\mathcal{L}}(\widetilde{t}_0(\tfrac{1}{4},\cdot))&=\exp\Big[i\int\limits_{[-\pi,\pi]}\gamma^*\alpha\Big]=\exp\Big[-\tfrac{1}{2}\int_{-\pi}^{\pi}\mathrm{tr}\,\sigma_z\,(\gamma^{-1}d\gamma)(k_2)\Big]\\&=\exp\Big[-\tfrac{1}{2}\int_{-\pi}^{0}\mathrm{tr}\,\sigma_z\Big(\widetilde{w}^{-1}d\widetilde{w})(a,-k_2)+\begin{cases}0&\text{if}\quad\widetilde{w}(a,\pi)=\widetilde{w}(a,0)\\-i\sigma_z\,dk_2&\text{if}\quad\widetilde{w}(a,\pi)=-\widetilde{w}(a,0)\end{cases}\Big)\Big].\end{aligned}\tag{4.120}$$

After changing sign of the integration variable, this gives:

$$\begin{aligned}\mathrm{Hol}_{\mathcal{L}}(\widetilde{t}_0(\tfrac{1}{4},\cdot))&=\exp\Big[\tfrac{1}{2}\int_0^{\pi}\mathrm{tr}\,\sigma_z\,(\widetilde{w}^{-1}d\widetilde{w})(a,k_2)+\begin{cases}0&\text{if}\quad\widetilde{w}(a,\pi)=\widetilde{w}(a,0)\\\pi i&\text{if}\quad\widetilde{w}(a,\pi)=-\widetilde{w}(a,0)\end{cases}\Big]\\&=\exp\Big[\pi i\,\mathcal{B}(\widetilde{t}_a)+\begin{cases}0&\text{if}\quad\widetilde{w}(a,\pi)=\widetilde{w}(a,0)\\\pi i&\text{if}\quad\widetilde{w}(a,\pi)=-\widetilde{w}(a,0)\end{cases}\Big],\end{aligned}\tag{4.121}$$

see (4.109). On the other hand, the contributions of the integrals of the pullbacks of 2-forms $B_i$ to (4.119) vanish for the dimensional reasons (as $\gamma$ depends only on $k_2$). Hence

$$e^{i\,S_{WZ}(\widetilde{t}_a)}=\exp\Big[\pi i\,\mathcal{B}(\widetilde{t}_a)+\begin{cases}0&\text{if}\quad\widetilde{w}(a,\pi)=\widetilde{w}(a,0)\\\pi i&\text{if}\quad\widetilde{w}(a,\pi)=-\widetilde{w}(a,0)\end{cases}\Big].\tag{4.122}$$

Upon the substitution of that relation to (4.106), the boundary terms $\mathcal{B}(\widetilde{t}_a)$ cancel out resulting in the formula

$$(-1)^{K[\mathcal{E}]}=\begin{cases}1&\text{if}\quad\widetilde{w}(0,\pi)=\widetilde{w}(0,0)\\-1&\text{if}\quad\widetilde{w}(0,\pi)=-\widetilde{w}(0,0)\end{cases}\times\begin{cases}1&\text{if}\quad\widetilde{w}(\pi,\pi)=\widetilde{w}(\pi,0)\\-1&\text{if}\quad\widetilde{w}(\pi,\pi)=-\widetilde{w}(\pi,0)\end{cases}.\tag{4.123}$$

Recalling that the Pfaffian $\mathrm{pf}\,\widetilde{w}(a,a')$ is equal to the sign in Eq. (4.107), we may rewrite result (4.123) in the form

$$(-1)^{K[\mathcal{E}]}=\prod_{\substack{\mathrm{TRIM}\\k^*}}\frac{1}{\mathrm{pf}\,\widetilde{w}(k^*)}=\prod_{\substack{\mathrm{TRIM}\\k^*}}\frac{\sqrt{\det w(k^*)}}{\mathrm{pf}\,w(k^*)}\tag{4.124}$$

with the products over four TRIM $k^*=(a.a')\in\mathrm{BZ}$. The latter expression coincides with the multiplicative formula for the Kane–Mele invariant $(-1)^{\mathrm{KM}[\mathcal{E}]}$ obtained in [9], ending the proof of the equality between the $\mathbb{Z}_2$-valued indices $K[\mathcal{E}]$ and $\mathrm{KM}[\mathcal{E}]$ for bundles $\mathcal{E}$ of rank 2.

## 5. Case of general rank

### 5.1. Wess–Zumino amplitudes of $SU(n)$-valued fields

We shall need some knowledge about constructing WZ amplitudes of fields with values in $SU(n)$ by gluing local expressions, see [16] for more details. Consider the diagonal $n \times n$ matrices

$$\begin{aligned}\lambda_1^\vee &= \mathrm{diag}\Big(\tfrac{n-1}{n}, -\tfrac{1}{n}, -\tfrac{1}{n}, \dots, -\tfrac{1}{n}, -\tfrac{1}{n}\Big)\\ \lambda_2^\vee &= \mathrm{diag}\Big(\tfrac{n-2}{n}, \tfrac{n-2}{n}, -\tfrac{2}{n}, \dots, -\tfrac{2}{n}, -\tfrac{2}{n}\Big)\\ &\vdots\\ \lambda_{n-1}^\vee &= \mathrm{diag}\Big(\tfrac{1}{n}, \tfrac{1}{n}, \tfrac{1}{n}, \dots, \tfrac{1}{n}, -\tfrac{n-1}{n}\Big).\end{aligned} \tag{5.1}$$

They form the simple (co)weights of Lie algebra $su(n)$. We shall add to them the vanishing matrix $\lambda_0^\vee = 0$. Let

$$\tau = \sum_{j=0}^{n-1} t_j \lambda_j^\vee \in su(n) \tag{5.2}$$

be a convex combination of such diagonal matrices with $0 \le t_j \le 1$ and $\sum_{j=0}^{n-1} t_j = 1$. Group $SU(n)$ may be covered by $n$ open subsets

$$O_i = \big\{ g = \gamma\, \mathrm{e}^{2\pi i \tau} \gamma^{-1} \mid \gamma \in SU(n),\ \ \tau \text{ with } t_i > 0 \big\}. \tag{5.3}$$

For $n = 2$, $O_0$ and $O_1$ coincide with the open subsets of $SU(2)$ considered in Section 4.8. Note that $O_i$ are invariant under the adjoint action of $SU(n)$ on itself. On subsets $O_i$ there exist smooth 2-forms

$$B_i = \tfrac{1}{4\pi} \mathrm{tr}\Big(\gamma^{-1}(\mathrm{d}\gamma)\, \mathrm{e}^{2\pi i \tau}\, \gamma^{-1}(\mathrm{d}\gamma)\, \mathrm{e}^{-2\pi i \tau}\Big) + i\, \mathrm{tr}\Big((\tau - \lambda_i^\vee)\big(\gamma^{-1}\mathrm{d}\gamma\big)^2\Big) \tag{5.4}$$

such that $dB_i = \frac{1}{12\pi}\, \chi$. On the double intersections $O_i \cap O_j \equiv O_{ij}$,

$$B_j - B_i = -\mathrm{i}\, \mathrm{tr}\, \lambda_{ij}^\vee \big(\gamma^{-1}(\mathrm{d}\gamma)\big)^2 \equiv B_{ij} \tag{5.5}$$

for $\lambda_{ij}^\vee = \lambda_j^\vee - \lambda_i^\vee$ are closed 2-forms and there exist line bundles $\mathcal{L}_{ij}$ with unitary connection whose curvature is equal to $B_{ij}$ and isomorphisms[8]

$$t_{ijk} : \mathcal{L}_{ij} \otimes \mathcal{L}_{jk} \to \mathcal{L}_{ik} \tag{5.6}$$

over the triple intersections $O_{ijk}$ that are associative over $O_{ijkl}$. For $i \le j$ there exists a smooth map

$$O_{ij} \ni g = \gamma\, \mathrm{e}^{2\pi i \tau} \gamma^{-1} \overset{r_{ij}}{\longmapsto} \gamma(V_{ij}) \in Gr_{j-i,n} \tag{5.7}$$

[8] All line bundles that we consider come equipped with a hermitian structure (a scalar product in the fibers) and a unitary connection and their isomorphisms are assumed to preserve those structures.

into the Grassmannian of $(j-i)$ dimensional subspaces in $\mathbb{C}^n$, where $V_{ij}$ is the subspace spanned by the canonical-basis vectors $f_{i+1}, \dots, f_j$ and $V_{ii} = \{0\}$. Let

$$G_{ij} = \{\gamma_0 \in SU(n) \mid \gamma_0(V_{ij}) = V_{ij}\} = \{\gamma_0 \in SU(n) \mid \gamma_0 \lambda_{ij}^\vee \gamma_0^{-1} = \lambda_{ij}^\vee\} = G_{ji}\,. \tag{5.8}$$

Grassmannian $Gr_{j-i,n}$ may be canonically identified with the homogeneous space $SU(n)/G_{ij}$. On $Gr_{j-i,n}$ there exist a natural line bundle $L_{ij}$ whose fiber at $\gamma(V_{ij}) \in Gr_{j-i,n}$ is composed of $(j-i)$-vectors $\zeta(\gamma f_{i+1}) \wedge \cdots \wedge (\gamma f_j)$ with $\zeta \in \mathbb{C}$ (for $i = j$, $L_{ii} = \{0\} \times \mathbb{C}$). One may define $L_{ji}$ as the dual bundle $L_{ij}^{-1}$ to $L_{ij}$. Such bundles may be also viewed as quotients of a trivial line-bundle over $SU(n)$:

$$L_{ij} \cong \big(SU(n) \times \mathbb{C}\big)/G_{ij} \tag{5.9}$$

for the action of $\gamma_0 \in G_{ij}$ given by

$$\gamma_0(\gamma, \zeta) = (\gamma\gamma_0^{-1}, \chi_{ij}(\gamma_0)\zeta)\,, \tag{5.10}$$

where

$$\chi_{ij}(\gamma_0) = \begin{cases} \det(\gamma_0|_{V_{ij}}) & \text{if} \quad i < j\,, \\ \det(\gamma_0|_{V_{ji}})^{-1} & \text{if} \quad i > j\,, \\ 1 & \text{if} \quad i = j\,, \end{cases} \tag{5.11}$$

$$\mathrm{d}\ln\chi_{ij}(\gamma_0) = \mathrm{tr}\big(\lambda_{ij}\gamma_0^{-1}\mathrm{d}\gamma_0\big), \tag{5.12}$$

$$\chi_{ij}(\overline{\gamma_0}) = \chi_{ij}(\gamma_0^{-1}) = \chi_{ij}(\gamma_0)^{-1} = \chi_{ji}(\gamma_0)\,. \tag{5.13}$$

The orbit of $(\gamma, \zeta)$ under this action will be denoted $[\gamma, \zeta]_{ij}$. We equip $L_{ij}$ with the hermitian connection that descends from the connection on the trivial bundle given by the 1-form $\alpha_{ij} \equiv \mathrm{i}\,\mathrm{tr}\big(\lambda_{ij}^\vee \gamma^{-1} d\gamma\big)$ on $SU(n)$. One then takes for line bundles $\mathcal{L}_{ij}$ over $O_{ij}$ the pullbacks $r_{ij}^* L_{ij}$ of line bundles on the Grassmannians. If $i < j < k$ then the canonical isomorphism $t_{ijk} : \mathcal{L}_{ij} \otimes \mathcal{L}_{jk} \to \mathcal{L}_{ik}$ is defined by

$$t_{ijk}\Big((\gamma f_{i+1}) \wedge \cdots \wedge (\gamma f_j) \otimes (\gamma f_{j+1}) \wedge \cdots \wedge (\gamma f_k)\Big) = (\gamma f_{i+1}) \wedge \cdots \wedge (\gamma f_k)\,. \tag{5.14}$$

It extends to other configurations of $i, j, k$ by duality in such a way that

$$t_{ijk}\Big([\gamma, \zeta]_{ij} \otimes [\gamma, \zeta']_{jk}\Big) = [\gamma, \zeta\zeta']_{ik} \tag{5.15}$$

in terms of the realization (5.9).

Let $\Sigma$ be a closed oriented two-dimensional surface and $G : \Sigma \to SU(2m)$. If we triangulate $\Sigma$ so that for each triangle $c$ there exists $i_c$ such that $G(c) \subset O_{i_c}$ then the WZ amplitude of $G$ may be expressed in the form [16]

$$\mathrm{e}^{\mathrm{i}S_{\mathrm{WZ}}(G)} = \exp\Big[\mathrm{i}\sum_c \int_c G^* B_{i_c}\Big] \bigotimes_{b \subset c} \mathrm{Hol}_{\mathcal{L}_{i_c i_b}}(G|_b)\,, \tag{5.16}$$

where the tensor product over $b \subset c$ runs over the edges of $\partial c$ for which we choose index $i_b$ such that $G(b) \subset O_{i_b}$ and

$$\mathrm{Hol}_{\mathcal{L}_{ij}}(\mathcal{I}) \in (\mathcal{L}_{ij})_{\mathcal{I}_+} \otimes (\mathcal{L}_{ij})_{\mathcal{I}_-}^{-1} \tag{5.17}$$

denotes (somewhat misleadingly) the parallel transport in line bundle $\mathcal{L}_{ij}$ along an oriented line $\mathcal{I} \subset O_{ij} \subset SU(n)$ joining points $\mathcal{I}_-$ and $\mathcal{I}_+$. The left-hand side of (5.16) is a complex number whereas the right-hand side is an element of a tensor product of 1-dimensional vector spaces so to give sense to that formula we have to show that the latter tensor product may be canonically identified with $\mathbb{C}$. This is done using the subsequently isomorphisms $t_{ijk}$ (the order will not matter because of their associativity property). Indeed,

$$\bigotimes_{b\subset c} \mathrm{Hol}_{\mathcal{L}_{i_c i_b}}(G|_b) \in \bigotimes_{v\in b\subset c} (\mathcal{L}_{i_c i_b})^{\pm 1}_{G(v)}, \tag{5.18}$$

where $v$ runs through the vertices of the triangulation of $\Sigma$ and the dual fiber is taken if $v$ is the starting point of the edge $b \subset c$ taken with the orientation inherited from $c$. Recalling that $\mathcal{L}_{ij}^{-1} = \mathcal{L}_{ji}$, the tensor product of fibers at a given vertex $v$, see Fig. 14, may be rewritten as the fiber

$$\Big(\mathcal{L}_{i_{b_1} i_{c_1}} \otimes \mathcal{L}_{i_{c_1} i_{b_2}} \otimes \mathcal{L}_{i_{b_2} i_{c_2}} \otimes \mathcal{L}_{i_{c_2} i_{b_3}} \otimes \cdots \otimes \mathcal{L}_{i_{b_n} i_{c_n}} \otimes \mathcal{L}_{i_{c_n} i_{b_1}}\Big)_{G(v)} \tag{5.19}$$

of a tensor product of line bundles which the subsequent use of isomorphisms $t_{ijk}$ allows to canonically trivialize. If $D : \Sigma_1 \to \Sigma_2$ is a diffeomorphism of closed oriented surfaces then

$$\mathrm{e}^{\mathrm{i}S_{\mathrm{WZ}}(\Phi\circ D)} = \left(\mathrm{e}^{\mathrm{i}S_{\mathrm{WZ}}(\Phi)}\right)^{\pm 1} \tag{5.20}$$

where the inverse is taken if $D$ is orientation changing.

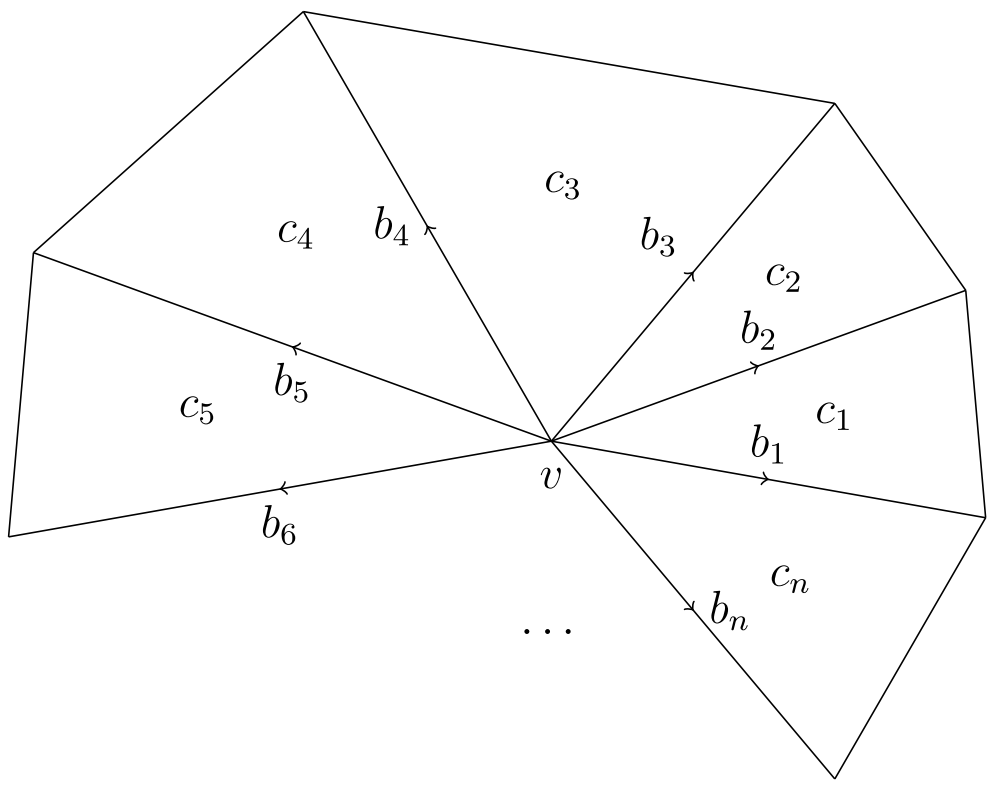

Fig. 14. Triangulation around a vertex $v$.

If $\Sigma$ has a boundary $\partial\Sigma$ composed of one circle $\mathcal{C}$ then the expression on the right-hand side of (5.16) takes values in the line

$$\bigotimes_{v\in b\in\mathcal{C}} (\mathcal{L}^{\pm 1}_{i_v i_b})_{G(v)} \equiv (\mathbb{L})_{G|_{\mathcal{C}}}\,, \tag{5.21}$$

where $i_v$ is chosen so that $G(v) \in O_{i_v}$. The above lines are canonically isomorphic for any choice of a triangulation (splitting into edges meeting at vertices) of the boundary loop and any consistent choice of indices $i_v, i_b$. They depend only on the loop $G|_{\mathcal{C}}$ in $SU(n)$. Lines over loops differing by an orientation-preserving reparameterization may be canonically identified. Those over loops differing by orientation-changing reparameterization are canonically dual to each other. Collected together, such lines form a line bundle $\mathbb{L}$ over the loop space $LSU(n)$ so that

$$\mathrm{e}^{\mathrm{i}S_{\mathrm{WZ}}(G)} \in (\mathbb{L})_{G|_{\mathcal{C}}}\,. \tag{5.22}$$

More generally, if $\partial\Sigma$ is a disjoint union of circles $\mathcal{C}_\alpha$ then

$$e^{iS_{WZ}(G|_\Sigma)} \in \bigotimes_\alpha (\mathbb{L})_{G|_{\mathcal{C}_\alpha}}. \tag{5.23}$$

Relation (5.20) still holds for surfaces with boundary if the inverse of an element in $(\mathbb{L})_\phi$ is interpreted as the element in the dual line bundle that pairs with the original element to 1. Line bundle $\mathbb{L}$ inherits a hermitian structure from line bundles $\mathcal{L}_{i_v i_b}$ and may be equipped with a unitary connection such that for $\Sigma = [0,1]\times S^1$

$$e^{iS_{WZ}(G)} \in \mathbb{L}_{G|_{\{1\}\times S^1}} \otimes \mathbb{L}^{-1}_{G|_{\{0\}\times S^1}} \tag{5.24}$$

is the parallel transport in $\mathbb{L}$ along the 1-parameter family of loops $[0,1] \ni s \mapsto G|_{\{s\}\times S^1}$.

We shall need below the following result about the WZ amplitudes for fields $(H,G) : \Sigma \to SU(2m)$ such that $H\overline{G}H^{-1} = G$ on $\partial\Sigma = \sqcup_\alpha \mathcal{C}_\alpha$:

$$\Big(\prod_\alpha \mathrm{Hol}_{\widehat{\mathcal{J}}}((H,G)|_{\mathcal{C}_\alpha})\Big)\, e^{iS_{WZ}(H\overline{G}H^{-1})} = \exp\Big[\tfrac{i}{4\pi}(H,G)^*\widehat{\beta}\Big]\, e^{iS_{WZ}(G)}, \tag{5.25}$$

where both WZ amplitudes are viewed as taking values in $\bigotimes_\alpha (\mathbb{L})_{G|_{\mathcal{C}_\alpha}}$ and $\widehat{\beta}(h,g) = \beta(h,\bar{g})$ with the 2-form $\beta$ given by Eq. (A.5) in Appendix A. The first term on the left-hand side is the product of holonomies along loops $(H,G)|_{\mathcal{C}_\alpha}$ in a hermitian line bundle $\widehat{\mathcal{J}}$ with a unitary connection of curvature $\widehat{\beta}$ over the variety

$$F = \{(h,g)\,|\,h\bar{g}h^{-1} = g\} \subset SU(2m)\times SU(2m)\,. \tag{5.26}$$

This term, that is absent for surfaces without boundary, may be viewed as describing a version of the boundary "gauge-anomaly". (An identity similar to (5.25) but relating $e^{iS_{WZ}(HGH^{-1})}$ to $e^{iS_{WZ}(G)}$ holds in the case when $HGH^{-1} = G$ on $\partial\Sigma$.) Result (5.25) is most naturally obtained in the language of gerbes, but we present in Section 6 its more elementary proof split into several steps.

### 5.2. *WZ amplitudes of boundary fields $\widetilde{t}_a$*

We shall use result (5.25) to calculate the WZ amplitudes $e^{iS_{WZ}(\widetilde{t}_a)}$ that enter the multiplicative expression (4.106) for $K[\mathcal{E}]$. Such amplitudes may be split into contributions from the restrictions of $\widetilde{t}_a$ to surfaces with boundary

$$\Sigma_+ = [0,\tfrac12]\times[0,\pi] \qquad \text{and} \qquad \Sigma_- = [0,\tfrac12]\times[-\pi,0] \tag{5.27}$$

(it does not matter that the boundaries $\partial\Sigma_\pm$ have corners). As discussed in the previous subsection,

$$e^{iS_{WZ}(\widetilde{t}_a|_{\Sigma_\pm})} \in (\mathbb{L})_{\widetilde{t}|_{\partial\Sigma_\pm}}, \tag{5.28}$$

i.e. to fibers of the line bundle $\mathbb{L}$ over the loop space $LSU(2m)$. The boundary loops $\widetilde{t}_a|_{\partial\Sigma^\pm}$ differ only by an orientation-changing reparameterization so that the corresponding fibers of $\mathbb{L}$ are canonically dual to each other and

$$e^{iS_{WZ}(\widetilde{t}_a)} = \Big\langle e^{iS_{WZ}(\widetilde{t}_a|_{\Sigma_-})}, e^{iS_{WZ}(\widetilde{t}_a|_{\Sigma_+})}\Big\rangle. \tag{5.29}$$

By (4.86),

$$\widetilde{t}_a|_{\Sigma_-} = \big(\widetilde{w}_a\overline{\widetilde{t}_a}\widetilde{w}_a^{-1}\big)|_{\Sigma_+}\circ\vartheta \tag{5.30}$$

where $\widetilde{w}_a(r,k_2)\equiv\widetilde{w}(a,k_2)$ (and is $r$-independent) and $\vartheta$, sending $(r,k_2)$ to $(r,-k_2)$, is an orientation-changing diffeomorphism from $\Sigma_-$ to $\Sigma_+$. Hence, by (5.20),

$$\mathrm{e}^{\mathrm{i}S_{WZ}(\widetilde{t}_a|_{\Sigma_-})} = \Big(\mathrm{e}^{\mathrm{i}S_{WZ}((\widetilde{w}_a\overline{\widetilde{t}_a}\widetilde{w}_a^{-1})|_{\Sigma_+})}\Big)^{-1}. \tag{5.31}$$

On the other hand, $\widetilde{w}_a\overline{\widetilde{t}_a}\widetilde{w}_a^{-1}=\widetilde{t}_a$ on $\partial\Sigma_+$ so that, by (5.25),

$$\mathrm{e}^{\mathrm{i}S_{WZ}((\widetilde{w}_a\overline{\widetilde{t}_a}\widetilde{w}_a^{-1})|_{\Sigma_+})} = \Big(\mathrm{Hol}_{\widehat{\mathcal{J}}}((\widetilde{w}_a,\widetilde{t}_a)|_{\partial\Sigma_+})\Big)^{-1}\exp\Big[\tfrac{\mathrm{i}}{4\pi}(\widetilde{w}_a,\widetilde{t}_a)^*\widehat{\beta}\Big]\,\mathrm{e}^{\mathrm{i}S_{WZ}(\widetilde{t}_a|_{\Sigma_+})}, \tag{5.32}$$

where

$$\begin{aligned}\exp\Big[-\tfrac{\mathrm{i}}{4\pi}\int\limits_{\Sigma_+}(\widetilde{w}_a,\widetilde{t}_a)^*\widehat{\beta}\Big] &= \exp\Big[\tfrac{\mathrm{i}}{4\pi}\int\limits_{\Sigma_+}\mathrm{tr}\Big((\widetilde{w}_a^{-1}d\widetilde{w}_a)\,\overline{(\widetilde{t}_a^{-1}(\mathrm{d}\widetilde{t}_a)+(\mathrm{d}\widetilde{t}_a)\widetilde{t}_a^{-1})}\Big)\Big]\\ &= \exp\big[\pi\mathrm{i}\,\mathcal{B}(\widetilde{t}_a)\big],\end{aligned} \tag{5.33}$$

see (A.5) and (4.100). Relations (5.32), (5.31) and (5.33), together with the tautological equality $\big\langle(\mathrm{e}^{\mathrm{i}S_{WZ}(\widetilde{t}_a|_{\Sigma_+})})^{-1},\,\mathrm{e}^{\mathrm{i}S_{WZ}(\widetilde{t}_a|_{\Sigma_+})}\big\rangle=1$, imply that

$$\mathrm{e}^{\mathrm{i}S_{WZ}(\widetilde{t}_a)} = \mathrm{Hol}_{\widehat{\mathcal{J}}}((\widetilde{w}_a,\widetilde{t}_a)|_{\partial\Sigma_+})\,\exp\big[\pi\mathrm{i}\,\mathcal{B}(\widetilde{t}_a)\big] \tag{5.34}$$

and permit to rewrite Eq. (4.106) as the identity

$$(-1)^{K[\mathcal{E}]} = \frac{\mathrm{Hol}_{\widehat{\mathcal{J}}}((\widetilde{w}_0,\widetilde{t}_0)|_{\partial\Sigma_+})}{\mathrm{Hol}_{\widehat{\mathcal{J}}}((\widetilde{w}_\pi,\widetilde{t}_\pi)|_{\partial\Sigma_+})}. \tag{5.35}$$

### 5.3. *Calculation of the loop holonomy in line bundle $\widehat{\mathcal{J}}$*

We are left with the calculation of the holonomies $\mathrm{Hol}_{\widehat{\mathcal{J}}}((\widetilde{w}_a,\widetilde{t}_a)|_{\partial\Sigma_+})$. As shown in Section 6, line bundle $\widehat{\mathcal{J}}$ over subvariety $F\subset SU(2m)\times SU(2m)$, see (5.26), is obtained by gluing local line bundles $\widehat{\mathcal{J}}_i$ defined over open subsets $\widehat{O}_i=\widehat{O}_{2m-i}$, where $\widehat{O}_i=(SU(2m)\times O_i)\cap F$. Let $p_2$ denote the projection on the second factor in $SU(2m)\times SU(2m)$. Bundles $\widehat{\mathcal{J}}_i$ are the restrictions to $\widehat{O}_i$ of line bundles $\widehat{\mathcal{L}}_{i(2m-i)}$ equal to $p_2^*\mathcal{L}_{i(2m-i)}$ but having the connections modified by the addition of 1-forms $\widehat{\Pi}_i$ of Eq. (6.49). For $(h,g)\in\widehat{O}_i$ with $g=\gamma\,\mathrm{e}^{2\pi\mathrm{i}\,\tau}\gamma^{-1}$,

$$\gamma_0=\gamma^{-1}h\,\overline{\gamma}\,\omega^{-1}\in G_{i(2m-i)} \tag{5.36}$$

and the gluing isomorphisms $\widehat{\jmath}_{ij}:\widehat{\mathcal{J}}_j\to\widehat{\mathcal{J}}_i$ defined over open subsets $\widehat{O}_{ij}$ act by the formula

$$\widehat{\jmath}_{ij}\,[\gamma,\zeta]_{j(2m-j)} = \chi_{(2m-j)(2m-i)}(\gamma_0)\,[\gamma,\zeta]_{i(2m-i)}\,. \tag{5.37}$$

Let us decompose $\partial\Sigma_+$ into segments $e_0$, $e_{\frac12}$, $\ell_0$, $\ell_\pi$,

$$e_0=\{0\}\times[0,\pi],\qquad e_{\frac12}=\{\tfrac12\}\times[0,\pi],\qquad \ell_0=[1,\tfrac12]\otimes\{0\},\qquad \ell_\pi[1,\tfrac12]\times\{0\}, \tag{5.38}$$

that we shall consider with orientations inherited from $\partial\Sigma_+$, see Fig. 15. We first note that $\widetilde{w}_a|_{\ell_{a'}}=\widetilde{w}(a,a')$ is constant so that $\widetilde{t}_a|_{\ell_{a'}}$ takes values in the subgroup $Sp_{\widetilde{w}(a,a')}(2m)$ defined by Eq. (4.80) that is simply connected. Besides, the curvature form $\widehat{\beta}$ of bundle $\widehat{\mathcal{J}}$ vanishes when restricted to $\{\widetilde{w}_{a,a'}\} \times Sp_{\widetilde{w}(a,a')}(2m) \subset F$ so that $\mathrm{Hol}_{\widehat{\mathcal{J}}}((\widetilde{w}_a, \widetilde{t}_a)|_{\partial\Sigma_+})$ does not depend on the choices of the boundary values $\widetilde{t}_a|_{\ell_{a'}}$ provided they respect the constraints

$$\widetilde{t}_a(0,a') = I\,, \qquad\qquad \widetilde{t}_a(\tfrac{1}{2},a') = -I\,, \tag{5.39}$$

$$\widetilde{w}(a,a')\,\overline{\widetilde{t}_a(r,a')}\,\widetilde{w}(a,a')^{-1} = \widetilde{t}_a(r,a')\,. \tag{5.40}$$

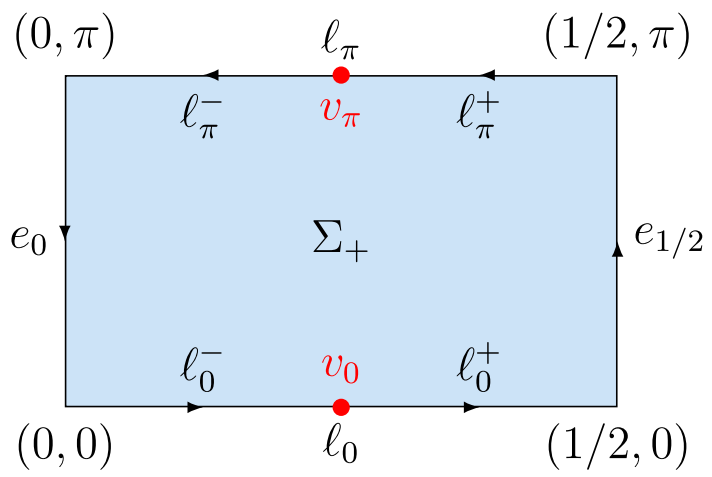

Fig. 15. Surface $\Sigma_+$ and the triangulation of its boundary.

We shall use the local expression for holonomy corresponding to a sufficiently fine triangulation of $\partial\Sigma_+$:

$$\mathrm{Hol}_{\widehat{\mathcal{J}}}((\widetilde{w}_a,\widetilde{t}_a)|_{\partial\Sigma_+}) = \bigotimes_{v\in b\subset\partial\Sigma_+} \widehat{\mathcal{J}}^{\pm 1}_{i_v i_b}\Big(\bigotimes_{b\subset\partial\Sigma_+} \mathrm{Hol}_{\widehat{\mathcal{J}}_{i_b}}((\widetilde{w}_a,\widetilde{t}_a)|_b)\Big), \tag{5.41}$$

where $\mathrm{Hol}_{\widehat{\mathcal{J}}_{i_b}}((\widetilde{w}_a,\widetilde{t}_a)|_b)$ are parallel transports in line bundles $\widehat{\mathcal{J}}_{i_b}$ along open curves $(\widetilde{w}_a,\widetilde{t}_a)_b$ lying in subsets $\widehat{O}_{i_b}$ and isomorphisms $\bigotimes\limits_{v\in b\subset\partial\Sigma_+} \widehat{\mathcal{J}}^{\pm 1}_{i_v i_b}$ are used to glue such local parallel transports together, see Eq. (6.64). We may assume that the triangulation of $\partial\Sigma_+$ contains the segments $e_0$ and $e_{\frac{1}{2}}$ among its edges $b$. For the corresponding indices $i_b$ we shall choose $i_{e_0} = 0 = 2m - i_{e_0}$ and $i_{e_{\frac{1}{2}}} = m = 2m - i_{e_{\frac{1}{2}}}$. Indeed, $\widetilde{t}_a|_{e_0} = I \in \widehat{O}_0$ and $\widetilde{t}_a|_{e_{\frac{1}{2}}} = -I \in \widehat{O}_m$. The corresponding line bundle $\widehat{\mathcal{J}}_0 = \widehat{\mathcal{L}}_{00}$ is trivial and its connection form $\widehat{\Pi}_0$ vanishes along $(\widetilde{w}_a,\widetilde{t}_a)|_{e_0}$. In particular, $\mathrm{Hol}_{\widehat{\mathcal{J}}_0}((\widetilde{w}_a,\widetilde{t}_a)|_{e_0})$ maps element $[\gamma,\zeta]_{00}$ of $(\widehat{\mathcal{L}}_{00})_{(\widetilde{w}(a,\pi),I)}$ into the same element in $(\widehat{\mathcal{L}}_{00})_{(\widetilde{w}(a,0),I)}$ where we wrote $I = \gamma\, \mathrm{e}^{2\pi \mathrm{i}\cdot 0}\gamma^{-1}$. Similarly, the corresponding line bundle $\widehat{\mathcal{J}}_m = \widehat{\mathcal{L}}_{mm}$ is trivial, its connection form $\widehat{\Pi}_m$ vanishes along $(\widetilde{w}_a,\widetilde{t}_a)|_{e_{\frac{1}{2}}}$, and $\mathrm{Hol}_{\widehat{\mathcal{J}}_m}((\widetilde{w}_a,\widetilde{t}_a)|_{e_{\frac{1}{2}}})$ maps element $[\gamma,\zeta]_{mm}$ of $(\widehat{\mathcal{L}}_{mm})_{(\widetilde{w}(a,0),-I)}$ into the same element in $(\widehat{\mathcal{L}}_{mm})_{(\widetilde{w}(a,\pi),-I)}$ if $-I = \gamma\, \mathrm{e}^{2\pi \mathrm{i}\,\lambda_m^\vee}\gamma^{-1}$. In other words, the segments $e_0$ and $e_{\frac{1}{2}}$ do not contribute to (5.41). Let us use the freedom of choice of the boundary values $\widetilde{t}_a|_{\ell_{a'}}$ that allows us not to bother how to extend such restrictions to the interior of $\Sigma_+$. Decomposing

$$\widetilde{w}(a,a') = u_0(a,a')\,\omega\, u_0(a,a')^T \tag{5.42}$$

according to (4.79) with $\det u_0(a,a') = \mathrm{pf}\,\widetilde{w}(a,a')$, we may take

$$\widetilde{t}_a(r,a') = u(a,a')\,\mathrm{e}^{4\pi \mathrm{i}\, r\, \lambda_m^\vee}\, u(a,a')^{-1} \tag{5.43}$$

for

$$u(a,a') = \begin{cases} u_0(a,a') & \text{if} \quad \mathrm{pf}\,\widetilde{w}(a,a') = 1\,, \\ \mathrm{e}^{\frac{\pi \mathrm{i}}{2m}}\, u_0(a,a') & \text{if} \quad \mathrm{pf}\,\widetilde{w}(a,a') = -1 \end{cases} \tag{5.44}$$

so that $u(a,a') \in SU(2m)$. That choice satisfies the 1st constraint (5.39). Since

$$
\begin{aligned}
&\widetilde{w}(a,a')\,\overline{\widetilde{t}_a(r,a')}\,\widetilde{w}(a,a')^{-1}\\
&\quad = u_0(a,a')\,\omega\, u_0(a,a')^T\,\overline{u(a,a')\,\mathrm{e}^{4\pi\mathrm{i}\,r\,\lambda_m^\vee}\,u(a,a')^{-1}}\,(u_0(a,a')^T)^{-1}\omega^{-1}u_0(a,a')^{-1}\\
&\quad = u_0(a,a')\,\omega\,\mathrm{e}^{-4\pi\mathrm{i}\,r\,\lambda_m^\vee}\,\omega^{-1}u_0(a,a')^{-1} = u(a,a')\,\mathrm{e}^{4\pi\mathrm{i}\,r\,\lambda_m^\vee}\,u(a,a')^{-1}\\
&\quad = \widetilde{t}_a(r,a')\,,
\end{aligned}
\tag{5.45}
$$

also the 2nd constraint (5.40) demanding that $(\widetilde{w}(a,a'),\widetilde{t}_a(r,a')) \in F$ is satisfied. Besides

$$
\begin{aligned}
&(\widetilde{w}(a,a'),\widetilde{t}_a(r,a')) \in \widehat{O}_0 \quad \text{for} \quad 0 \le r < \tfrac{1}{2}\,,\\
&(\widetilde{w}(a,a'),\widetilde{t}_a(r,a')) \in \widehat{O}_m \quad \text{for} \quad 0 < r \le \tfrac{1}{2}\,.
\end{aligned}
\tag{5.46}
$$

Note that the corresponding $SU(2m)$ elements (5.36) are $r$-independent and take the values

$$
\gamma_0(a,a') = u(a,a')^{-1}\widetilde{w}(a,a')\,\overline{u(a,a')}\,\omega^{-1} = \begin{cases} I & \text{if} \quad \mathrm{pf}\,\widetilde{w}(a,a') = 1\,,\\ \mathrm{e}^{-\frac{\pi\mathrm{i}}{m}}I & \text{if} \quad \mathrm{pf}\,\widetilde{w}(a,a') = -1\,. \end{cases}
\tag{5.47}
$$

Let us take $\ell_0^\pm$ and $\ell_\pi^\pm$ with

$$
\ell_{a'}^- = [0,\tfrac{1}{4}]\times\{a'\}\,, \qquad \ell_{a'}^+ = [\tfrac{1}{4},\tfrac{1}{2}]\times\{a'\}
\tag{5.48}
$$

as the remaining edges of the triangulation of $\partial\Sigma_+$, again with the orientations inherited from this boundary, setting $i_{\ell_{a'}^-} = 0$ and $i_{\ell_{a'}^+} = m$. Then, similarly as before, $\mathrm{Hol}_{\widehat{\mathcal{J}}_0}((\widetilde{w}_a,\widetilde{t}_a)|_{\ell_{a'}^-}$ are identity maps sending $[\gamma,\zeta]_{00}$ to the same element and $\mathrm{Hol}_{\widehat{\mathcal{J}}_m}((\widetilde{w}_a,\widetilde{t}_a)|_{\ell_{a'}^+}$ are identity maps sending $[\gamma,\zeta]_{mm}$ to itself. Thus the only contribution to the holonomy comes from the gluing isomorphisms (5.37):

$$
\begin{aligned}
&J_{i_{v_0}i_{\ell_0^-}} \otimes J^{-1}_{i_{v_0}i_{\ell_0^+}} = J_{i_{\ell_0^+},i_{\ell_0^-}} = J_{m0}\,, \qquad J_{m0}\,[\gamma,\zeta]_{00} = \chi_{0m}(\gamma_0(a,0))\,[\gamma,\zeta]_{mm}\,,\\
&J_{i_{v_\pi}i_{\ell_\pi^+}} \otimes J^{-1}_{i_{v_\pi}i_{\ell_\pi^-}} = J_{i_{\ell_\pi^-},i_{\ell_\pi^+}} = J_{0m}\,, \qquad J_{0m}\,[\gamma,\zeta]_{mm} = \chi_{m0}(\gamma_0(a,\pi))\,[\gamma,\zeta]_{00}
\end{aligned}
\tag{5.49}
$$

so that

$$
\mathrm{Hol}_{\widehat{\mathcal{J}}}(\widetilde{t}_a|_{\partial\Sigma_+}) \;=\; \chi_{0m}(\gamma_0(a,0))\,\chi_{m0}(\gamma_0(a,\pi))\,.
\tag{5.50}
$$

For $\gamma_0(a,a')$ given by (5.47),

$$
\chi_{0m}(\gamma_0(a,a')) = \chi_{m0}(\gamma_0(a,a')^{-1}) = \mathrm{pf}\,\widetilde{w}(a,a')\,,
\tag{5.51}
$$

see (5.11). Hence

$$
\mathrm{Hol}_{\widehat{\mathcal{J}}}(\widetilde{t}_a|_{\partial\Sigma_+}) \;=\; \frac{\mathrm{pf}\,\widetilde{w}(a,0)}{\mathrm{pf}\,\widetilde{w}(a,\pi)}\,.
\tag{5.52}
$$

The substitution of this relation to Eq. (5.35) permits to rewrite that identity in the form (4.124), completing the proof of equality between the $\mathbb{Z}_2$-valued index $K[\mathcal{E}]$ and the Kane–Mele invariant for vector bundles $\mathcal{E}$ of general rank.

## 6. Proof of the boundary gauge anomaly identity (5.25)

We shall proceed in several steps.

*6.1. Gauge transformations of Wess–Zumino amplitudes*

Suppose now that $H, G : \Sigma \to SU(n)$. If $\partial\Sigma = \emptyset$ then

$$\mathrm{e}^{\mathrm{i}S_{WZ}(HGH^{-1})} = \exp\Big[\tfrac{\mathrm{i}}{4\pi}\int\limits_{\Sigma}(H,G)^*\beta\Big]\,\mathrm{e}^{\mathrm{i}S_{WZ}(G)} \tag{6.1}$$

where $\beta$ is the 2-form on $SU(n)\times SU(n)$ given by Eq. (A.5) from Appendix A. The above formula specifies the behavior of the WZ amplitudes under the gauge transformation $G \mapsto Ad_H(G)$ of fields. It follows easily from Eq. (A.4) if we use formula (4.18) to define the WZ amplitudes but we shall reprove it here using the local expressions (5.16) for the amplitudes because such a proof will be easy to extend to the case of surfaces with boundary that we shall subsequently need.

In terms of the local expressions for the WZ amplitudes, identity (6.1) takes the form

$$\begin{aligned}&\exp\Big[\mathrm{i}\sum_c\int\limits_c(HGH^{-1})^*B_{i_c}\Big]\bigotimes_{b\subset c}\mathrm{Hol}_{\mathcal{L}_{i_ci_b}}(HGH^{-1}|_b)\\&\quad=\exp\Big[\tfrac{\mathrm{i}}{4\pi}\int\limits_{\Sigma}(H,G)^*\beta\Big]\exp\Big[\mathrm{i}\sum_c\int\limits_c G^*B_{i_c}\Big]\bigotimes_{b\subset c}\mathrm{Hol}_{\mathcal{L}_{i_ci_b}}(G|_b)\,.\end{aligned} \tag{6.2}$$

Consider two maps from $SU(n)\times SU(n)$ to $SU(n)$ defined by

$$(h,g)\longmapsto hgh^{-1}\equiv\Phi(h,g)\,,\qquad (h,g)\longmapsto g\equiv p_2(h,g)\,. \tag{6.3}$$

Both send $SU(n)\times O_i$ onto $O_i$. An easy calculation done in Appendix B shows that

$$\Phi^*B_i = p_2^*B_i + \tfrac{1}{4\pi}\beta - d\Pi_i\,, \tag{6.4}$$

where $\Pi_i$ are 1-forms on $SU(n)\times O_i$ given by formula

$$\Pi_i(h,g) = \mathrm{i}\,\mathrm{tr}\big(\gamma(\tau-\lambda_i^\vee)\gamma^{-1}h^{-1}dh\big)\,, \tag{6.5}$$

which is consistent because the maps

$$O_i \ni g = \gamma\,\mathrm{e}^{2\pi\mathrm{i}\tau}\gamma^{-1}\longmapsto\gamma(\tau-\lambda_i^\vee)\gamma^{-1}\in su(2m) \tag{6.6}$$

are well defined and smooth, see [16] and [37]. Eq. (6.4) and the Stokes theorem permit to rewrite

$$\begin{aligned}\exp\Big[\mathrm{i}\sum_c\int\limits_c(HGH^{-1})^*B_{i_c}\Big] &= \exp\Big[\tfrac{\mathrm{i}}{4\pi}\int\limits_{\Sigma}(H,G)^*\beta - \mathrm{i}\sum_{b\subset c}\int\limits_b(H,G)^*\Pi_{i_c}\Big]\\&\qquad\times\exp\Big[\mathrm{i}\sum_c\int\limits_c G^*B_{i_c}\Big].\end{aligned} \tag{6.7}$$

Suppose now that there is a family of line bundles $\mathcal{N}_i$ over sets $SU(n)\times O_i$ such that there exists a family of isomorphisms

$$\iota_{ij} : \Phi^*\mathcal{L}_{ij}\longmapsto p_2^*\mathcal{L}_{ij}\otimes\mathcal{N}_i\otimes\mathcal{N}_j^{-1} \tag{6.8}$$

of line bundles over $SU(2m)\times O_{ij}$ that is compatible with isomorphisms $t_{ijk}$ so that

$$t_{ijk}\circ\iota_{ij}\otimes\iota_{jk} = \iota_{ik}\circ t_{ijk} \tag{6.9}$$

(keep in mind that all line bundles are equipped with a hermitian structure and a unitary connection and their isomorphisms are assumed to preserve that structure). Such a family can be easily constructed. One takes for $\mathcal{N}_i$ the trivial line bundles $(SU(n)\times O_i)\times\mathbb{C}$ with the connection 1-forms $\Pi_i$ given by (6.5). For isomorphisms $\iota_{ij}$ one takes the maps defined by

$$\iota_{ij}\,[h\gamma,\zeta]_{ij}=[\gamma,\zeta]_{ij}\,. \tag{6.10}$$

That $\iota_{ij}$ preserve the connections follows from the following calculation:

$$\begin{aligned}\alpha_{ij}(hgh^{-1})&=\mathrm{i}\,\mathrm{tr}\big(\lambda_{ij}^{\vee}(h\gamma)^{-1}\mathrm{d}(h\gamma)\big)\\&=\mathrm{i}\,\mathrm{tr}\big(\lambda_{ij}^{\vee}\gamma^{-1}\mathrm{d}\gamma)\big)+\mathrm{i}\,\mathrm{tr}\big(\gamma\lambda_{ij}^{\vee}\gamma^{-1}h^{-1}\mathrm{d}h\big)\\&=\alpha_{ij}+\Pi_i(h,g)-\Pi_j(h,g)\,.\end{aligned} \tag{6.11}$$

The compatibility of $\iota_{ij}$ with isomorphisms $t_{ijk}$ is also straightforward to check. Recall that

$$\bigotimes_{b\subset c}\mathrm{Hol}_{\mathcal{L}_{i_ci_b}}(HGH^{-1}|_b)=\bigotimes_{b\subset c}\mathrm{Hol}_{\Phi^*\mathcal{L}_{i_ci_b}}((H,G)|_b)\in\bigotimes_{v\in b\subset c}(\Phi^*\mathcal{L}_{i_ci_b}^{\pm1})_{(H,G)(v)} \tag{6.12}$$

Now, denoting by $\iota_{ij}^{-1}$ the inverse of the dual of isomorphism $\iota_{ij}$,

$$\begin{aligned}&\bigotimes_{v\in b\subset c}\iota_{i_ci_b}^{\pm1}:\bigotimes_{v\in b\subset c}(\Phi^*\mathcal{L}_{i_ci_b}^{\pm1})_{(H,G)(v)}\\&\quad\longrightarrow\bigotimes_{v\in b\subset c}\Big((\mathcal{L}_{i_ci_b}^{\pm1})_{G(v)}\otimes(\mathcal{N}_{i_c}^{\pm1})_{(H,G)(v)}\otimes(\mathcal{N}_{i_b}^{\mp1})_{(H,G)(v)}\Big).\end{aligned} \tag{6.13}$$

Besides

$$\begin{aligned}&\bigotimes_{v\in b\subset c}\iota_{i_ci_b}^{\pm1}\Big(\bigotimes_{b\subset c}\mathrm{Hol}_{\Phi^*\mathcal{L}_{i_ci_b}}((H,G)|_b)\Big)=\Big(\bigotimes_{b\subset c}\mathrm{Hol}_{\mathcal{L}_{i_ci_b}}(G|_b)\Big)\\&\quad\otimes\Big(\bigotimes_{b\subset c}\mathrm{Hol}_{\mathcal{N}_{i_c}}((H,G)|_b)\Big)\otimes\Big(\bigotimes_{b\subset c}\mathrm{Hol}_{\mathcal{N}_{i_b}^{-1}}((H,G)|_b)\Big).\end{aligned} \tag{6.14}$$

Note that

$$\bigotimes_{b\subset c}\mathrm{Hol}_{\mathcal{N}_{i_c}}((H,G)|_b)\in\bigotimes_{v\in b\subset c}(\mathcal{N}_{i_c}^{\pm1})_{(H,G)(v)} \tag{6.15}$$

which is canonically a trivial line since each factor $(\mathcal{N}_{i_c})_{(H,g)(v)}$ is accompanied by $(\mathcal{N}_{i_c}^{-1})_{(H,g)(v)}$ for another $b$, but also because the individual factors are canonically trivial as fibers of trivial bundles (but with non-trivial connection forms). The two trivializations clearly agree. Using the second trivialization, we see that

$$\bigotimes_{b\subset c}\mathrm{Hol}_{\mathcal{N}_{i_c}}((H,G)|_b)=\prod_{b\subset c}\exp\Big[\mathrm{i}\int_b(H,G)^*\Pi_{i_c}\Big]. \tag{6.16}$$

Similarly,

$$\bigotimes_{b\subset c}\mathrm{Hol}_{\mathcal{N}_{i_b}^{-1}}((H,G)|_b)\in\bigotimes_{v\in b\subset c}(\mathcal{N}_{i_b}^{\mp1})((H,G)(v)) \tag{6.17}$$

which is also canonically a trivial line since each factor $(\mathcal{N}_{i_b})_{(H,G)(v)}$ is accompanied by $(\mathcal{N}_{i_b}^{-1})_{(H,G)(v)}$ for another $c$, but also because the individual factors are fibers of trivial line bundles. The two trivializations again agree. Using the second trivialization, we have

$$\bigotimes_{b\subset c}\mathrm{Hol}_{\mathcal{N}_{i_b}^{-1}}((H,G)|_b)=\prod_{b\subset c}\exp\Big[-\mathrm{i}\int_b(H,G)^*\Pi_{i_b}\Big]=1\,. \tag{6.18}$$

Now, the property (6.9) implies that the isomorphism $\bigotimes\limits_{v\in b\subset c}\iota^{\pm1}_{i_c i_b}$ commutes with the trivializations of lines in (5.19) and, assuming that such trivializations are performed, we infer that, as complex numbers,

$$\bigotimes_{b\subset c}\mathrm{Hol}_{\mathcal{L}_{i_c i_b}}(HGH^{-1}|_b)=\prod_{b\subset c}\exp\Big[\mathrm{i}\int_b(H,G)^*\Pi_{i_c}\Big]\bigotimes_{b\subset c}\mathrm{Hol}_{\mathcal{L}_{i_c i_b}}(G|_b) \tag{6.19}$$

Multiplying relations (6.7) and (6.19), we obtain identity (6.2) establishing the gauge transformation law (6.1).

If $\partial\Sigma=\mathcal{C}$ then the WZ amplitudes take values in the line bundle $\mathbb{L}$ over the loop space $LSU(n)$. In particular,

$$\mathrm{e}^{\mathrm{i}S_{\mathrm{WZ}}(HGH^{-1})}\in(\mathbb{L})_{HGH^{-1}|_{\mathcal{C}}}\quad\text{and}\quad\mathrm{e}^{\mathrm{i}S_{\mathrm{WZ}}(G)}\in(\mathbb{L})_{G|_{\mathcal{C}}} \tag{6.20}$$

and $(\mathbb{L})_{HGH^{-1}|_{\mathcal{C}}}$ and $(\mathbb{L})_{G|_{\mathcal{C}}}$ are different fibers of $\mathbb{L}$ so that relation (6.1) *a priory* does not make sense unless we specify a way to identify such fibers. The required identification is produced by the map

$$\mathbb{I}=\bigotimes_{v\in b\subset\mathcal{C}}\iota^{\pm1}_{i_v i_b} \tag{6.21}$$

that establishes an isomorphism of the lines

$$\begin{aligned}\mathbb{I}:&\bigotimes_{v\in b\subset\mathcal{C}}(\mathcal{L}^{\pm1}_{i_v i_b})_{HGH^{-1}(v)}\\ &\longrightarrow\bigotimes_{v\in b\subset\mathcal{C}}\Big((\mathcal{L}^{\pm1}_{i_v i_b})_{G(v)}\otimes(\mathcal{N}^{\pm1}_{i_v})_{(H,G)(v)}\otimes(\mathcal{N}^{\mp1}_{i_b})_{(H,G)(v)}\Big).\end{aligned} \tag{6.22}$$

The line

$$\bigotimes_{v\in b\subset\mathcal{C}}(\mathcal{N}^{\pm1}_{i_v})_{(H,G)(v)} \tag{6.23}$$

is canonically trivial because each factor $(\mathcal{N}_{i_v})_{(H,G)(v)}$ is accompanied by its dual. On the other hand,

$$\bigotimes_{b\subset\mathcal{C}}\mathrm{Hol}_{\mathcal{N}_{i_b}}((H,G)|_b)\in\bigotimes_{v\in b\subset\mathcal{C}}(\mathcal{N}^{\pm1}_{i_b})_{(H,G)(v)} \tag{6.24}$$

is a non-zero element that canonically trivializes the latter tensor product of lines. Hence $\mathbb{I}$ may be viewed as providing an isomorphism between lines

$$\bigotimes_{v\in b\subset\mathcal{C}}(\mathcal{L}^{\pm1}_{i_v i_b})_{HGH^{-1}(v)}\quad\text{and}\quad\bigotimes_{v\in b\subset\mathcal{C}}(\mathcal{L}^{\pm1}_{i_v i_b})_{G(v)} \tag{6.25}$$

which, besides, is compatible with the identifications when one changes the triangulation of $\mathcal{C}$ and the assignments $i_v, i_b$ so that.[9]

$$\mathbb{I} : (\mathbb{L})_{HGH^{-1}|_{\mathcal{C}}} \longrightarrow (\mathbb{L})_{G|_{\mathcal{C}}} . \tag{6.26}$$

The gauge transformation formula for the WZ amplitudes on the surface $\Sigma$ with $\partial\Sigma = \mathcal{C}$ takes now the form

$$\mathbb{I}\left(\mathrm{e}^{\mathrm{i}S_{WZ}(HGH^{-1})}\right) = \exp\left[\tfrac{\mathrm{i}}{4\pi}\int\limits_{\Sigma}(H,G)^*\beta\right]\mathrm{e}^{\mathrm{i}S_{WZ}(G)} , \tag{6.27}$$

proven the same way as for the surface without boundary, with the isomorphism $\mathbb{I}$ providing the identification of the lines along the boundary of $\Sigma$.

### 6.2. *Wess–Zumino amplitudes of complex conjugate fields*

We shall also need to compare the WZ amplitudes for field $G : \Sigma \to SU(2m)$ and for its complex conjugate $\overline{G}$. If $g = \gamma\, \mathrm{e}^{2\pi\mathrm{i}\,\tau}\gamma^{-1} \in O_i$ then

$$\overline{\gamma\, \mathrm{e}^{2\pi\mathrm{i}\,\tau}\gamma^{-1}} = \overline{\gamma}\, \mathrm{e}^{-2\pi\mathrm{i}\tau}\overline{\gamma}^{-1} = \overline{\gamma}\,\omega^{-1}\mathrm{e}^{2\pi\mathrm{i}\widehat{\tau}}\omega\,\overline{\gamma}^{-1}, \tag{6.28}$$

where $\omega$ is given by (4.79) and, for $\tau = \sum\limits_{i=0} t_i\lambda_i^\vee$,

$$\widehat{\tau} = -\omega\tau\omega^{-1} = \sum_{i=0}^{2m-1} t_i\lambda_{2m-i}^\vee \tag{6.29}$$

setting $\lambda_{2m}^\vee \equiv \lambda_0^\vee = 0$ because

$$\omega\lambda_i^\vee\omega^{-1} = -\lambda_{2m-i}^\vee . \tag{6.30}$$

Denoting by $K$ the complex conjugation map on $SU(2m)$, we then infer that $\bar{g} \in O_{2m-i}$, i.e. that

$$K(O_i) = O_{2m-i} , \tag{6.31}$$

again with the convention that $O_{2m} \equiv O_0$. A straightforward check shows that

$$K^* B_{2m-i} = B_i . \tag{6.32}$$

There are natural isomorphisms

$$\kappa_{ij} : K^*\mathcal{L}_{(2m-i)(2m-j)} \longrightarrow \mathcal{L}_{ij} \tag{6.33}$$

of line bundles over $O_{ij}$ given by

$$\kappa_{ij}\, [\overline{\gamma}\,\omega^{-1}, \zeta]_{2m-i,2m-j} = [\gamma, \zeta]_{ij} . \tag{6.34}$$

This is well defined since if $\gamma_0 \in G_{(2m-i)(2m-j)}$ then $\overline{\omega^{-1}\gamma_0\,\omega} \in G_{ij}$ and

$$\chi_{(2m-i)(2m-j)}(\gamma_0) = \chi_{ij}(\overline{\omega^{-1}\gamma_0\,\omega}) , \tag{6.35}$$

[9] In the language of gerbes, the above construction describes the transgression from 1-isomorphisms of gerbes to isomorphisms of line bundles over the loop space.

see Eqs. (5.11) and (5.13). That $\kappa_{ij}$ preserve connections follows from the identity

$$\begin{aligned}\alpha_{(2m-i)(2m-j)}(\bar{g}) &= \mathrm{i}\,\mathrm{tr}\big(\omega^{-1}\lambda^\vee_{(2m-i)(2m-j)}\omega\,\bar{\gamma}^{-1}\mathrm{d}\bar{\gamma}\big)\\ &= -\mathrm{i}\,\mathrm{tr}\big(\lambda^\vee_{ij}\bar{\gamma}^{-1}\mathrm{d}\bar{\gamma}\big) = \mathrm{i}\,\mathrm{tr}\big(\lambda^\vee_{ij}\gamma^{-1}\mathrm{d}\gamma\big) = \alpha_{ij}(g)\,. \end{aligned} \tag{6.36}$$

Isomorphisms $\kappa_{ij}$ are also compatible with $t_{ijk}$:

$$t_{ijk}\circ\kappa_{ij}\otimes\kappa_{jk} = \kappa_{ik}\circ t_{(2m-i)(2m-j)(2m-k)}\,. \tag{6.37}$$

If $\partial\Sigma = \emptyset$ then

$$\begin{aligned} \mathrm{e}^{\mathrm{i}S_{\mathrm{WZ}}(\bar{G})} &= \exp\Big[\mathrm{i}\sum_c\int_c \bar{G}^*B_{2m-i_c}\Big]\bigotimes_{b\subset c}\mathrm{Hol}_{\mathcal{L}_{(2m-i_c)(2m-i_b)}}(\bar{G}|_b)\\ &= \exp\Big[\mathrm{i}\sum_c\int_c G^*B_{i_c}\Big]\bigotimes_{b\subset c}\mathrm{Hol}_{K^*\mathcal{L}_{(2m-i_c)(2m-i_b)}}(G|_b)\\ &= \exp\Big[\mathrm{i}\sum_c\int_c G^*B_{i_c}\Big]\bigotimes_{v\in b\subset c}\kappa^{\pm1}_{i_c i_b}\Big(\bigotimes_{b\subset c}\mathrm{Hol}_{K^*\mathcal{L}_{(2m-i_c)(2m-i_b)}}(G|_b)\Big)\\ &= \exp\Big[\mathrm{i}\sum_c\int_c G^*B_{i_c}\Big]\bigotimes_{b\subset c}\mathrm{Hol}_{\mathcal{L}_{i_c i_b}}(G|_b) \;=\; \mathrm{e}^{\mathrm{i}S_{\mathrm{WZ}}(G)}, \end{aligned} \tag{6.38}$$

where we used the compatibility (6.37) which guaranties that the isomorphism $\bigotimes_{v\in b\subset c}\kappa^{\pm1}_{i_c i_b}$ commutes with canonical trivializations of lines

$$\bigotimes_{v\in b\subset c}\big(K^*\mathcal{L}^{\pm1}_{(2m-i_c)(2m-i_b)}\big)_{G(v)} \quad\text{and}\quad \bigotimes_{v\in b\subset c}\big(\mathcal{L}^{\pm1}_{i_c i_b}\big)_{G(v)}. \tag{6.39}$$

In the case of surface with boundary $\partial\Sigma = \mathcal{C}$, when the WZ amplitudes take values in a line bundle over the loop space $LSU(2m)$,

$$\mathrm{e}^{\mathrm{i}S_{\mathrm{WZ}}(\bar{G})} \in (\mathbb{L})_{\bar{G}|_{\mathcal{C}}} \quad\text{and}\quad \mathrm{e}^{\mathrm{i}S_{\mathrm{WZ}}(G)} \in (\mathbb{L})_{G|_{\mathcal{C}}}. \tag{6.40}$$

We may use isomorphisms $\kappa_{ij}$ to compare the two lines. Indeed, for

$$\mathbb{K} = \bigotimes_{v\in b\subset\mathcal{C}}\kappa^{\pm1}_{i_v i_b}, \tag{6.41}$$

$$\mathbb{K}:\ \bigotimes_{v\in b\subset\mathcal{C}}\big(K^*\mathcal{L}_{(2m-i_v)(2m-i_b)}\big)_{G(v)} \longrightarrow \bigotimes_{v\in b\subset\mathcal{C}}\big(\mathcal{L}_{i_v i_b}\big)_{G(v)} \tag{6.42}$$

Property (6.37) assures that the isomorphisms are compatible with the identifications when one changes the triangulation of $\mathcal{C}$ and the assignments $i_v, i_b$ so that

$$\mathbb{K}: (\mathbb{L})_{\bar{G}|_{\mathcal{C}}} \longrightarrow (\mathbb{L})_{G|_{\mathcal{C}}}. \tag{6.43}$$

Relation (6.38) is now replaced by

$$\mathbb{K}\Big(\mathrm{e}^{\mathrm{i}S_{\mathrm{WZ}}(\bar{G})}\Big) = \mathrm{e}^{\mathrm{i}S_{\mathrm{WZ}}(G)}. \tag{6.44}$$

### 6.3. Case of fields satisfying $H\overline{G}H^{-1} = G$ on $\partial\Sigma$

By composing relations (6.44) and (6.27), we obtain for maps $H, G : \Sigma \to SU(2m)$ and $\partial\Sigma = \mathcal{C}$ the identity

$$\mathbb{J}\Big(\mathrm{e}^{\mathrm{i}S_{\mathrm{WZ}}(H\overline{G}H^{-1})}\Big) = \exp\Big[\tfrac{\mathrm{i}}{4\pi}(H,\overline{G})^*\beta\Big]\,\mathrm{e}^{\mathrm{i}S_{\mathrm{WZ}}(G)}\,, \tag{6.45}$$

where

$$\mathbb{J} : (\mathbb{L})_{(H\overline{G}H^{-1})|_{\mathcal{C}}} \longrightarrow (\mathbb{L})_{G|_{\mathcal{C}}} \tag{6.46}$$

has the local representation

$$\mathbb{J} = \bigotimes_{v\in b\subset\mathcal{C}} J^{\pm 1}_{i_v i_b}. \tag{6.47}$$

Here

$$J_{ij} : \widehat{\Phi}^*\mathcal{L}_{(2m-i)(2m-j)} \longrightarrow p_2^*\mathcal{L}_{ij}\otimes\widehat{\mathcal{N}}_i\otimes\widehat{\mathcal{N}}_j^{-1}, \tag{6.48}$$

are isomorphisms of line bundles on $SU(2m)\times O_{ij}$ with $\widehat{\Phi} = \Phi\circ(\mathrm{Id}, K)$, i.e. $\widehat{\Phi}(h, g) = h\overline{g}h^{-1}$, and with $\widehat{\mathcal{N}}_i = (\mathrm{Id}, K)^*\mathcal{N}_{2m-i}$ being trivial line bundles over $SU(2m)\times O_i$ with connection forms $\widehat{\Pi}_i = (\mathrm{Id}, K)^*\Pi_i$, i.e.

$$\widehat{\Pi}_i(h, g) = -\mathrm{i}\,\mathrm{tr}\big(\overline{\gamma}(\tau - \lambda_i^\vee)\overline{\gamma}^{-1}h^{-1}\mathrm{d}h\big)\,. \tag{6.49}$$

Recall relation (6.28). One has

$$J_{ij}\,[h\overline{\gamma}\,\omega^{-1}, \zeta]_{(2m-i)(2m-j)} = [\gamma, \zeta]_{ij}\,. \tag{6.50}$$

Now suppose that $H\overline{G}H^{-1} = G$ on $\partial\Sigma = \mathcal{C}$. Then the lines $(\mathbb{L})_{(H\overline{G}H^{-1})|_{\mathcal{C}}}$ and $(\mathbb{L})_{G|_{\mathcal{C}}}$ coincide and the isomorphism (6.46) has to act by the multiplication by a phase in the same line. We would like to find that phase. Let us consider the subvariety of $SU(2m)\times SU(2m)$

$$F = \{\,(h, g)\in SU(2m)\times SU(2m)\mid h\overline{g}h^{-1} = g\,\}\,. \tag{6.51}$$

We shall cover $F$ by open sets $\widehat{O}_i = (SU(2m)\times O_i)\cap F$. On subvariety $F$, $h\overline{\gamma}\,\omega^{-1}\mathrm{e}^{2\pi i\widehat{\tau}}\times(h\overline{\gamma}\,\omega^{-1})^{-1} = \gamma\,\mathrm{e}^{2\pi i\tau}\gamma^{-1}$ so that $\tau = \widehat{\tau}$ and

$$h\overline{\gamma}\,\omega^{-1} = \gamma\gamma_0\,, \tag{6.52}$$

where $\gamma_0\,\mathrm{e}^{2\pi i\tau}\gamma_0^{-1} = \gamma_0\mathrm{e}^{2\pi i\widehat{\tau}}\gamma_0^{-1} = 1$. Besides, if $g\in O_i$ then also $g\in O_{2m-i}$ and

$$\gamma_0(\tau - \lambda_i^\vee)\gamma_0^{-1} = \tau - \lambda_i\,,\qquad \gamma_0(\tau - \lambda_{2m-i}^\vee)\gamma_0^{-1} = \tau - \lambda_{2m-i}^\vee\,, \tag{6.53}$$

see [16]. Consequently, $\gamma_0\in G_{i(2m-i)}$, see (5.8). Over $\widehat{O}_{ij}$, one may naturally identify line bundle $\widehat{\Phi}^*\mathcal{L}_{(2m-i)(2m-j)}$ with $p_2^*\mathcal{L}_{(2m-i)(2m-j)}$ by the relation

$$[h\overline{\gamma}\,\omega^{-1}, \zeta]_{(2m-i)(2m-j)} = [\gamma\gamma_0, \zeta]_{(2m-i)(2m-j)}\,. \tag{6.54}$$

With this identification, the isomorphism $J_{ij}$ descends to

$$\widetilde{J}_{ij} : p_2^*\mathcal{L}_{(2m-i)(2m-j)} \longrightarrow p_2^*\mathcal{L}_{ij}\otimes\widehat{\mathcal{N}}_i\otimes\widehat{\mathcal{N}}_j^{-1} \tag{6.55}$$

over $\widehat{O}_{ij}$ such that

$$\widetilde{J}_{ij}\,[\gamma\gamma_0, \zeta]_{(2m-i)(2m-j)} = [\gamma, \zeta]_{ij} \tag{6.56}$$

or

$$\widetilde{\jmath}_{ij}\,[\gamma,\zeta]_{(2m-i)(2m-j)}=\chi_{(2m-i)(2m-j)}(\gamma_0)^{-1}\,[\gamma,\zeta]_{ij}\,. \tag{6.57}$$

It induces an isomorphism $\widehat{\jmath}_{ij}:\widehat{\mathcal{J}}_j\longmapsto\widehat{\mathcal{J}}_i$ of line bundles $\widehat{\mathcal{J}}_i=\big(p_2^*\mathcal{L}_{i(2m-i)}\otimes\widehat{\mathcal{N}}_i\big)|_{\widehat{O}_i}$ such that

$$\widehat{\jmath}_{ij}\,[\gamma,\zeta]_{j(2m-j)}\;=\;\chi_{(2m-j)(2m-i)}(\gamma_0)\,[\gamma,\zeta]_{i(2m-i)}\,. \tag{6.58}$$

Such isomorphisms permit to construct a global line bundle $\widehat{\mathcal{J}}$ over subvariety $F$. The relation

$$\widehat{\Phi}^*B_{2m-i}=p_2^*B_i+\tfrac{1}{4\pi}\widehat{\beta}-\mathrm{d}\widehat{\Pi}_i\,, \tag{6.59}$$

where $\widehat{\beta}=(\mathrm{Id},K)^*\beta$, see Appendix B, implies that on $\widehat{O}_i$

$$p_2^*B_{i(2m-i)}+d\widehat{\Pi}_i=\tfrac{1}{4\pi}\widehat{\beta} \tag{6.60}$$

so that the 2-form on the right-hand side is the curvature of line bundle $\widehat{\mathcal{J}}$. We want to interpret

$$\begin{aligned}&\bigotimes_{v\in b\subset\mathcal{C}}\widetilde{\jmath}^{\,\pm1}_{i_vi_b}:\bigotimes_{v\in b\subset\mathcal{C}}(\mathcal{L}^{\pm1}_{(2m-i_v)(2m-i_b)})_{G(v)}\\&\quad\longrightarrow\bigotimes_{v\in b\subset\mathcal{C}}\Big((\mathcal{L}^{\pm1}_{i_vi_b})_{G(v)}\otimes(\widehat{\mathcal{N}}^{\pm1}_{i_v})_{(H,G)(v)}\otimes(\widehat{\mathcal{N}}^{\mp1}_{i_b})_{(H,G)(v)}\Big)\end{aligned} \tag{6.61}$$

as a linear map of line $(\mathbb{L})_{G|_{\mathcal{C}}}$ into itself that necessarily acts as multiplication by a phase. Equivalently, we may look at

$$\bigotimes_{v\in b\subset\mathcal{C}}\widehat{\jmath}^{\,\pm1}_{i_vi_b}:\bigotimes_{v\in b\subset\mathcal{C}}(\widehat{\mathcal{J}}^{\pm1}_{i_b})_{(H,G)(v)}\longrightarrow\bigotimes_{v\in b\subset\mathcal{C}}(\widehat{\mathcal{J}}^{\pm1}_{i_v})_{(H,G)(v)}=\mathbb{C}\,. \tag{6.62}$$

Note that the first tensor product of lines is trivialized by the canonical element

$$\bigotimes_{b\subset\mathcal{C}}\mathrm{Hol}_{\widehat{\mathcal{J}}_{i_b}}((H,G)|_b)\in\bigotimes_{v\in b\subset\mathcal{C}}(\widehat{\mathcal{J}}^{\pm1}_{i_b})_{(H,G)(v)} \tag{6.63}$$

so that the phase that we search for will be equal to the image of that element under the linear map (6.62). But

$$\bigotimes_{v\in b\subset\mathcal{C}}\widehat{\jmath}^{\,\pm1}_{i_vi_b}\Big(\bigotimes_{b\subset\mathcal{C}}\mathrm{Hol}_{\widehat{\mathcal{J}}_{i_b}}((H,G)|_b)\Big)=\mathrm{Hol}_{\widehat{\mathcal{J}}}((H,G)|_{\mathcal{C}}) \tag{6.64}$$

as the isomorphisms $\widehat{\jmath}^{\,\pm1}_{i_vi_b}$ provide the gluing of the local parallel transports along $(H,G)|_b$ in line bundles $\widehat{\mathcal{J}}_{i_b}$.

Let us summarize the above discussion. For fields $H,G:\Sigma\to SU(2m)$ such that $H\overline{G}H^{-1}=G$ on $\partial\Sigma=\mathcal{C}$, we may rewrite relation (6.45) in the form

$$\mathrm{Hol}_{\widehat{\mathcal{J}}}((H,G)|_{\mathcal{C}})\,\mathrm{e}^{\mathrm{i}S_{\mathrm{WZ}}(H\overline{G}H^{-1})}=\exp\Big[\tfrac{\mathrm{i}}{4\pi}(H,G)^*\widehat{\beta}\Big]\,\mathrm{e}^{\mathrm{i}S_{\mathrm{WZ}}(G)} \tag{6.65}$$

if both WZ amplitudes are considered as taking values in the same line $(\mathbb{L})_{G|_{\mathcal{C}}}$. Result (5.25) for surfaces $\Sigma$ with an arbitrary number of boundary components is a straightforward generalization of (6.65).

## 7. Conclusions

The present paper gave mathematical details of the construction of a new topological invariant for periodically driven lattice two-dimensional systems with time-reversal symmetry and quasienergy gaps. The existence of such an invariant and its basic properties were recently announced in [5] by some of the authors. The invariant constructed here was represented by a gap-dependent $\mathbb{Z}_2$-valued index denoted $K_\epsilon[U]$. Index $K_\epsilon[U]$ was designed to replace in the presence of time-reversal symmetry the $\mathbb{Z}$-valued index $W_\epsilon[U]$ introduced in [44] which vanishes in this case, similarly as the $\mathbb{Z}_2$-valued Kane–Mele index replaces the first Chern numbers vanishing for time-reversal symmetric band insulators. The difference of indices $W_\epsilon[U]$ for two quasienergy gaps was shown in [44] to give the first Chern number of the band in-between the gaps. We showed that a similar relation holds in the time-reversal symmetric case with the difference of $K_\epsilon[U]$ for two gaps giving the Kane–Mele invariant of the band delimited by the gaps. The proof of the latter relation, however, appeared to be considerably harder than that of the former one. We based it on a new representation of the Kane–Mele invariant of a vector bundle composed of ranges of a smooth family $\mathrm{BZ} \ni k \mapsto P(k)$ of projectors as a square root of the topological Wess–Zumino amplitude[10] of the associated family of unitary matrices $U_P(k) = I - 2P(k)$. Such a representation, in the spirit of dimensional reduction discussed in [43] or [45], has allowed us to use powerful techniques, linked to geometry of bundle gerbes [19,16], that were developed for calculating Wess–Zumino amplitudes. In particular, a crucial role in our argument was played by a formula for the gauge transformation of the Wess–Zumino amplitudes on surfaces with boundary which permitted us to relate the Kane–Mele invariant to the boundary gauge anomaly. We expect that similar techniques will be useful to establish the bulk-boundary correspondence for the periodically driven time-reversal systems conjectures and numerically checked in [5]. This, as well as the extension of our construction to periodically driven systems in other symmetry classes with $\mathbb{Z}_2$-valued static invariants and relating such invariants to the response to external fields [45] remain the important open problems on the mathematical side that we leave to a future research.

On the physical side, although significant progress have been made, unambiguous observations of nontrivial $W$ indices have still not occurred. When this milestone is achieved, designing driven time-reversal invariant systems where a nontrivial $K$ index could be measured in the bulk will be conceivable. Beyond the observation of bulk indices, one can hope to realize this kind of out-of-equilibrium topological transition in electronic systems, where transport experiments would directly probe edge states. As opposed to Chern insulators, (strong) Kane–Mele insulators also exist in three dimensions, where they display two-dimensional surface states. One can therefore expect the existence of a corresponding three-dimensional out-of-equilibrium topological phase whose characterization remains an open problem. In order to fully connect the theoretical description to a potential experimental realization, which would inevitably be subject of some amount of disorder, a formulation of both $K$ and $W$ invariants avoiding the use of a Brillouin zone would be necessary, and could be attempted within the framework of noncommutative geometry. An important question related to the present work is the applicability of the closed-system description using unitary evolution operators in periodically driven regimes. Such a description does not account for the dissipation of the energy pumped into the system and supposes small frequencies of forcing. The question of the influence of the back-reaction of the environment (e.g. of phonons) on the topological properties of driven systems could be addressed on various levels of modeling of open dynamics. Despite important work on this topic, this question remains open.

[10] The strong Kane–Mele invariant in three dimensions is known to be related to the Chern–Simons amplitude, see [11].

## Acknowledgements

The work of D.C. and P.D. was supported by a grant from Agence Nationale de la Recherche (ANR Blanc-2010 IsoTop). C.T. thanks G. Panati, D. Monaco and D. Fiorenza for invitation and fruitful discussions.

## Appendix A. Some formulae for 3-forms

We first show that on $U(N)\times U(N)$,

$$\chi(u_1u_2)=\chi(u_1)+\chi(u_2)+3\,\mathrm{d}\,\mathrm{tr}\left(u_1^{-1}(\mathrm{d}u_1)u_2(\mathrm{d}u_2^{-1})\right). \tag{A.1}$$

This follows by inserting the relation

$$(u_1u_2)^{-1}\mathrm{d}(u_1u_2)=u_2^{-1}u_1^{-1}(du_1)u_2+u_2^{-1}\mathrm{d}u_2 \tag{A.2}$$

into the left-hand side of (A.1), grouping the terms using the cyclicity of the trace and observing that

$$\begin{aligned}&\mathrm{d}\,\mathrm{tr}\left(u_1^{-1}(\mathrm{d}u_1)u_2(\mathrm{d}u_2^{-1})\right)\\&\quad=-\mathrm{tr}\left(u_1^{-1}(\mathrm{d}u_1)u_1^{-1}(\mathrm{d}u_1)u_2(\mathrm{d}u_2^{-1})\right)-\mathrm{tr}\left(u_1^{-1}(\mathrm{d}u_1)u_2(\mathrm{d}u_2^{-1})u_2(\mathrm{d}u_2^{-1})\right)\end{aligned} \tag{A.3}$$

and that $u_2\mathrm{d}u_2^{-1}=-(du_2)u_2^{-1}$. Next, the iteration of formula (A.1) shows that

$$\chi(u_1u_2u_1^{-1})=\chi(u_2)+3\,\mathrm{d}\beta(u_1,u_2), \tag{A.4}$$

where 2-form $\beta$ on $U(N)\times U(N)$ is given by

$$\beta(u_1,u_2)=-\mathrm{tr}\Big(u_2(u_1^{-1}\mathrm{d}u_1)u_2^{-1}(u_1^{-1}\mathrm{d}u_1)+u_1^{-1}(\mathrm{d}u_1)\big(u_2^{-1}(\mathrm{d}u_2)+(\mathrm{d}u_2)u_2^{-1}\big)\Big). \tag{A.5}$$

## Appendix B. Proof of formulae (6.4) and (6.59)

We first prove (6.4). For $h\in SU(n)$ and $g=\gamma\,\mathrm{e}^{2\pi\mathrm{i}\tau}\gamma^{-1}\in O_i$, one has

$$\begin{aligned}(\Phi^*B_i)(h,g)&=\tfrac{1}{4\pi}\,\mathrm{tr}\Big(\gamma^{-1}h^{-1}\mathrm{d}(h\gamma)\,\mathrm{e}^{2\pi i\tau}\,\gamma^{-1}h^{-1}\mathrm{d}(h\gamma)\,\mathrm{e}^{-2\pi i\tau}\Big)\\&\qquad+\mathrm{i}\,\mathrm{tr}\Big((\tau-\lambda_i^\vee)\big(\gamma^{-1}h^{-1}\mathrm{d}(h\gamma)\big)^2\Big)\\&=B_i(g)+\tfrac{1}{4\pi}\,\mathrm{tr}\Big(\gamma^{-1}h^{-1}(\mathrm{d}h)\gamma\,\mathrm{e}^{2\pi i\tau}\,\gamma^{-1}(\mathrm{d}\gamma)\,\mathrm{e}^{-2\pi i\tau}\Big)\\&\qquad+\mathrm{i}\,\mathrm{tr}\Big((\tau-\lambda_i^\vee)\big(\gamma^{-1}h^{-1}(\mathrm{d}h)(\mathrm{d}\gamma)\big)\Big)\\&\qquad+\tfrac{1}{4\pi}\,\mathrm{tr}\Big(\gamma^{-1}(\mathrm{d}\gamma)\,\mathrm{e}^{2\pi i\tau}\,\gamma^{-1}h^{-1}(\mathrm{d}h)\gamma\,\mathrm{e}^{-2\pi i\tau}\Big)\\&\qquad+\mathrm{i}\,\mathrm{tr}\Big((\tau-\lambda_i^\vee)\big(\gamma^{-1}(\mathrm{d}\gamma)\gamma^{-1}h^{-1}(\mathrm{d}h)\gamma\big)\Big)\\&\qquad+\tfrac{1}{4\pi}\,\mathrm{tr}\Big(\gamma^{-1}h^{-1}(\mathrm{d}h)\gamma\,\mathrm{e}^{2\pi i\tau}\,\gamma^{-1}h^{-1}(\mathrm{d}h)\gamma\,\mathrm{e}^{-2\pi i\tau}\Big)\end{aligned}$$

$$
\begin{aligned}
&+ \mathrm{i}\,\mathrm{tr}\Big((\tau-\lambda_i^\vee)\big(\gamma^{-1}h^{-1}(\mathrm{d}h)\gamma\big)^2\Big)\\
&= B_i(g) + \tfrac{1}{4\pi}\,\mathrm{tr}\Big(h^{-1}(\mathrm{d}h)\big(g(d\gamma)\gamma^{-1}g^{-1} - g^{-1}(\mathrm{d}\gamma)\gamma^{-1}g\big)\Big)\\
&\quad - \tfrac{1}{4\pi}\,\mathrm{tr}\Big(g\,h^{-1}(\mathrm{d}h)g^{-1}h^{-1}(\mathrm{d}h)\Big) - \mathrm{i}\,\mathrm{tr}\Big((\mathrm{d}\gamma)(\tau-\lambda_i^\vee)\gamma^{-1}h^{-1}(\mathrm{d}h)\Big)\\
&\quad + \mathrm{i}\,\mathrm{tr}\Big(\gamma(\tau-\lambda_i^\vee)\gamma^{-1}(\mathrm{d}\gamma)\gamma^{-1}h^{-1}(\mathrm{d}h)\Big)\\
&\quad + \mathrm{i}\,\mathrm{tr}\Big(\gamma(\tau-\lambda_i^\vee)\gamma^{-1}(h^{-1}\mathrm{d}h)^2\Big)
\end{aligned}
\tag{B.1}
$$

Since

$$
\begin{aligned}
d\Pi_i(h,g) = &-\mathrm{i}\,\mathrm{tr}\Big(\gamma(\tau-\lambda_i^\vee)\gamma^{-1}(h^{-1}\mathrm{d}h)^2\Big) + \mathrm{i}\,\mathrm{tr}\Big((\mathrm{d}\gamma)(\tau-\lambda_i^\vee)\gamma^{-1}h^{-1}(\mathrm{d}h)\Big)\\
&- \mathrm{i}\,\mathrm{tr}\Big(\gamma(\tau-\lambda_i^\vee)\gamma^{-1}(\mathrm{d}\gamma)\gamma^{-1}h^{-1}(\mathrm{d}h)\Big) + \mathrm{i}\,\mathrm{tr}\Big(\gamma(\mathrm{d}\tau)\gamma^{-1}h^{-1}(\mathrm{d}h)\Big),
\end{aligned}
\tag{B.2}
$$

we obtain

$$
\begin{aligned}
(\Phi^*B_i)(h,g) = B_i(g) &+ \tfrac{1}{4\pi}\,\mathrm{tr}\Big(h^{-1}(\mathrm{d}h)\big(g(d\gamma)\gamma^{-1}g^{-1} - g^{-1}(\mathrm{d}\gamma)\gamma^{-1}g\\
&- 4\pi\mathrm{i}\gamma(\mathrm{d}\tau)\gamma^{-1}\big)\Big) - \tfrac{1}{4\pi}\,\mathrm{tr}\Big(\gamma^{-1}h^{-1}\mathrm{d}(h\gamma)\,\mathrm{e}^{2\pi i\tau}\,\gamma^{-1}h^{-1}\mathrm{d}(h\gamma)\,\mathrm{e}^{-2\pi i\tau}\Big)\\
&- \tfrac{1}{4\pi}\,\mathrm{tr}\Big(g\,h^{-1}(\mathrm{d}h)g^{-1}h^{-1}(\mathrm{d}h)\Big) - d\Pi_i(h,g)\,.
\end{aligned}
\tag{B.3}
$$

Finally, using the identity

$$
g^{-1}(\mathrm{d}g) + (\mathrm{d}g)g^{-1} = g^{-1}(\mathrm{d}\gamma)\gamma^{-1}g - g(\mathrm{d}\gamma)\gamma^{-1}g^{-1} + 4\pi\mathrm{i}\,\gamma(\mathrm{d}\tau)\gamma^{-1},
\tag{B.4}
$$

we infer that

$$
\begin{aligned}
(\Phi^*B_i)(h,g) &= B_i(g) - \tfrac{1}{4\pi}\,\mathrm{tr}\Big(h^{-1}(\mathrm{d}h)\big(g^{-1}(\mathrm{d}g) + (\mathrm{d}g)g^{-1}\big)\Big)\\
&\quad - \tfrac{1}{4\pi}\,\mathrm{tr}\Big(\gamma^{-1}h^{-1}\mathrm{d}(h\gamma)\,\mathrm{e}^{2\pi i\tau}\,\gamma^{-1}h^{-1}\mathrm{d}(h\gamma)\,\mathrm{e}^{-2\pi i\tau}\Big)\\
&\quad - \tfrac{1}{4\pi}\,\mathrm{tr}\Big(g\,h^{-1}(\mathrm{d}h)g^{-1}h^{-1}(\mathrm{d}h)\Big) - d\Pi_i(h,g)\\
&= B_i(g) + \tfrac{1}{4\pi}\beta(h,g) - \mathrm{d}\Pi_i(h,g)
\end{aligned}
\tag{B.5}
$$

which proves (6.4). Applying the pullback by $(\mathrm{Id}, K)$ to (6.4) for index $i$ replaced by $(2m-i)$ and using identity (6.32) and the relation $\widehat{\Pi}_i = (\mathrm{Id}, K)^*\Pi_{2m-i}$, one obtains Eq. (6.59).

## Appendix C. Relation to Rudner et al.'s formulation

Ref. [44] used a somewhat different periodization of the evolution operator $U(t,k)$ defined by

$$
M_\epsilon(t,k) = \begin{cases} U(2t,k) & \text{for } 0 \le t \le T/2\,,\\ \mathrm{e}^{\mathrm{i}H_\epsilon^{\mathrm{eff}}(k)(2T-2t)} & \text{for } T/2 \le t \le T\,, \end{cases}
\tag{C.1}
$$

whereas in the present paper we worked with

$$
V_\epsilon(t,k) = U(t,k)\,\mathrm{e}^{\mathrm{i}\,tH_\epsilon^{\mathrm{eff}}(k)}\,.
\tag{C.2}
$$

Let us show that the two choices are homotopic so that $W_\epsilon[U] = \deg(M_\epsilon) = \deg(V_\epsilon)$. Consider the map

$$N_\epsilon(s,t,k) = \begin{cases} U(2t,k)\,\mathrm{e}^{2\mathrm{i}st H^{\mathrm{eff}}_\epsilon(k)} & \text{for } 0 \le t \le T/2\,, \\ \mathrm{e}^{-2\mathrm{i}(1-s)(T-t)H^{\mathrm{eff}}_\epsilon(k)} & \text{for } T/2 \le t \le T \end{cases} \tag{C.3}$$

which is well defined continuous and periodic in $t$ for all $s$ because the two determinations agree for $t = T/2$ and $N(s,0,k) = I = N(s,T,k)$. One has

$$N_\epsilon(0,t,k) = M_\epsilon(t,k) \qquad \text{and} \qquad N_\epsilon(1,t,k) = \begin{cases} V_\epsilon(2t,k) & \text{for } 0 \le t \le T/2 \\ 1 & \text{for } T/2 \le t \le T\,, \end{cases} \tag{C.4}$$

and the last map is clearly homotopic with $V_\epsilon$.

## Appendix D. A simple case where spin is conserved

In Ref. [5], we mentioned the somewhat artificial but enlightening case with conserved spin, where both the evolution operator $U$ and its periodized version $V_\epsilon$ are block-diagonal in the $(\uparrow,\downarrow)$ basis, the two blocks being related by time reversal. In this case, the $K$ index can be related to the $W$ index from [44] of one of the spin blocks, namely

$$K_\epsilon\left[\begin{pmatrix} U_\uparrow & 0 \\ 0 & U_\downarrow \end{pmatrix}\right] = \frac{W_\epsilon[U_\uparrow] - W_\epsilon[U_\downarrow]}{2} \quad \text{mod } 2\,, \tag{D.1}$$

where $W_\epsilon[U_\uparrow] = W_\epsilon[\Theta U_\downarrow \Theta^{-1}] = -W_\epsilon[U_\downarrow]$.

Let us provide a demonstration of the preceding equality. Consider the periodized evolution operator $V_\epsilon$ it its block-diagonal form

$$V_\epsilon = \begin{pmatrix} V_{\epsilon,\uparrow} & 0 \\ 0 & V_{\epsilon,\downarrow} \end{pmatrix} \equiv V_{\epsilon,\uparrow} \oplus V_{\epsilon,\downarrow}. \tag{D.2}$$

Time-reversal invariance of $V_\epsilon$ gives

$$V_{\epsilon,\downarrow} \circ \vartheta_3 = \overline{V_{\epsilon,\uparrow}} \tag{D.3}$$

where $\vartheta_3$ is an orientation-reversing diffeomorphism of $T^3$ induced by the map $(t,k) \mapsto (T-t,-k)$. One may choose to define

$$\widehat{V}_\epsilon(t,k) \equiv \widehat{V}_{\epsilon,\uparrow}(t,k) \oplus \widehat{V}_{\epsilon,\downarrow}(t,k) = \begin{cases} V_{\epsilon,\uparrow}(t,k) \oplus V_{\epsilon,\downarrow}(t,k) & \text{for } 0 \le t \le T/2 \\ V_{\epsilon,\uparrow}(t,k) \oplus \overline{V_{\epsilon,\uparrow}(t,k)} & \text{for } T/2 \le t \le T \end{cases} \tag{D.4}$$

As the degree factorizes over the blocks,

$$\deg(\widehat{V}_\epsilon) = \deg(\widehat{V}_{\epsilon,\uparrow}) + \deg(\widehat{V}_{\epsilon,\downarrow}). \tag{D.5}$$

Whilst the former term is simply $\deg(V_{\epsilon,\uparrow}) = W_\epsilon[U_\uparrow]$, the latter vanishes from time-reversal invariance (D.3), hence the result.

Indeed, one could choose another $\widehat{V}_\epsilon$ to show that $K_\epsilon[U]$ is also $-W_\epsilon[U_\downarrow]$, or use the equality $0 = W_\epsilon[U] = W_\epsilon[U_\uparrow] + W_\epsilon[U_\downarrow]$ (see Eq. (2.42)).

## References

[1] V.V. Albert, L.I. Glazman, L. Jiang, Topological properties of linear circuit lattices, arXiv:1410.1243, 2014.
[2] M. Bellec, U. Kuhl, G. Montambaux, F. Mortessagne, Manipulation of edge states in microwave artificial graphene, New J. Phys. 16 (11) (2014) 113023.
[3] R. Bott, J.R. Seeley, Some remarks on the paper of Callias: "Axial anomalies and index theorems on open spaces", Commun. Math. Phys. 62 (1978) 235–245.

[4] R. Bott, L.W. Tu, Differential Forms in Algebraic Topology, 3rd edition, Springer, 1982.
[5] D. Carpentier, P. Delplace, M. Fruchart, K. Gawędzki, Topological index for periodically driven time-reversal invariant 2d systems, Phys. Rev. Lett. 114 (2015) 106806.
[6] B.A. Dubrovin, A.T. Fomenko, S.P. Novikov, Modern Geometry – Methods and Applications: Part II: The Geometry and Topology of Manifolds, Springer, 1985.
[7] G. De Nittis, K. Gomi, Classification of "Quaternionic" Bloch-bundles: topological quantum systems of type AII, arXiv:1404.5804, 2014.
[8] M. Fruchart, D. Carpentier, An introduction to topological insulators, C. R. Phys. 14 (2013) 779–815.
[9] L. Fu, C.L. Kane, Time reversal polarization and a $Z_2$ adiabatic spin pump, Phys. Rev. B 74 (2006) 195312.
[10] L. Fu, C.L. Kane, Topological insulators with inversion symmetry, Phys. Rev. B 76 (2007) 045302.
[11] D.S. Freed, G.W. Moore, Twisted equivariant matter, Ann. Henri Poincaré 14 (2013) 1927–2023.
[12] D. Fiorenza, D. Monaco, G. Panati, $Z_2$ invariants of topological insulators as geometric obstructions, arXiv:1408.1030, 2014.
[13] R.H. Fox, Homotopy groups and torus homotopy groups, Ann. Math. 49 (1948) 471–510.
[14] K. Gawędzki, Topological actions in two-dimensional quantum field theory, in: G. 't Hooft, A. Jaffe, G. Mack, P. Mitter, R. Stora (Eds.), Nonperturbative Quantum Field Theory, Plenum Press, 1988, pp. 101–141.
[15] K. Gawędzki, Conformal field theory: a case study, in: Y. Nutku, C. Saclioglu, T. Turgut (Eds.), New Non-Perturbative Methods in String and Field Theory, Perseus Publishing, 2000, pp. 1–55.
[16] K. Gawędzki, N. Reis, WZW branes and gerbes, Rev. Math. Phys. 14 (2002) 1281–1334.
[17] K. Gawędzki, R.R. Suszek, K. Waldorf, Bundle gerbes for orientifold sigma models, Adv. Theor. Math. Phys. 15 (2011) 621–688.
[18] M. Hafezi, E.A. Demler, M.D. Lukin, J.M. Taylor, Robust optical delay lines with topological protection, Nat. Phys. 7 (2011) 907–912.
[19] Nigel J. Hitchin, What is ... a gerbe? Not. Am. Math. Soc. 50 (2003) 218–219.
[20] M. Hafezi, S. Mittal, J. Fan, A. Migdall, J.M. Taylor, Imaging topological edge states in silicon photonics, Nat. Photonics 7 (2013) 1001–1005.
[21] W. Hu, J.C. Pillay, K. Wu, M. Pasek, P.P. Shum, Y.D. Chong, Measurement of a topological edge invariant in a microwave network, Phys. Rev. X 5 (1) (2015) 011012.
[22] F.D.M. Haldane, S. Raghu, Possible realization of directional optical waveguides in photonic crystals with broken time-reversal symmetry, Phys. Rev. Lett. 100 (2008) 013904.
[23] D. Husemoller, Fibre Bundles, 3rd edition, Springer, 1993.
[24] J.-I. Inoue, A. Tanaka, Photoinduced transition between conventional and topological insulators in two-dimensional electronic systems, Phys. Rev. Lett. 105 (1) (2010) 017401.
[25] G. Jotzu, M. Messer, R. Desbuquois, M. Lebrat, T. Uehlinger, D. Greif, T. Esslinger, Experimental realization of the topological Haldane model with ultracold fermions, Nature 515 (2014) 237–240.
[26] N. Jia, A. Sommer, D. Schuster, J. Simon, Time reversal invariant topologically insulating circuits, arXiv:1309.0878, 2013.
[27] T. Kitagawa, M.A. Broome, A. Fedrizzi, M.S. Rudner, E. Berg, I. Kassal, A. Aspuru-Guzik, E. Demler, A.G. White, Observation of topologically protected bound states in photonic quantum walks, Nat. Commun. 3 (2012) 882.
[28] T. Kitagawa, E. Berg, M. Rudner, E. Demler, Topological characterization of periodically driven quantum systems, Phys. Rev. B 82 (23) (2010) 235114.
[29] K.v. Klitzing, G. Dorda, M. Pepper, New method for high-accuracy determination of the fine-structure constant based on quantized Hall resistance, Phys. Rev. Lett. 45 (1980) 494–497.
[30] A. Kitaev, Periodic table for topological insulators and superconductors, AIP Conf. Proc. 1134 (1) (2009) 22–30.
[31] C.L. Kane, T.C. Lubensky, Topological boundary modes in isostatic lattices, Nat. Phys. 10 (2014) 39–45.
[32] C.L. Kane, E.J. Mele, $Z_2$ topological order and the quantum spin Hall effect, Phys. Rev. Lett. 95 (2005) 146802.
[33] T. Kitagawa, T. Oka, A. Brataas, L. Fu, E. Demler, Transport properties of nonequilibrium systems under the application of light: photoinduced quantum Hall insulators without Landau levels, Phys. Rev. B 84 (23) (2011) 235108.
[34] S.-S. Lee, S. Ryu, Many-body generalization of the $Z_2$ topological invariant for the quantum spin Hall effect, Phys. Rev. Lett. 100 (2008) 186807.
[35] N.H. Lindner, G. Refael, V. Galitski, Floquet topological insulator in semiconductor quantum wells, Nat. Phys. 7 (2011) 490–495.
[36] J.E. Moore, L. Balents, Topological invariants of time-reversal-invariant band structures, Phys. Rev. B 75 (2007) 121–306.
[37] E. Meinrenken, The basic gerbe over a compact simple Lie group, Enseign. Math. 49 (2003) 307–333.
[38] J.W. Milnor, Topology from the Differentiable Viewpoint, The University of Virginia Press, 1965.

[39] N. Manton, P. Sutcliffe, Topological Solitons, Cambridge University Press, 2004.
[40] T. Oka, H. Aoki, Photovoltaic Hall effect in graphene, Phys. Rev. B 79 (8) (2009) 081406.
[41] G. Panati, Triviality of Bloch and Bloch–Dirac bundles, Ann. Henri Poincaré 8 (2007) 995–1011.
[42] A. Polyakov, P.B. Wiegmann, Theory of nonabelian Goldstone bosons in two dimensions, Phys. Lett. B 131 (1983) 121–126.
[43] X.-L. Qi, T.L. Hughes, S.-C. Zhang, Topological field theory of time-reversal invariant insulators, Phys. Rev. B 78 (2008) 195424.
[44] M.S. Rudner, N.H. Lindner, E. Berg, M. Levin, Anomalous edge states and the bulk-edge correspondence for periodically driven two-dimensional systems, Phys. Rev. X 3 (2013) 031005.
[45] S. Ryu, A.P. Schnyder, A. Furusaki, A.W.W. Ludwig, Topological insulators and superconductors: tenfold way and dimensional hierarchy, New J. Phys. 12 (6) (2010) 065010.
[46] M.C. Rechtsman, J.M. Zeuner, Y. Plotnik, Y. Lumer, D. Podolsky, F. Dreisow, S. Nolte, M. Segev, A. Szameit, Photonic Floquet topological insulators, Nature 496 (April 2013) 196–200.
[47] A.P. Schnyder, S. Ryu, A. Furusaki, A.W.W. Ludwig, Classification of topological insulators and superconductors, AIP Conf. Proc. 1134 (1) (2009) 10–21.
[48] U. Schreiber, C. Sweigert, K. Waldorf, Unoriented WZW models and holonomy of bundle gerbes, Commun. Math. Phys. 274 (2007) 31–64.
[49] E. Witten, Non-abelian bosonization in two dimensions, Commun. Math. Phys. 92 (1984) 455–472.